\documentclass[showpacs,aps,prd]{revtex4}
\input epsf

\begin{document}
\title{Heavy baryon spectroscopy in QCD}
\author{Jian-Rong Zhang and Ming-Qiu Huang}
\affiliation{Department of Physics, National University of Defense
Technology, Hunan 410073, China}

\begin{abstract} We perform a systematic study of the masses of
charmed and bottom baryons in the framework of the QCD sum rule
approach. Contributions of the operators up to dimension six are
included in operator product expansion. The resulting heavy baryon
masses from the calculations are well consistent with the
experimental values, and predictions to the spectroscopy of the
unobserved bottom baryons are also presented.

\end{abstract}
\maketitle

\section{Introduction}\label{sec1}
  During the past several years there has been tremendous progress in the
experimental investigations of the heavy baryon spectroscopy. With
the precise measurement for the mass of $\Lambda_{b}$ by the CDF
collaboration \cite{lamdar-b}, the D0 collaboration proclaimed the
observation of $\Xi_{b}$ \cite{ksi-b}, which was quickly confirmed
by the CDF collaboration \cite{ksi-b1}. The first observations of
$\Sigma_{b}$ and $\Sigma_{b}^{*}$ have been reported by the CDF
\cite{sigma-b}. The BABAR collaboration announced the observation of
$\Omega_{c}^{*}$ \cite{omega-c star} and  the production of
$\Omega_{c}$ from B decays \cite{omega-c}. The $\Xi_{c}$ as well as
the excited states of $\Xi_{c}$ were set forth by the Belle and the
BABAR collaborations \cite{ksi-c}. Therefore, a large amount of
experimental data on charmed and bottom baryons has become
available.

On the other hand, various theoretical models have been used to
study heavy baryon masses, such as quark models \cite{quark
model,quark model 1}, mass formulas \cite{mass formular}, and
lattice QCD calculations \cite{lattice}. From QCD sum rules
\cite{svzsum}, masses of the heavy baryons were primarily calculated
in the heavy-quark limit \cite{evs}, and subsequently in the heavy
quark effective theory \cite{alfa,HQET,mqhuang}. In Refs.
\cite{EBagan}, the calculations for the heavy baryons began with the
full theory. Recently the masses of $\Xi_{Q}$ and $\Omega_{Q}^{(*)}$
were tested in QCD sum rules \cite{ksi,zhigang}. Other renewed works
were inspired by the current significant observations continually
\cite{new,zhu,jrzhang}. Proceeding from the motivation to evaluate
the spectroscopy of the heavy baryons systematically in full QCD, we
shall study mass sum rules for the heavy baryons, with the technique
developed in \cite{kim,bracco}. For the cases of $\Xi_{Q}$ and
$\Omega_{Q}^{(*)}$, the interpolating currents utilized in this work
are not entirely the same as the ones used in Refs
\cite{ksi,zhigang}.

The paper is organized as follows. In Sec \ref{sec2}, QCD sum rules
for the heavy baryons are derived. Section \ref{sec3} contains
numerical analysis, a brief summary, and some discussions.
\section{ QCD sum rules for the heavy baryons}\label{sec2}
It is of interest to apply the QCD sum rule to study the heavy
baryons composed of a heavy quark ($Q=c,b$) and two light ($u$, $d$,
or $s$) quarks. The basic point is to choose the suitable
interpolating current.  For the ground states, the currents are
correlated with the spin-parity quantum numbers
$J_{\ell}^{P_{\ell}}=0^{+}$ and $J_{\ell}^{P_{\ell}}=1^{+}$ for the
light diquark system, along with the heavy quark forming the state
with $J^{P}=\frac{1}{2}^{+}$ and the pair of degenerate states with
$J^{P}=\frac{1}{2}^{+}$ and $J^{P}=\frac{3}{2}^{+}$, which may
determine the choice of $\Gamma_{k}$ and $\Gamma_{k}^{'}$ matrices
in baryonic currents \cite{evs}. For the $\Gamma_{k}^{'}$ matrices
for the excited baryons with $I(J^{P})=0(\frac{1}{2}^{-})$, it might
be referred to the current of the heavy-light vector meson
\cite{reinders}. For the baryons with $J=\frac{3}{2}$, the currents
may be gained from those of baryons with $J=\frac{1}{2}$ using SU(3)
symmetry relations \cite{Ioffe}. Thus, the following forms of
currents are adopted \cite{evs,EBagan,reinders,Ioffe,cohen}:
\begin{eqnarray}
j_{\Lambda_{1Q}}&=&\varepsilon_{abc}(q_{1a}^{T}C\Gamma_{k}q_{2b})\Gamma_{k}^{'}Q_{c},\nonumber\\
j_{\Lambda_{1Q}^{*}}&=&\varepsilon_{abc}[\frac{2}{\sqrt{3}}(q_{1a}^{T}C\Gamma_{k}Q_{b})\Gamma_{k}^{'}q_{2c}
+\frac{1}{\sqrt{3}}(q_{1a}^{T}C\Gamma_{k}q_{2b})\Gamma_{k}^{'}Q_{c}],\nonumber\\
j_{\Xi_{Q}}&=&\varepsilon_{abc}(q_{1a}^{T}C\Gamma_{k}s_{b})\Gamma_{k}^{'}Q_{c},\nonumber\\
j_{\Xi_{Q}^{'}}&=&\varepsilon_{abc}(q_{1a}^{T}C\Gamma_{k}s_{b})\Gamma_{k}^{'}Q_{c},\nonumber\\
j_{\Xi_{Q}^{'*}}&=&\varepsilon_{abc}[\frac{2}{\sqrt{3}}(q_{1a}^{T}C\Gamma_{k}Q_{b})\Gamma_{k}^{'}s_{c}
+\frac{1}{\sqrt{3}}(q_{1a}^{T}C\Gamma_{k}s_{b})\Gamma_{k}^{'}Q_{c}],\\
j_{\Xi_{1Q}}&=&\varepsilon_{abc}(q_{1a}^{T}C\Gamma_{k}s_{b})\Gamma_{k}^{'}Q_{c},\nonumber\\
j_{\Xi_{1Q}^{*}}&=&\varepsilon_{abc}[\frac{2}{\sqrt{3}}(q_{1a}^{T}C\Gamma_{k}Q_{b})\Gamma_{k}^{'}s_{c}
+\frac{1}{\sqrt{3}}(q_{1a}^{T}C\Gamma_{k}s_{b})\Gamma_{k}^{'}Q_{c}],\nonumber\\
j_{\Omega_{Q}}&=&\varepsilon_{abc}(s^{T}_{a}C\Gamma_{k}s_{b})\Gamma_{k}^{'}Q_{c},\nonumber\\
j_{\Omega_{Q}^{*}}&=&\varepsilon_{abc}[\frac{2}{\sqrt{3}}(s^{T}_{a}C\Gamma_{k}Q_{b})\Gamma_{k}^{'}s_{c}
+\frac{1}{\sqrt{3}}(s^{T}_{a}C\Gamma_{k}s_{b})\Gamma_{k}^{'}Q_{c}].\nonumber
\end{eqnarray}
Here the index $T$ means matrix transposition, $C$ is the charge
conjugation matrix, $a$, $b$, $c$ are color indices, and $q_{1}$,
$q_{2}$ are $u$ or $d$ depending on the concrete quark contents of
the corresponding heavy baryons. The choice of $\Gamma_{k}$ and
$\Gamma_{k}^{'}$ matrices  are shown in TABLE \ref{table:1}.
\begin{table}[htb!]\caption{The choice of $\Gamma_{k}$ and $\Gamma_{k}^{'}$ matrices in baryonic currents}
 \centerline{\begin{tabular}{ p{3.6cm} p{1.8cm} p{1.8cm} p{1.8cm} p{1.8cm} p{1.8cm} p{1.8cm}} \hline\hline
State (quark content)                          &$J^{P}$               &  $S_{\ell}$  &  $L_{\ell}$  &  $J_{\ell}^{P_{\ell}}$   &   $\Gamma_{k}$     &     $\Gamma_{k}^{'}$           \\
\hline
$\Xi_{Q}(qsQ)$                                 &$\frac{1}{2}^{+} $    &      0       &      0       &        $0^{+}$           &   $\gamma_{5}$     &     $1$                        \\
$\Omega_{Q}(ssQ)$,~~$\Xi_{Q}^{'}(qsQ)$         &$\frac{1}{2}^{+} $    &      1       &      0       &        $1^{+}$           &   $\gamma_{\mu}$   &     $\gamma_{\mu}\gamma_{5}$   \\
$\Omega_{Q}^{*}(ssQ)$,~~$\Xi_{Q}^{'*}(qsQ)$    &$\frac{3}{2}^{+} $    &      1       &      0       &        $1^{+}$           &   $\gamma_{\mu}$   &     $\gamma_{\mu}\gamma_{5}$   \\
$\Lambda_{1Q}(qqQ)$,~~$\Xi_{1Q}(qsQ)$          &$\frac{1}{2}^{-}$     &      0       &      1       &        $1^{-}$           &   $\gamma_{5}$     &     $\gamma_{\mu}$             \\
$\Lambda_{1Q}^{*}(qqQ)$,~~$\Xi_{1Q}^{*}(qsQ)$  &$\frac{3}{2}^{-}$     &      0       &      1       &        $1^{-}$           &   $\gamma_{5}$     &     $\gamma_{\mu}$             \\
\hline\hline
\end{tabular}}
\label{table:1}
\end{table}

The QCD sum rules for the heavy baryons are constructed from the
two-point correlation function
\begin{eqnarray}\label{correlator}
\Pi(q^{2})=i\int
d^{4}x\mbox{e}^{iq.x}\langle0|T[j(x)\overline{j}(0)]|0\rangle.
\end{eqnarray}
Lorentz covariance implies that the two-point correlation function
has the form
\begin{eqnarray}
\Pi(q^{2})=\Pi_{1}(q^{2})+\rlap/q\Pi_{2}(q^{2}).
\end{eqnarray}
For each invariant function $\Pi_{1}$ and $\Pi_{2}$, a sum rule can
be obtained, which is shown below. Phenomenologically, the
correlator can be expressed as a dispersion integral over a physical
spectral function
\begin{eqnarray}
\Pi(q^{2})=\lambda^{2}_H\frac{\rlap/q+M_{H}}{M_{H}^{2}-q^{2}}+\frac{1}{\pi}\int_{s_{0}}
^{\infty}ds\frac{\mbox{Im}\Pi^{\mbox{phen}}(s)}{s-q^{2}}+\mbox{subtractions},
\end{eqnarray}
where $M_{H}$ denotes the mass of the heavy baryon. In obtaining the
above expression, the Dirac and Rarita-Schwinger spinor sums
\begin{eqnarray}
\sum_{s}N(q,s)\bar{N}(q,s)=\rlap/q+M_{H},
\end{eqnarray}
for spin-$\frac{1}{2}$ baryon, and
\begin{eqnarray}
\sum_{s}N_{\mu}(q,s)\bar{N}_{\nu}(q,s)=(\rlap/q+M_{H})(g_{\mu\nu}-\frac{1}{3}\gamma_{\mu}\gamma_{\nu}+\frac{q_{\mu}\gamma_{\nu}-q_{\nu}\gamma_{\mu}}{3M_{H}}-\frac{2q_{\mu}q_{\nu}}{3M_{H}^{2}}),
\end{eqnarray}
for spin-$\frac{3}{2}$ baryon have been used. In the operator
product expansion (OPE) side, short-distance effects are given by
Wilson coefficients, while long-distance confinement effects are
attributed to power corrections and parameterized in terms of vacuum
condensates. Hence
\begin{eqnarray}\label{ope}
\Pi_{i}(q^{2})&=&\Pi_{i}^{\mbox{pert}}(q^{2})+\Pi_{i}^{\mbox{cond}}(q^{2}),
 i=1,2.
\end{eqnarray}
We work at leading order in $\alpha_{s}$ and consider condensates up
to dimension six.  The strange quark is dealt as a light one and the
diagrams  are considered up to order $m_{s}$. To keep the
heavy-quark mass finite, the momentum-space expression for the
heavy-quark propagator is used. We follow Refs. \cite{kim,bracco}
and calculate the light-quark part of the correlation function in
the coordinate space, which is then Fourier-transformed to the
momentum space in $D$ dimension. The resulting light-quark part is
combined with the heavy-quark part before it is dimensionally
regularized at $D=4$. For the heavy-quark propagator with two and
three gluons attached, the momentum-space expressions given in Ref.
\cite{reinders} are used. With Eq. (\ref{ope}), the correlation
function in the OPE side in terms of a dispersion relation  can be
written as
\begin{eqnarray}
\Pi_{i}(q^{2})=\int_{m_{Q}^{2}}^{\infty}ds\frac{\rho_{i}(s)}{s-q^{2}}+\Pi_{i}^{\mbox{cond}}(q^{2}),
\end{eqnarray}
where the spectral density is given by the imaginary part of the
correlation function
\begin{eqnarray}
\rho_{i}(s)=\frac{1}{\pi}\mbox{Im}\Pi_{i}^{\mbox{OPE}}(s).
\end{eqnarray}
After equating the two expressions for $\Pi(q^{2})$, assuming
quark-hadron duality, and making a Borel transform, the sum rules
can be written as
\begin{eqnarray}\label{sumrule1}
\lambda_{H}^{2}M_{H}e^{-M_{H}^{2}/M^{2}}&=&\int_{m_{Q}^{2}}^{s_{0}}ds\rho_{1}(s)e^{-s/M^{2}}+\hat{B}\Pi_{1}^{\mbox{cond}},
\end{eqnarray}
\begin{eqnarray}\label{sumrule2}
\lambda_{H}^{2}e^{-M_{H}^{2}/M^{2}}&=&\int_{m_{Q}^{2}}^{s_{0}}ds\rho_{2}(s)e^{-s/M^{2}}+\hat{B}\Pi_{2}^{\mbox{cond}}.
\end{eqnarray}
To eliminate the baryon coupling constant $\lambda_H$ and extract
the $M_{H}$, first take the derivative of Eq. (\ref{sumrule1}) with
respect to $-\frac{1}{M^{2}}$, divide the result by Eq.
(\ref{sumrule1}) itself, and similarly deal with Eq.
(\ref{sumrule2}) to yield
\begin{eqnarray}\label{sum rule m}
M_{H}^{2}&=&\{\int_{m_{Q}^{2}}^{s_{0}}ds\rho_{1}(s)s
e^{-s/M^{2}}+d/d(-\frac{1}{M^{2}})\hat{B}\Pi_{1}^{\mbox{cond}}(s)\}/
\{\int_{m_{Q}^{2}}^{s_{0}}ds\rho_{1}(s)e^{-s/M^{2}}
+\hat{B}\Pi_{1}^{\mbox{cond}}(s)\},
\end{eqnarray}
\begin{eqnarray}\label{sum rule q}
M_{H}^{2}&=&\{\int_{m_{Q}^{2}}^{s_{0}}ds\rho_{2}(s)s
e^{-s/M^{2}}+d/d(-\frac{1}{M^{2}})\hat{B}\Pi_{2}^{\mbox{cond}}(s)\}/
\{\int_{m_{Q}^{2}}^{s_{0}}ds\rho_{2}(s)e^{-s/M^{2}}
+\hat{B}\Pi_{2}^{\mbox{cond}}(s)\},
\end{eqnarray}
where
\begin{eqnarray}
\rho_{i}(s)=\rho_{i}^{\mbox{pert}}(s)+\rho_{i}^{\langle\bar{q}q\rangle}(s)+\rho_{i}^{\langle\bar{s}s\rangle}(s)+\rho_{i}^{\langle
G^{2}\rangle}(s),
\end{eqnarray}
with
\begin{eqnarray}
\rho_{1}^{\mbox{pert}}(s)&=&\frac{3}{2^{5}\pi^{4}}m_{Q}\int^{1}_{\Lambda}d\alpha(\frac{1-\alpha}{\alpha})^{2}(m_{Q}^{2}-s\alpha)^{2},\\
\rho_{1}^{\langle G^{2}\rangle}(s)&=&\frac{\langle
g^{2}G^{2}\rangle}{2^{7}\pi^{4}}m_{Q}\int_{\Lambda}^{1}d\alpha[(\frac{1-\alpha}{\alpha})^{2}+2],\\
\hat{B}\Pi_{1}^{\mbox{cond}}&=&
-\frac{\langle g^{2}G^{2}\rangle}{3\cdot2^{7}\pi^{4}}m_{Q}^{3}\int^{1}_{0}d\alpha\frac{(1-\alpha)^{2}}{\alpha^{3}}e^{-m_{Q}^{2}/(\alpha M^{2})}\nonumber\\
&
&{}+\frac{2\langle\bar{q}q\rangle^{2}}{3}m_{Q}e^{-m_{Q}^{2}/M^{2}}\nonumber\\
& & {}-\frac{\langle g^{3}G^{3}\rangle}{3\cdot2^{8}\pi^{4}}m_{Q}
\int_{0}^{1}
d\alpha\frac{(1-\alpha)^{2}}{\alpha^{3}}(3-\frac{m_{Q}^{2}}{\alpha
M^{2}})e^{-m_{Q}^{2}/(\alpha M^{2})},\\
\rho_{2}^{\mbox{pert}}(s)&=&-\frac{3}{2^{6}\pi^{4}}\int^{1}_{\Lambda}d\alpha\frac{(1-\alpha)^{2}}{\alpha}(m_{Q}^{2}-s\alpha)^{2},\\
\rho_{2}^{\langle G^{2}\rangle}(s)&=&-\frac{\langle
g^{2}G^{2}\rangle}{2^{8}\pi^{4}}[1-(\frac{m_{Q}^{2}}{s})^{2}],\\
\hat{B}\Pi_{2}^{\mbox{cond}}&=&
\frac{\langle g^{2}G^{2}\rangle}{3\cdot2^{8}\pi^{4}}m_{Q}^{2}\int^{1}_{0}d\alpha(\frac{1-\alpha}{\alpha})^{2}e^{-m_{Q}^{2}/(\alpha M^{2})}\nonumber\\
&
&{}-\frac{\langle\bar{q}q\rangle^{2}}{3}e^{-m_{Q}^{2}/M^{2}}\nonumber\\
& & {}+\frac{\langle g^{3}G^{3}\rangle}{3\cdot2^{10}\pi^{4}}
\int_{0}^{1}d\alpha(\frac{1-\alpha}{\alpha})^{2}(1-\frac{2m_{Q}^{2}}{\alpha
M^{2}})e^{-m_{Q}^{2}/(\alpha M^{2})},
\end{eqnarray}
for $\Lambda_{1Q}$ baryons,
\begin{eqnarray}
\rho_{1}^{\mbox{pert}}(s)&=&\frac{1}{2^{5}\pi^{4}}m_{Q}\int^{1}_{\Lambda}d\alpha(\frac{1-\alpha}{\alpha})^{2}(m_{Q}^{2}-s\alpha)^{2},\\
\rho_{1}^{\langle\bar{q}q\rangle}(s)&=&\frac{2}{3\pi^{2}}\langle\bar{q}q\rangle
\int_{\Lambda}^{1}d\alpha(m_{Q}^{2}-s\alpha),\\
\rho_{1}^{\langle G^{2}\rangle}(s)&=&\frac{\langle
g^{2}G^{2}\rangle}{3\cdot2^{7}\pi^{4}}m_{Q}\int_{\Lambda}^{1}d\alpha[(\frac{1-\alpha}{\alpha})^{2}+2],\\
\hat{B}\Pi_{1}^{\mbox{cond}}&=&
-\frac{\langle g^{2}G^{2}\rangle}{3^{2}\cdot2^{7}\pi^{4}}m_{Q}^{3}\int^{1}_{0}d\alpha\frac{(1-\alpha)^{2}}{\alpha^{3}}e^{-m_{Q}^{2}/(\alpha M^{2})}\nonumber\\
&
&{}+\frac{5\cdot2\langle\bar{q}q\rangle^{2}}{3^{2}}m_{Q}e^{-m_{Q}^{2}/M^{2}}\nonumber\\
& & {}-\frac{\langle g^{3}G^{3}\rangle}{3^{2}\cdot2^{8}\pi^{4}}m_{Q}
\int_{0}^{1}
d\alpha\frac{(1-\alpha)^{2}}{\alpha^{3}}(3-\frac{m_{Q}^{2}}{\alpha
M^{2}})e^{-m_{Q}^{2}/(\alpha M^{2})},\\
\rho_{2}^{\mbox{pert}}(s)&=&\frac{1}{2^{6}\pi^{4}}\int_{\Lambda}^{1}d\alpha\frac{(1-\alpha)^{2}(1-3\alpha)}{\alpha^{2}}(m_{Q}^{2}-s\alpha)^{2},\\
\rho_{2}^{\langle\bar{q}q\rangle}(s)&=&\frac{\langle\bar{q}q\rangle}{3\cdot2\pi^{2}}
m_{Q}(1-\frac{m_{Q}^{2}}{s})^{2},\\
\rho_{2}^{\langle G^{2}\rangle}(s)&=&\frac{\langle
g^{2}G^{2}\rangle}{3\cdot2^{7}\pi^{4}}[-\frac{1}{2}+\frac{(m_{Q}^{2}/s)^2}{2}+\int_{\Lambda}^{1}d\alpha(\alpha-1)(m_{Q}^{2}-s\alpha)],\\
\hat{B}\Pi_{2}^{\mbox{cond}}&=&-\frac{\langle
g^{2}G^{2}\rangle}{3^{2}\cdot2^{8}\pi^{4}}m_{Q}^{2}\int_{0}^{1}d\alpha\frac{(1-\alpha)^{2}(1-3\alpha)}{\alpha^{3}}e^{-m_{Q}^{2}/(\alpha
M^{2})}\nonumber\\& & {}-\frac{\langle g\bar{q}\sigma\cdot G
q\rangle}{3\cdot2^{2}\pi^{2}}m_{Q}\int_{0}^{1}d\alpha
e^{-m_{Q}^{2}/(\alpha
M^{2})}\nonumber\\
&
&{}-\frac{\langle\bar{q}q\rangle^{2}}{3^{2}}e^{-m_{Q}^{2}/M^{2}}\nonumber\\&
& {} -\frac{\langle
g^{3}G^{3}\rangle}{3^{2}\cdot2^{10}\pi^{4}}\int_{0}^{1}
d\alpha\frac{1-\alpha}{\alpha^{4}}[\alpha(1-4\alpha+9\alpha^{2})
-2(1-4\alpha+5\alpha^{2})\frac{m_{Q}^{2}}{M^{2}}]e^{-m_{Q}^{2}/(\alpha
M^{2})},
\end{eqnarray}
for $\Lambda_{1Q}^{*}$ baryons,
\begin{eqnarray}
\rho_{1}^{\mbox{pert}}(s)&=&\frac{3}{2^{7}\pi^{4}}m_{Q}\int_{\Lambda}^{1}d\alpha(\frac{1-\alpha}{\alpha})^{2}(m_{Q}^{2}-s\alpha)^{2},\\
\rho_{1}^{\langle\bar{q}q\rangle}(s)&=&-\frac{\langle\bar{q}q\rangle}{2^{3}\pi^{2}}m_{s}m_{Q}(1-\frac{m_{Q}^{2}}{s}),\\
\rho_{1}^{\langle\bar{s}s\rangle}(s)&=&\frac{\langle\bar{s}s\rangle}{2^{4}\pi^{2}}m_{s}m_{Q}(1-\frac{m_{Q}^{2}}{s}),\\
\rho_{1}^{\langle G^{2}\rangle}(s)&=&\frac{\langle
g^{2}G^{2}\rangle}{2^{9}\pi^{4}}m_{Q}\int_{\Lambda}^{1}d\alpha[(\frac{1-\alpha}{\alpha})^{2}+2],\\
\hat{B}\Pi_{1}^{\mbox{cond}}&=&-\frac{\langle
g^{2}G^{2}\rangle}{3\cdot2^{9}\pi^{4}}m_{Q}^{3}\int_{0}^{1}d\alpha\frac{(1-\alpha)^{2}}{\alpha^{3}}e^{-m_{Q}^{2}/(\alpha
M^{2})}\nonumber\\
& &{}+\frac{\langle g\bar{q}\sigma\cdot G
q\rangle}{2^{5}\pi^{2}}m_{s}m_{Q}e^{-m_{Q}^{2}/M^{2}}\nonumber\\&&{}+\frac{\langle
g\bar{s}\sigma\cdot G
s\rangle}{3\cdot2^{4}\pi^{2}}m_{s}m_{Q}e^{-m_{Q}^{2}/M^{2}}\nonumber\\
& &{}
+\frac{\langle\bar{q}q\rangle\langle\bar{s}s\rangle}{6}m_{Q}e^{-m_{Q}^{2}/M^{2}}\nonumber\\&&{}-\frac{\langle
g^{3}G^{3}\rangle}{3\cdot2^{10}\pi^{4}}m_{Q}\int_{0}^{1}
d\alpha\frac{(1-\alpha)^{2}}{\alpha^{3}}(3-\frac{m_{Q}^{2}}{\alpha
M^{2}})e^{-m_{Q}^{2}/(\alpha M^{2})},\\
\rho_{2}^{\mbox{pert}}(s)&=&\frac{3}{2^{7}\pi^{4}}\int_{\Lambda}^{1}d\alpha\frac{(1-\alpha)^{2}}{\alpha}(m_{Q}^{2}-s\alpha)^{2},\\
\rho_{2}^{\langle\bar{q}q\rangle}(s)&=&-\frac{\langle\bar{q}q\rangle}{2^{4}\pi^{2}}m_{s}[1-(\frac{m_{Q}^{2}}{s})^{2}],\\
\rho_{2}^{\langle\bar{s}s\rangle}(s)&=&\frac{\langle\bar{s}s\rangle}{2^{5}\pi^{2}}m_{s}[1-(\frac{m_{Q}^{2}}{s})^{2}],\\
\rho_{2}^{\langle G^{2}\rangle}(s)&=&\frac{\langle
g^{2}G^{2}\rangle}{2^{9}\pi^{4}}[1-(\frac{m_{Q}^{2}}{s})^{2}],\\
\hat{B}\Pi_{2}^{\mbox{cond}}&=&-\frac{\langle
g^{2}G^{2}\rangle}{3\cdot2^{9}\pi^{4}}m_{Q}^{2}\int_{0}^{1}d\alpha(\frac{1-\alpha}{\alpha})^{2}e^{-m_{Q}^{2}/(\alpha
M^{2})}\nonumber\\
& &{}+\frac{\langle g\bar{q}\sigma\cdot G
q\rangle}{2^{5}\pi^{2}}m_{s}e^{-m_{Q}^{2}/M^{2}}\nonumber\\&&{}+\frac{\langle
g\bar{s}\sigma\cdot G
s\rangle}{3\cdot2^{4}\pi^{2}}m_{s}e^{-m_{Q}^{2}/M^{2}}\nonumber\\
&
&{}+\frac{\langle\bar{q}q\rangle\langle\bar{s}s\rangle}{6}e^{-m_{Q}^{2}/M^{2}}\nonumber\\
&&{}-\frac{\langle
g^{3}G^{3}\rangle}{3\cdot2^{11}\pi^{4}}\int_{0}^{1}
d\alpha(\frac{1-\alpha}{\alpha})^{2}(1-\frac{2m_{Q}^{2}}{\alpha
M^{2}})e^{-m_{Q}^{2}/(\alpha M^{2})},
\end{eqnarray}
for $\Xi_{Q}$ baryons,
\begin{eqnarray}
\rho_{1}^{\mbox{pert}}(s)&=&\frac{3}{2^{4}\pi^{4}}m_{Q}\int^{1}_{\Lambda}d\alpha(\frac{1-\alpha}{\alpha})^{2}(m_{Q}^{2}-s\alpha)^{2},\\
\rho_{1}^{\langle\bar{q}q\rangle}(s)&=&-\frac{2\langle\bar{q}q\rangle}{\pi^{2}}m_{s}m_{Q}(1-\frac{m_{Q}^{2}}{s}),\\
\rho_{1}^{\langle\bar{s}s\rangle}(s)&=&\frac{\langle\bar{s}s\rangle}{2\pi^{2}}m_{s}m_{Q}(1-\frac{m_{Q}^{2}}{s}),\\
\rho_{1}^{\langle G^{2}\rangle}(s)&=&\frac{\langle
g^{2}G^{2}\rangle}{2^{6}\pi^{4}}m_{Q}\int_{\Lambda}^{1}d\alpha[(\frac{1-\alpha}{\alpha})^{2}-2],\\
\hat{B}\Pi_{1}^{\mbox{cond}}&=& -\frac{\langle
g^{2}G^{2}\rangle}{3\cdot2^{6}\pi^{4}}m_{Q}^{3}\int^{1}_{0}d\alpha\frac{(1-\alpha)^{2}}{\alpha^{3}}e^{-m_{Q}^{2}/(\alpha
M^{2})}\nonumber\\
& &{}+\frac{\langle g\bar{q}\sigma\cdot G
q\rangle}{2\pi^{2}}m_{s}m_{Q}e^{-m_{Q}^{2}/M^{2}}\nonumber\\
& & {}-\frac{\langle g\bar{s}\sigma\cdot G
s\rangle}{3\pi^{2}}m_{s}m_{Q}e^{-m_{Q}^{2}/M^{2}}\nonumber\\
&
&{}+\frac{8\langle\bar{q}q\rangle\langle\bar{s}s\rangle}{3}m_{Q}e^{-m_{Q}^{2}/M^{2}}\nonumber\\
& & {}-\frac{\langle g^{3}G^{3}\rangle}{3\cdot2^{7}\pi^{4}}m_{Q}
\int_{0}^{1}
d\alpha\frac{(1-\alpha)^{2}}{\alpha^{3}}(3-\frac{m_{Q}^{2}}{\alpha
M^{2}})e^{-m_{Q}^{2}/(\alpha M^{2})},\\
\rho_{2}^{\mbox{pert}}(s)&=&\frac{3}{2^{5}\pi^{4}}\int^{1}_{\Lambda}d\alpha\frac{(1-\alpha)^{2}}{\alpha}(m_{Q}^{2}-s\alpha)^{2},\\
\rho_{2}^{\langle\bar{q}q\rangle}(s)&=&-\frac{\langle\bar{q}q\rangle}{2\pi^{2}}m_{s}[1-(\frac{m_{Q}^{2}}{s})^{2}],\\
\rho_{2}^{\langle\bar{s}s\rangle}(s)&=&\frac{\langle\bar{s}s\rangle}{2^{3}\pi^{2}}m_{s}[1-(\frac{m_{Q}^{2}}{s})^{2}],\\
\rho_{2}^{\langle G^{2}\rangle}(s)&=&-\frac{\langle
g^{2}G^{2}\rangle}{2^{7}\pi^{4}}[1-(\frac{m_{Q}^{2}}{s})^{2}],\\
\hat{B}\Pi_{2}^{\mbox{cond}}&=& -\frac{\langle
g^{2}G^{2}\rangle}{3\cdot2^{7}\pi^{4}}m_{Q}^{2}\int^{1}_{0}d\alpha(\frac{1-\alpha}{\alpha})^{2}e^{-m_{Q}^{2}/(\alpha
M^{2})}\nonumber\\
& &{}+\frac{\langle g\bar{q}\sigma\cdot G
q\rangle}{2^{2}\pi^{2}}m_{s}e^{-m_{Q}^{2}/M^{2}}\nonumber\\
& & {}-\frac{\langle g\bar{s}\sigma\cdot G
s\rangle}{3\cdot2\pi^{2}}m_{s}e^{-m_{Q}^{2}/M^{2}}\nonumber\\
&
&{}+\frac{4\langle\bar{q}q\rangle\langle\bar{s}s\rangle}{3}e^{-m_{Q}^{2}/M^{2}}\nonumber\\
& & {}-\frac{\langle g^{3}G^{3}\rangle}{3\cdot2^{9}\pi^{4}}
\int_{0}^{1}
d\alpha(\frac{1-\alpha}{\alpha})^{2}(1-\frac{2m_{Q}^{2}}{\alpha
M^{2}})e^{-m_{Q}^{2}/(\alpha M^{2})},
\end{eqnarray}
for $\Xi_{Q}^{'}$ baryons,
\begin{eqnarray}
\rho_{1}^{\mbox{pert}}(s)&=&\frac{1}{2^{4}\pi^{4}}m_{Q}\int^{1}_{\Lambda}d\alpha(\frac{1-\alpha}{\alpha})^{2}(m_{Q}^{2}-s\alpha)^{2}
+\frac{1}{2\pi^{4}}m_{s}\int^{1}_{\Lambda}d\alpha\frac{1-\alpha}{\alpha}(m_{Q}^{2}-s\alpha)^{2},\\
\rho_{1}^{\langle\bar{q}q\rangle}(s)&=&-\frac{10\langle\bar{q}q\rangle}{3\pi^{2}}m_{s}m_{Q}(1-\frac{m_{Q}^{2}}{s}),\\
\rho_{1}^{\langle\bar{s}s\rangle}(s)&=&\frac{2^{2}\langle\bar{s}s\rangle}{3\pi^{2}}
\int_{\Lambda}^{1}d\alpha(m_{Q}^{2}-s\alpha)+\frac{\langle\bar{s}s\rangle}{3\cdot2\pi^{2}}m_{s}m_{Q}(1-\frac{m_{Q}^{2}}{s}),\\
\rho_{1}^{\langle G^{2}\rangle}(s)&=&\frac{\langle
g^{2}G^{2}\rangle}{3\cdot2^{6}\pi^{4}}m_{Q}\int_{\Lambda}^{1}d\alpha[(\frac{1-\alpha}{\alpha})^{2}-2]-\frac{\langle g^{2}G^{2}\rangle}{3\cdot2^{3}\pi^{4}}m_{s}\int_{\Lambda}^{1}d\alpha(m_{Q}^{2}-s\alpha),\\
\hat{B}\Pi_{1}^{\mbox{cond}}&=&
-\frac{\langle g^{2}G^{2}\rangle}{3^{2}\cdot2^{6}\pi^{4}}m_{Q}^{3}\int^{1}_{0}d\alpha\frac{(1-\alpha)^{2}}{\alpha^{3}}e^{-m_{Q}^{2}/(\alpha M^{2})}\nonumber\\
&
&{}-\frac{\langle g^{2}G^{2}\rangle}{3^{2}\cdot2^{3}\pi^{4}}m_{s}m_{Q}^{2}\int^{1}_{0}d\alpha\frac{1-\alpha}{\alpha^{2}}e^{-m_{Q}^{2}/(\alpha M^{2})}\nonumber\\
& &{}+\frac{5\langle g\bar{q}\sigma\cdot G
q\rangle}{6\pi^{2}}m_{s}m_{Q}e^{-m_{Q}^{2}/ M^{2}}\nonumber\\
&&{}-\frac{\langle g\bar{s}\sigma\cdot G
s\rangle}{3^{2}\pi^{2}}m_{s}m_{Q}e^{-m_{Q}^{2}/M^{2}}\nonumber\\
&&{}+\frac{5\cdot2^{3}\langle\bar{q}q\rangle\langle\bar{s}s\rangle}{3^{2}}m_{Q}e^{-m_{Q}^{2}/M^{2}}\nonumber\\
&&{}-\frac{\langle g^{3}G^{3}\rangle}{3^{2}\cdot2^{7}\pi^{4}}m_{Q}
\int_{0}^{1}
d\alpha\frac{(1-\alpha)^{2}}{\alpha^{3}}(3-\frac{m_{Q}^{2}}{\alpha
M^{2}})e^{-m_{Q}^{2}/(\alpha M^{2})}\nonumber\\
&&{}-\frac{\langle g^{3}G^{3}\rangle}{3^{2}\cdot2^{5}\pi^{4}}m_{s}
\int_{0}^{1}
d\alpha\frac{1}{\alpha^{2}}[(1+2\alpha)-\frac{2m_{Q}^{2}}{\alpha
M^{2}}]e^{-m_{Q}^{2}/(\alpha M^{2})},\\
\rho_{2}^{\mbox{pert}}(s)&=&-\frac{1}{2^{5}\pi^{4}}\int_{\Lambda}^{1}d\alpha(\frac{1-\alpha}{\alpha})^{2}(1-3\alpha)(m_{Q}^{2}-s\alpha)^{2},\\
\rho_{2}^{\langle\bar{q}q\rangle}(s)&=&-\frac{2\langle\bar{q}q\rangle}{3\pi^{2}}
m_{Q}(1-\frac{m_{Q}^{2}}{s})^{2}-\frac{\langle\bar{q}q\rangle}{3\cdot2\pi^{2}}m_{s}[1-(\frac{m_{Q}^{2}}{s})^{2}],\\
\rho_{2}^{\langle\bar{s}s\rangle}(s)&=&\frac{\langle\bar{s}s\rangle}{3\cdot2^{2}\pi^{2}}m_{s}
\int_{\Lambda}^{1}d\alpha(3-\alpha),\\
\rho_{2}^{\langle
G^{2}\rangle}(s)&=&\frac{\langle
g^{2}G^{2}\rangle}{3\cdot2^{6}\pi^{4}}[-\frac{1}{2}+\frac{(m_{Q}^{2}/s)^2}{2}+\int_{\Lambda}^{1}d\alpha(\alpha-1)(m_{Q}^{2}-s\alpha)],\\
\hat{B}\Pi_{2}^{\mbox{cond}}&=&\frac{\langle
g^{2}G^{2}\rangle}{3^{2}\cdot2^{7}\pi^{4}}m_{Q}^{2}\int_{0}^{1}d\alpha\frac{(1-\alpha)^{2}(1-3\alpha)}{\alpha^{3}}e^{-m_{Q}^{2}/(\alpha
M^{2})}\nonumber\\
&&{}+\frac{\langle g\bar{q}\sigma\cdot G
q\rangle}{3\pi^{2}}m_{Q}\int_{0}^{1}d\alpha e^{-m_{Q}^{2}/(\alpha
M^{2})}\nonumber\\
& &{}+\frac{\langle g\bar{q}\sigma\cdot G
q\rangle}{3\cdot2^{2}\pi^{2}}m_{s}e^{-m_{Q}^{2}/M^{2}}\nonumber\\
&&{}-\frac{\langle g\bar{s}\sigma\cdot G
s\rangle}{3\cdot2^{2}\pi^{2}}m_{s}\int_{0}^{1}d\alpha
e^{-m_{Q}^{2}/(\alpha M^{2})}\nonumber\\
& &{}-\frac{\langle g\bar{s}\sigma\cdot G
s\rangle}{3^{2}\cdot2\pi^{2}}m_{s}e^{-m_{Q}^{2}/M^{2}}\nonumber\\
&&{}+\frac{4\langle\bar{q}q\rangle\langle\bar{s}s\rangle}{3^{2}}(1-\frac{2m_{s}m_{Q}}{M^{2}})e^{-m_{Q}^{2}/M^{2}}\nonumber\\
& & {}+\frac{\langle
g^{3}G^{3}\rangle}{3^{2}\cdot2^{9}\pi^{4}}\int_{0}^{1}
d\alpha\frac{1-\alpha}{\alpha^{4}}[\alpha(1-4\alpha-3\alpha^{2})
-2(1-4\alpha+\alpha^{2})\frac{m_{Q}^{2}}{M^{2}}]e^{-m_{Q}^{2}/(\alpha
M^{2})},
\end{eqnarray}
for $\Xi_{Q}^{'*}$ baryons,
\begin{eqnarray}
\rho_{1}^{\mbox{pert}}(s)&=&\frac{3}{2^{5}\pi^{4}}m_{Q}\int^{1}_{\Lambda}d\alpha(\frac{1-\alpha}{\alpha})^{2}(m_{Q}^{2}-s\alpha)^{2},\\
\rho_{1}^{\langle\bar{q}q\rangle}(s)&=&-\frac{\langle\bar{q}q\rangle}{2\pi^{2}}m_{s}m_{Q}(1-\frac{m_{Q}^{2}}{s}),\\
\rho_{1}^{\langle\bar{s}s\rangle}(s)&=&\frac{\langle\bar{s}s\rangle}{2^{2}\pi^{2}}m_{s}m_{Q}(1-\frac{m_{Q}^{2}}{s}),\\
\rho_{1}^{\langle G^{2}\rangle}(s)&=&\frac{\langle
g^{2}G^{2}\rangle}{2^{7}\pi^{4}}m_{Q}\int_{\Lambda}^{1}d\alpha[(\frac{1-\alpha}{\alpha})^{2}+2],\\
\hat{B}\Pi_{1}^{\mbox{cond}}&=& -\frac{\langle
g^{2}G^{2}\rangle}{3\cdot2^{7}\pi^{4}}m_{Q}^{3}\int^{1}_{0}d\alpha\frac{(1-\alpha)^{2}}{\alpha^{3}}e^{-m_{Q}^{2}/(\alpha
M^{2})}\nonumber\\
&&{}+\frac{\langle g\bar{q}\sigma\cdot G
q\rangle}{2^{3}\pi^{2}}m_{s}m_{Q}e^{-m_{Q}^{2}/ M^{2}}\nonumber\\
& & {}+\frac{\langle g\bar{s}\sigma\cdot G
s\rangle}{3\cdot2^{2}\pi^{2}}m_{s}m_{Q}e^{-m_{Q}^{2}/
M^{2}}\nonumber\\
&&{}+\frac{2\langle\bar{q}q\rangle\langle\bar{s}s\rangle}{3}m_{Q}e^{-m_{Q}^{2}/M^{2}}\nonumber\\
& & {}-\frac{\langle g^{3}G^{3}\rangle}{3\cdot2^{8}\pi^{4}}m_{Q}
\int_{0}^{1}
d\alpha\frac{(1-\alpha)^{2}}{\alpha^{3}}(3-\frac{m_{Q}^{2}}{\alpha
M^{2}})e^{-m_{Q}^{2}/(\alpha M^{2})},\\
\rho_{2}^{\mbox{pert}}(s)&=&-\frac{3}{2^{6}\pi^{4}}\int^{1}_{\Lambda}d\alpha\frac{(1-\alpha)^{2}}{\alpha}(m_{Q}^{2}-s\alpha)^{2},\\
\rho_{2}^{\langle\bar{q}q\rangle}(s)&=&\frac{\langle\bar{q}q\rangle}{2^{3}\pi^{2}}m_{s}[1-(\frac{m_{Q}^{2}}{s})^{2}],\\
\rho_{2}^{\langle\bar{s}s\rangle}(s)&=&-\frac{\langle\bar{s}s\rangle}{2^{4}\pi^{2}}m_{s}[1-(\frac{m_{Q}^{2}}{s})^{2}],\\
\rho_{2}^{\langle G^{2}\rangle}(s)&=&-\frac{\langle
g^{2}G^{2}\rangle}{2^{8}\pi^{4}}[1-(\frac{m_{Q}^{2}}{s})^{2}],\\
\hat{B}\Pi_{2}^{\mbox{cond}}&=& \frac{\langle
g^{2}G^{2}\rangle}{3\cdot2^{8}\pi^{4}}m_{Q}^{2}\int^{1}_{0}d\alpha(\frac{1-\alpha}{\alpha})^{2}e^{-m_{Q}^{2}/(\alpha
M^{2})}\nonumber\\
&&{}-\frac{\langle g\bar{q}\sigma\cdot G
q\rangle}{2^{4}\pi^{2}}m_{s}e^{-m_{Q}^{2}/ M^{2}}\nonumber\\
& & {}-\frac{\langle g\bar{s}\sigma\cdot G
s\rangle}{3\cdot2^{3}\pi^{2}}m_{s}e^{-m_{Q}^{2}/
M^{2}}\nonumber\\
&&{}-\frac{\langle\bar{q}q\rangle\langle\bar{s}s\rangle}{3}e^{-m_{Q}^{2}/M^{2}}\nonumber\\
& & {}+\frac{\langle g^{3}G^{3}\rangle}{3\cdot2^{10}\pi^{4}}
\int_{0}^{1}
d\alpha(\frac{1-\alpha}{\alpha})^{2}(1-\frac{2m_{Q}^{2}}{\alpha
M^{2}})e^{-m_{Q}^{2}/(\alpha M^{2})},
\end{eqnarray}
for $\Xi_{1Q}$ baryons,
\begin{eqnarray}
\rho_{1}^{\mbox{pert}}(s)&=&\frac{1}{2^{5}\pi^{4}}m_{Q}\int^{1}_{\Lambda}d\alpha(\frac{1-\alpha}{\alpha})^{2}(m_{Q}^{2}-s\alpha)^{2}
+\frac{1}{2^{2}\pi^{4}}m_{s}\int^{1}_{\Lambda}d\alpha\frac{1-\alpha}{\alpha}(m_{Q}^{2}-s\alpha)^{2},\\
\rho_{1}^{\langle\bar{q}q\rangle}(s)&=&-\frac{5\langle\bar{q}q\rangle}{3\cdot2\pi^{2}}m_{s}m_{Q}(1-\frac{m_{Q}^{2}}{s}),\\
\rho_{1}^{\langle\bar{s}s\rangle}(s)&=&\frac{2\langle\bar{s}s\rangle}{3\pi^{2}}\int_{\Lambda}^{1}d\alpha(m_{Q}^{2}-s\alpha)+\frac{\langle\bar{s}s\rangle}{3\cdot2^{2}\pi^{2}}m_{s}m_{Q}(1-\frac{m_{Q}^{2}}{s}),\\
\rho_{1}^{\langle G^{2}\rangle}(s)&=&\frac{\langle
g^{2}G^{2}\rangle}{3\cdot2^{7}\pi^{4}}m_{Q}\int_{\Lambda}^{1}d\alpha[(\frac{1-\alpha}{\alpha})^{2}+2]+\frac{\langle g^{2}G^{2}\rangle}{3\cdot2^{4}\pi^{4}}m_{s}\int_{\Lambda}^{1}d\alpha(m_{Q}^{2}-s\alpha),\\
\hat{B}\Pi_{1}^{\mbox{cond}}&=& -\frac{\langle
g^{2}G^{2}\rangle}{3^{2}\cdot2^{7}\pi^{4}}m_{Q}^{3}\int^{1}_{0}d\alpha\frac{(1-\alpha)^{2}}{\alpha^{3}}e^{-m_{Q}^{2}/(\alpha
M^{2})}\nonumber\\
&&{} -\frac{\langle
g^{2}G^{2}\rangle}{3^{2}\cdot2^{4}\pi^{4}}m_{s}m_{Q}^{2}\int^{1}_{0}d\alpha\frac{1-\alpha}{\alpha^{2}}e^{-m_{Q}^{2}/(\alpha
M^{2})}\nonumber\\&&{}+\frac{5\langle g\bar{q}\sigma\cdot G
q\rangle}{3\cdot2^{3}\pi^{2}}m_{s}m_{Q}e^{-m_{Q}^{2}/
M^{2}}\nonumber\\
&&{}+\frac{\langle g\bar{s}\sigma\cdot G
s\rangle}{3^{2}\cdot2^{2}\pi^{2}}m_{s}m_{Q}e^{-m_{Q}^{2}/
M^{2}}\nonumber\\
&&{}+\frac{5\cdot2\langle\bar{q}q\rangle\langle\bar{s}s\rangle}{3^{2}}m_{Q}e^{-m_{Q}^{2}/M^{2}}\nonumber\\
&&{}-\frac{\langle g^{3}G^{3}\rangle}{3^{2}\cdot2^{8}\pi^{4}}m_{Q}
\int_{0}^{1}
d\alpha\frac{(1-\alpha)^{2}}{\alpha^{3}}(3-\frac{m_{Q}^{2}}{\alpha
M^{2}})e^{-m_{Q}^{2}/(\alpha M^{2})}\nonumber\\&&{}-\frac{\langle
g^{3}G^{3}\rangle}{3^{2}\cdot2^{6}\pi^{4}}m_{s} \int_{0}^{1}
d\alpha\frac{1}{\alpha^{2}}[(1-4\alpha)-\frac{2(1-2\alpha)m_{Q}^{2}}{\alpha
M^{2}}]e^{-m_{Q}^{2}/(\alpha M^{2})},\\
\rho_{2}^{\mbox{pert}}(s)&=&\frac{1}{2^{6}\pi^{4}}\int_{\Lambda}^{1}d\alpha(\frac{1-\alpha}{\alpha})^{2}(1-3\alpha)(m_{Q}^{2}-s\alpha)^{2},\\
\rho_{2}^{\langle\bar{q}q\rangle}(s)&=&\frac{\langle\bar{q}q\rangle}{3\cdot2\pi^{2}}
m_{Q}(1-\frac{m_{Q}^{2}}{s})^{2}+\frac{\langle\bar{q}q\rangle}{3\cdot2^{3}\pi^{2}}m_{s}[1-(\frac{m_{Q}^{2}}{s})^{2}],\\
\rho_{2}^{\langle\bar{s}s\rangle}(s)&=&-\frac{\langle\bar{s}s\rangle}{3\cdot2^{3}\pi^{2}}m_{s}
\int_{\Lambda}^{1}d\alpha(3-\alpha),\\
\rho_{2}^{\langle
G^{2}\rangle}(s)&=&\frac{\langle
g^{2}G^{2}\rangle}{3\cdot2^{7}\pi^{4}}[-\frac{1}{2}+\frac{(m_{Q}^{2}/s)^2}{2}+\int_{\Lambda}^{1}d\alpha(\alpha-1)(m_{Q}^{2}-s\alpha)],\\
\hat{B}\Pi_{2}^{\mbox{cond}}&=&-\frac{\langle
g^{2}G^{2}\rangle}{3^{2}\cdot2^{8}\pi^{4}}m_{Q}^{2}\int_{0}^{1}d\alpha\frac{(1-\alpha)^{2}(1-3\alpha)}{\alpha^{3}}e^{-m_{Q}^{2}/(\alpha
M^{2})}\nonumber\\
&&{}-\frac{\langle g\bar{q}\sigma\cdot G
q\rangle}{3\cdot2^{2}\pi^{2}}m_{Q}\int_{0}^{1}d\alpha
e^{-m_{Q}^{2}/(\alpha M^{2})}\nonumber\\& &{}-\frac{\langle
g\bar{q}\sigma\cdot G
q\rangle}{3\cdot2^{4}\pi^{2}}m_{s}e^{-m_{Q}^{2}/ M^{2}}\nonumber\\
&&{} +\frac{\langle g\bar{s}\sigma\cdot G
s\rangle}{3\cdot2^{3}\pi^{2}}m_{s}\int_{0}^{1}d\alpha
e^{-m_{Q}^{2}/(\alpha M^{2})}\nonumber\\& &{}-\frac{\langle
g\bar{s}\sigma\cdot G
s\rangle}{3^{2}\cdot2^{3}\pi^{2}}m_{s}e^{-m_{Q}^{2}/ M^{2}}\nonumber\\
&&{}-\frac{\langle\bar{q}q\rangle\langle\bar{s}s\rangle}{3^{2}}(1-\frac{2m_{Q}m_{s}}{M^{2}})e^{-m_{Q}^{2}/M^{2}}\nonumber\\
& & {}-\frac{\langle
g^{3}G^{3}\rangle}{3^{2}\cdot2^{10}\pi^{4}}\int_{0}^{1}
d\alpha\frac{1-\alpha}{\alpha^{4}}[\alpha(1-4\alpha+9\alpha^{2})
-2(1-4\alpha+5\alpha^{2})\frac{m_{Q}^{2}}{M^{2}}]e^{-m_{Q}^{2}/(\alpha
M^{2})},
\end{eqnarray}
for $\Xi_{1Q}^{*}$ baryons,
\begin{eqnarray}
\rho_{1}^{\mbox{pert}}(s)&=&\frac{3}{2^{4}\pi^{4}}m_{Q}\int^{1}_{\Lambda}d\alpha(\frac{1-\alpha}{\alpha})^{2}(m_{Q}^{2}-s\alpha)^{2},\\
\rho_{1}^{\langle\bar{s}s\rangle}(s)&=&-\frac{3\langle\bar{s}s\rangle}{\pi^{2}}m_{s}m_{Q}(1-\frac{m_{Q}^{2}}{s}),\\
\rho_{1}^{\langle G^{2}\rangle}(s)&=&\frac{\langle
g^{2}G^{2}\rangle}{2^{6}\pi^{4}}m_{Q}\int_{\Lambda}^{1}d\alpha[(\frac{1-\alpha}{\alpha})^{2}-2],\\
\hat{B}\Pi_{1}^{\mbox{cond}}&=&-\frac{\langle
g^{2}G^{2}\rangle}{3\cdot2^{6}\pi^{4}}m_{Q}^{3}\int^{1}_{0}d\alpha\frac{(1-\alpha)^{2}}{\alpha^{3}}e^{-m_{Q}^{2}/(\alpha
M^{2})}\nonumber\\
&&{}+\frac{\langle g\bar{s}\sigma\cdot G
s\rangle}{3\pi^{2}}m_{s}m_{Q}e^{-m_{Q}^{2}/M^{2}}\nonumber\\
& &
{}+\frac{8\langle\bar{s}s\rangle^{2}}{3}m_{Q}e^{-m_{Q}^{2}/M^{2}}\nonumber\\
&&{}-\frac{\langle g^{3}G^{3}\rangle}{3\cdot2^{7}\pi^{4}}m_{Q}
\int_{0}^{1}
d\alpha\frac{(1-\alpha)^{2}}{\alpha^{3}}(3-\frac{m_{Q}^{2}}{\alpha
M^{2}})e^{-m_{Q}^{2}/(\alpha M^{2})},\\
\rho_{2}^{\mbox{pert}}(s)&=&\frac{3}{2^{5}\pi^{4}}\int^{1}_{\Lambda}d\alpha\frac{(1-\alpha)^{2}}{\alpha}(m_{Q}^{2}-s\alpha)^{2},\\
\rho_{2}^{\langle\bar{s}s\rangle}(s)&=&-\frac{3\langle\bar{s}s\rangle}{2^{2}\pi^{2}}m_{s}[1-(\frac{m_{Q}^{2}}{s})^{2}],\\
\rho_{2}^{\langle G^{2}\rangle}(s)&=&-\frac{\langle
g^{2}G^{2}\rangle}{2^{7}\pi^{4}}[1-(\frac{m_{Q}^{2}}{s})^{2}],\\
\hat{B}\Pi_{2}^{\mbox{cond}}&=& -\frac{\langle
g^{2}G^{2}\rangle}{3\cdot2^{7}\pi^{4}}m_{Q}^{2}
\int^{1}_{0}d\alpha(\frac{1-\alpha}{\alpha})^{2}e^{-m_{Q}^{2}/(\alpha
M^{2})}\nonumber\\
&&{}+\frac{\langle g\bar{s}\sigma\cdot G
s\rangle}{3\cdot2\pi^{2}}m_{s}e^{-m_{Q}^{2}/M^{2}}\nonumber \\
&&{}+\frac{4\langle\bar{s}s\rangle^{2}}{3}e^{-m_{Q}^{2}/M^{2}}\nonumber\\
&&{}-\frac{\langle g^{3}G^{3}\rangle}{3\cdot2^{9}\pi^{4}}
\int_{0}^{1}
d\alpha(\frac{1-\alpha}{\alpha})^{2}(1-\frac{2m_{Q}^{2}}{\alpha
M^{2}})e^{-m_{Q}^{2}/(\alpha M^{2})},
\end{eqnarray}
for $\Omega_{Q}$ baryons, and
\begin{eqnarray}
\rho_{1}^{\mbox{pert}}(s)&=&\frac{1}{2^{4}\pi^{4}}m_{Q}\int^{1}_{\Lambda}d\alpha(\frac{1-\alpha}{\alpha})^{2}(m_{Q}^{2}-s\alpha)^{2}
+\frac{1}{2\pi^{4}}m_{s}\int^{1}_{\Lambda}d\alpha\frac{1-\alpha}{\alpha}(m_{Q}^{2}-s\alpha)^{2},\\
\rho_{1}^{\langle\bar{s}s\rangle}(s)&=&\frac{2^{2}\langle\bar{s}s\rangle}{3\pi^{2}}
\int_{\Lambda}^{1}d\alpha(m_{Q}^{2}-s\alpha)-\frac{19\langle\bar{s}s\rangle}{3\pi^{2}}m_{s}m_{Q}(1-\frac{m_{Q}^{2}}{s}),\\
\rho_{1}^{\langle G^{2}\rangle}(s)&=&\frac{\langle
g^{2}G^{2}\rangle}{3\cdot2^{6}\pi^{4}}m_{Q}\int_{\Lambda}^{1}d\alpha[(\frac{1-\alpha}{\alpha})^{2}-2]-\frac{\langle g^{2}G^{2}\rangle}{3\cdot2^{3}\pi^{4}}m_{s}\int_{\Lambda}^{1}d\alpha(m_{Q}^{2}-s\alpha),\\
\hat{B}\Pi_{1}^{\mbox{cond}}&=&
-\frac{\langle g^{2}G^{2}\rangle}{3^{2}\cdot2^{6}\pi^{4}}m_{Q}^{3}\int^{1}_{0}d\alpha\frac{(1-\alpha)^{2}}{\alpha^{3}}e^{-m_{Q}^{2}/(\alpha M^{2})}\nonumber\\
&&{}-\frac{\langle g^{2}G^{2}\rangle}{3^{2}\cdot2^{3}\pi^{4}}m_{s}m_{Q}^{2}\int^{1}_{0}d\alpha\frac{(1-\alpha)}{\alpha^{2}}e^{-m_{Q}^{2}/(\alpha M^{2})}\nonumber\\
& &{}+\frac{13\langle g\bar{s}\sigma\cdot G
s\rangle}{3^{2}\pi^{2}}m_{s}m_{Q}e^{-m_{Q}^{2}/M^{2}}\nonumber\\
&&{}+\frac{5\cdot2^{3}\langle\bar{s}s\rangle^{2}}{3^{2}}m_{Q}e^{-m_{Q}^{2}/M^{2}}\nonumber\\
&&{}-\frac{2^{3}\langle\bar{s}s\rangle^{2}}{3^{2}}m_{s}(1+\frac{m_{Q}^{2}}{M^{2}})e^{-m_{Q}^{2}/M^{2}}\nonumber\\
&&{}-\frac{\langle g^{3}G^{3}\rangle}{3^{2}\cdot2^{7}\pi^{4}}m_{Q}
\int_{0}^{1}
d\alpha\frac{(1-\alpha)^{2}}{\alpha^{3}}(3-\frac{m_{Q}^{2}}{\alpha
M^{2}})e^{-m_{Q}^{2}/(\alpha M^{2})}\nonumber\\
&&{}-\frac{\langle g^{3}G^{3}\rangle}{3^{2}\cdot2^{5}\pi^{4}}m_{s}
\int_{0}^{1}
d\alpha\frac{1}{\alpha^{2}}[(1+2\alpha)-\frac{2m_{Q}^{2}}{\alpha
M^{2}}]e^{-m_{Q}^{2}/(\alpha M^{2})},\\
\rho_{2}^{\mbox{pert}}(s)&=&-\frac{1}{2^{5}\pi^{4}}\int_{\Lambda}^{1}d\alpha(\frac{1-\alpha}{\alpha})^{2}(1-3\alpha)(m_{Q}^{2}-s\alpha)^{2}
-\frac{1}{2\pi^{4}}m_{s}m_{Q}\int^{1}_{\Lambda}d\alpha\frac{(1-\alpha)^{2}}{\alpha}(m_{Q}^{2}-s\alpha),\\
\rho_{2}^{\langle\bar{s}s\rangle}(s)&=&\frac{\langle\bar{s}s\rangle}{3\cdot2\pi^{2}}m_{s}
\int_{\Lambda}^{1}d\alpha(3-5\alpha)-\frac{2\langle\bar{s}s\rangle}{3\pi^{2}}
m_{Q}(1-\frac{m_{Q}^{2}}{s})^{2},\\
\rho_{2}^{\langle G^{2}\rangle}(s)&=&\frac{\langle
g^{2}G^{2}\rangle}{3\cdot2^{6}\pi^{4}}[-\frac{1}{2}+\frac{(m_{Q}^{2}/s)^2}{2}+\int_{\Lambda}^{1}d\alpha(\alpha-1)(m_{Q}^{2}-s\alpha)],\\
\hat{B}\Pi_{2}^{\mbox{cond}}&=&\frac{\langle
g^{2}G^{2}\rangle}{3^{2}\cdot2^{7}\pi^{4}}m_{Q}^{2}\int_{0}^{1}d\alpha\frac{(1-\alpha)^{2}(1-3\alpha)}{\alpha^{3}}e^{-m_{Q}^{2}/(\alpha
M^{2})}\nonumber\\
& &{}
+\frac{\langle g^{2}G^{2}\rangle}{3^{2}\cdot2^{4}\pi^{4}}m_{s}m_{Q}\int^{1}_{0}d\alpha(\frac{1-\alpha}{\alpha})^{2}(3-\frac{m_{Q}^{2}}{\alpha M^{2}})e^{-m_{Q}^{2}/(\alpha M^{2})}\nonumber \\
& & {}-\frac{\langle g\bar{s}\sigma\cdot G
s\rangle}{3\cdot2^{3}\pi^{2}}m_{s}\int_{0}^{1}d\alpha\frac{3+2\alpha}{\alpha}
e^{-m_{Q}^{2}/(\alpha M^{2})}\nonumber\\
& &{}+\frac{\langle g\bar{s}\sigma\cdot G
s\rangle}{3\pi^{2}}m_{Q}\int_{0}^{1}d\alpha e^{-m_{Q}^{2}/(\alpha
M^{2})}\nonumber\\
& &{}+\frac{\langle g\bar{s}\sigma\cdot G
s\rangle}{3^{2}\cdot2\pi^{2}}m_{s}e^{-m_{Q}^{2}/M^{2}}\nonumber\\
& &{}+\frac{4\langle\bar{s}s\rangle^{2}}{3^{2}}(1-\frac{2m_{Q}m_{s}}{M^{2}})e^{-m_{Q}^{2}/M^{2}}\nonumber\\
&&{}+\frac{\langle
g^{3}G^{3}\rangle}{3^{2}\cdot2^{9}\pi^{4}}\int_{0}^{1}
d\alpha\frac{1-\alpha}{\alpha^{4}}[\alpha(1-4\alpha-3\alpha^{2})
-2(1-4\alpha+\alpha^{2})\frac{m_{Q}^{2}}{M^{2}}]e^{-m_{Q}^{2}/(\alpha
M^{2})}\nonumber\\
& &{}-\frac{\langle
g^{3}G^{3}\rangle}{3^{2}\cdot2^{5}\pi^{4}}m_{s}m_{Q}\int_{0}^{1}
d\alpha(\frac{1-\alpha}{\alpha})^{2}[\frac{3}{\alpha
M^{2}}-\frac{m_{Q}^{2}}{\alpha^{2}(M^{2})^{2}}]e^{-m_{Q}^{2}/(\alpha
M^{2})},
\end{eqnarray}
for $\Omega_{Q}^{*}$ baryons. The lower limit of integration is
given by $\Lambda=m_{Q}^{2}/s$.

\section{Numerical results and discussions}\label{sec3}

 In the numerical analysis, the input values are taken as $m_{c}=1.25\pm0.09~\mbox{GeV},
m_{b}=4.20\pm0.07~\mbox{GeV}$ \cite{PDG} with
$m_{s}=0.13~\mbox{GeV},
\langle\bar{q}q\rangle=-(0.23)^{3}~\mbox{GeV}^{3},
\langle\bar{s}s\rangle=0.8\langle\bar{q}q\rangle,\langle
g\bar{q}\sigma\cdot G q\rangle=m_{0}^{2}\langle\bar{q}q\rangle,
m_{0}^{2}=0.8~\mbox{GeV}^{2}, \langle
g^{2}G^{2}\rangle=0.5~\mbox{GeV}^{4},$ and $\langle
g^{3}G^{3}\rangle=0.045~\mbox{GeV}^{6}$ \cite{bracco}. The proper
ranges of the thresholds can be determined, to which the stability
of the Borel curves should not be sensitive. The Borel windows are
fixed in this way: the lower limit constraint for $M^{2}$ is
obtained from the condition that the perturbative contribution
should be larger than the condensate contributions while the upper
one is got by the pole contribution larger than the continuum
contribution \cite{borel}.  Giving an illustration, the comparison
between pole and continuum contributions from Eq. (12) for
$\sqrt{s_{0}}=3.4~\mbox{GeV}$ for $\Omega_{c}$ is shown in Fig.
1(a), and its OPE convergence by comparing the different
contributions is shown in Fig. 1(b). The analysis for the others has
similarly been done, but the corresponding figures are not listed to
keep the paper from being too lengthy. Accordingly, the threshold
are taken as those values exhibited in corresponding figures, and
Borel windows are $M^{2}=1.5\sim3.0~\mbox{GeV}^{2}$ and
$4.5\sim6.0~\mbox{GeV}^{2}$ for charmed and bottom baryons,
respectively. The Borel curves for the dependence on $M^2$ of the
heavy baryon masses are shown in Figs. 2-6. In Table \ref{table:2},
we present our results for the masses of the heavy baryons and
compare with experimental data and other theoretical approaches. In
the numerical evaluation, the results gained from the two sum rules
have been averaged to decrease the systematic errors. It is worth
noting that the uncertainty in the results are merely owing to Borel
windows, not involving the ones rooting in the variation of the
quark masses and QCD parameters.

\begin{figure}[b]
\centerline{\epsfysize=3.5truecm\epsfbox{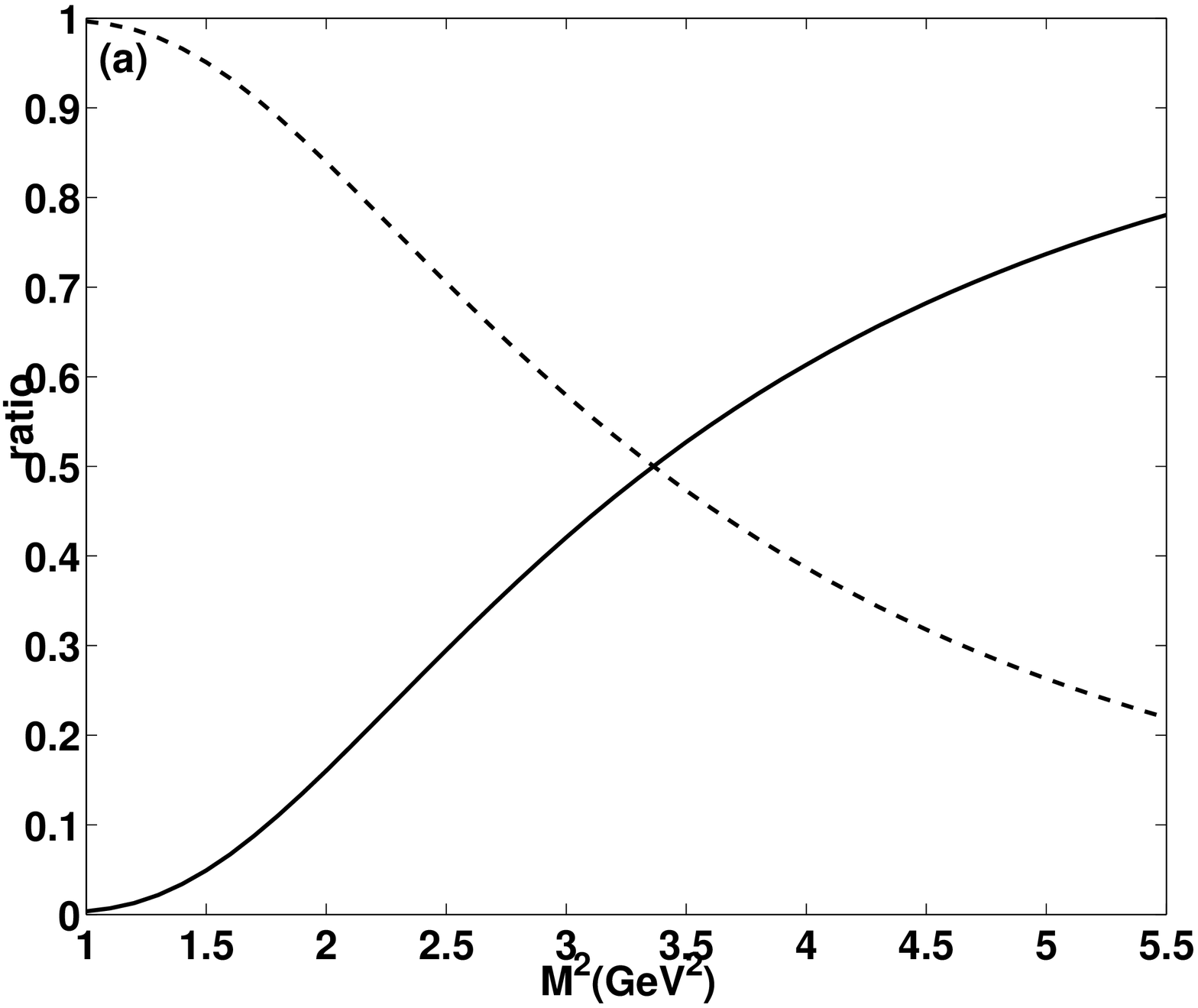}\epsfysize=3.5truecm\epsfbox{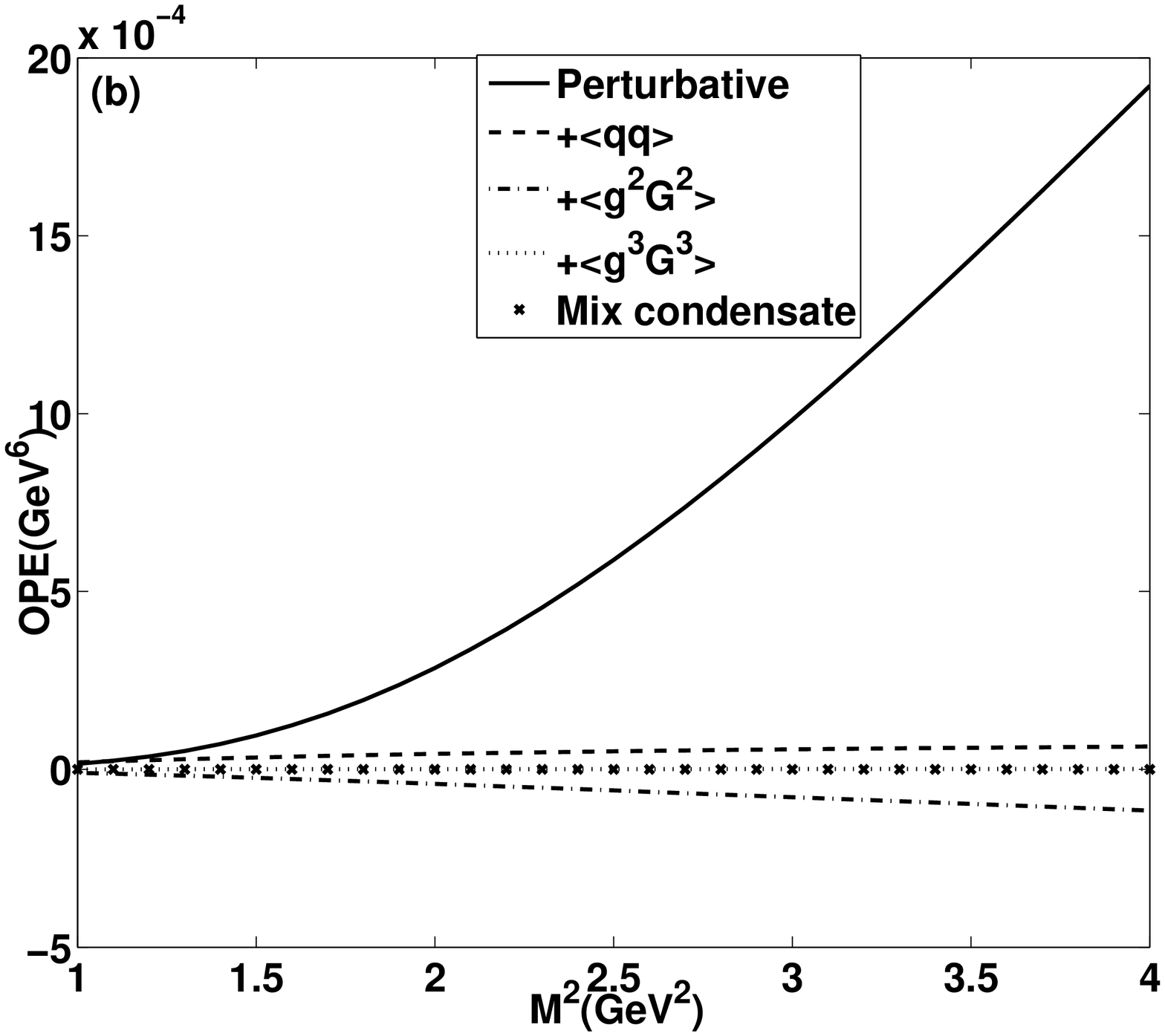}}
\caption{In (a), the dashed line shows the relative pole
contribution (the pole contribution divided by the total, pole plus
continuum contribution) and the solid line shows the relative
continuum contribution from Eq. (12) for
$\sqrt{s_{0}}=3.4~\mbox{GeV}$ for $\Omega_{c}$. Its OPE convergence
is shown in (b) by comparing the perturbative, quark condensate,
two-gluon condensate, three-gluon condensate, and mixed condensate
contributions.} \label{fig:1}
\end{figure}

\begin{figure}
\centerline{\epsfysize=3.5truecm
\epsfbox{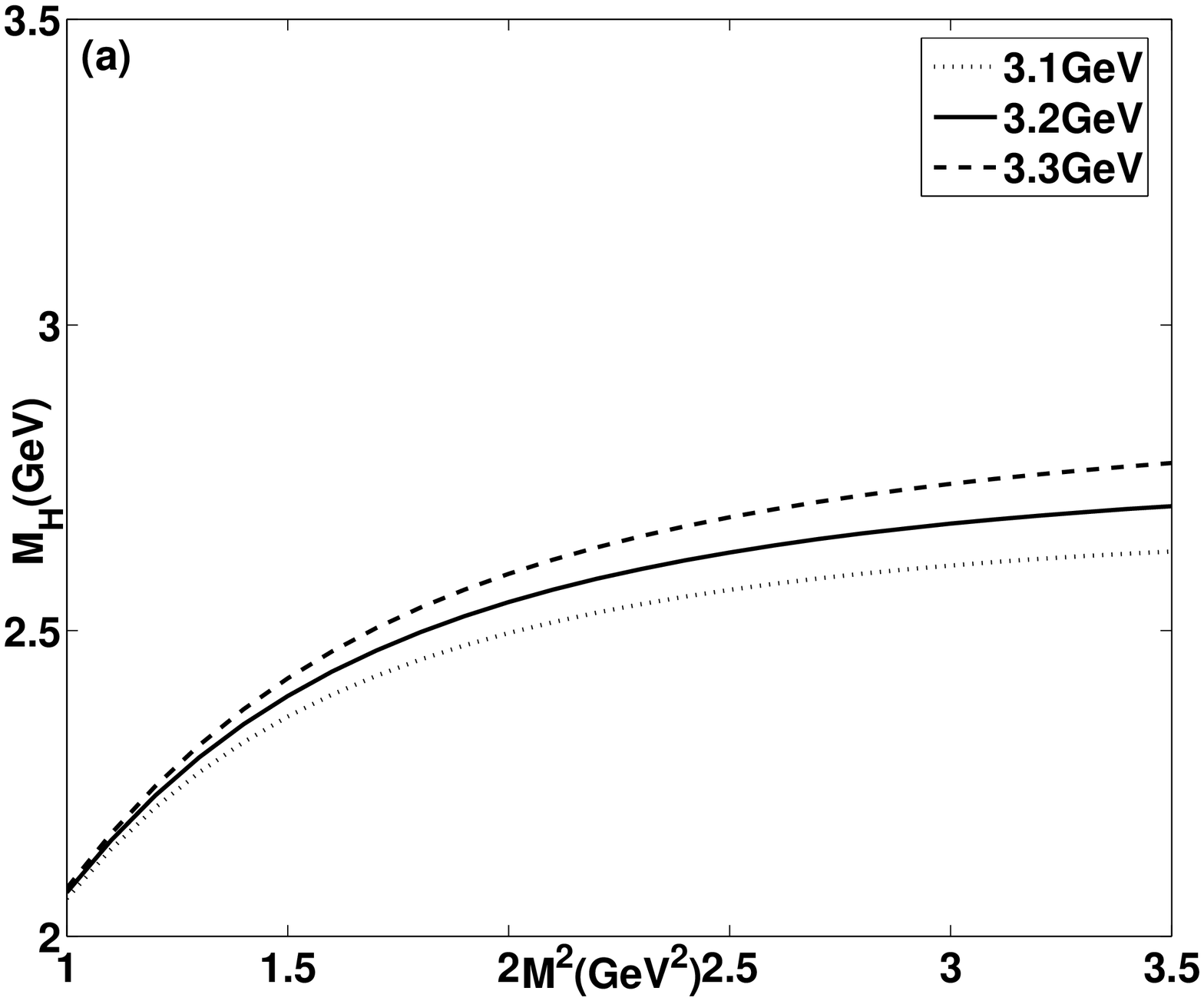}\epsfysize=3.5truecm\epsfbox{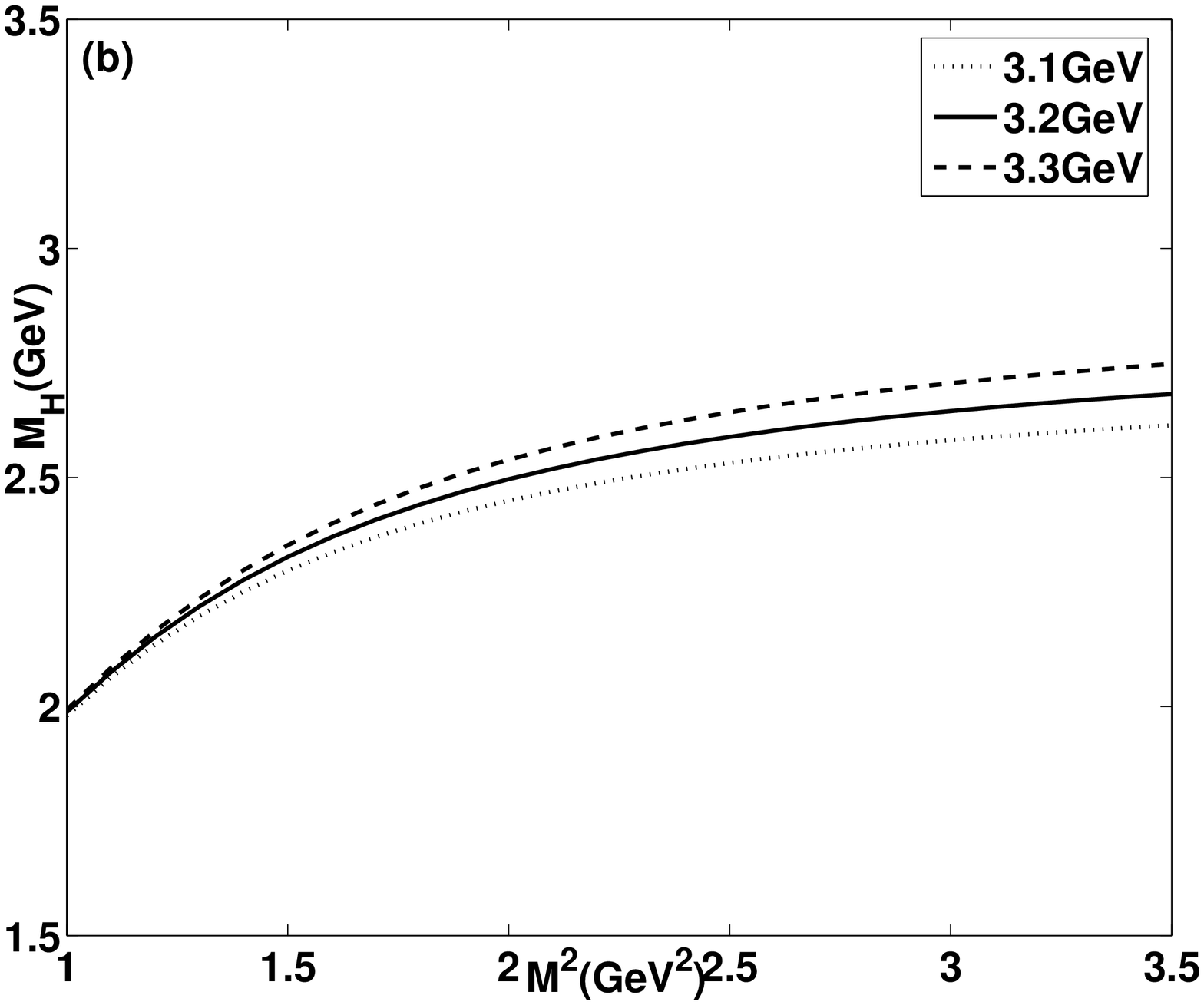}\epsfysize=3.5truecm
\epsfbox{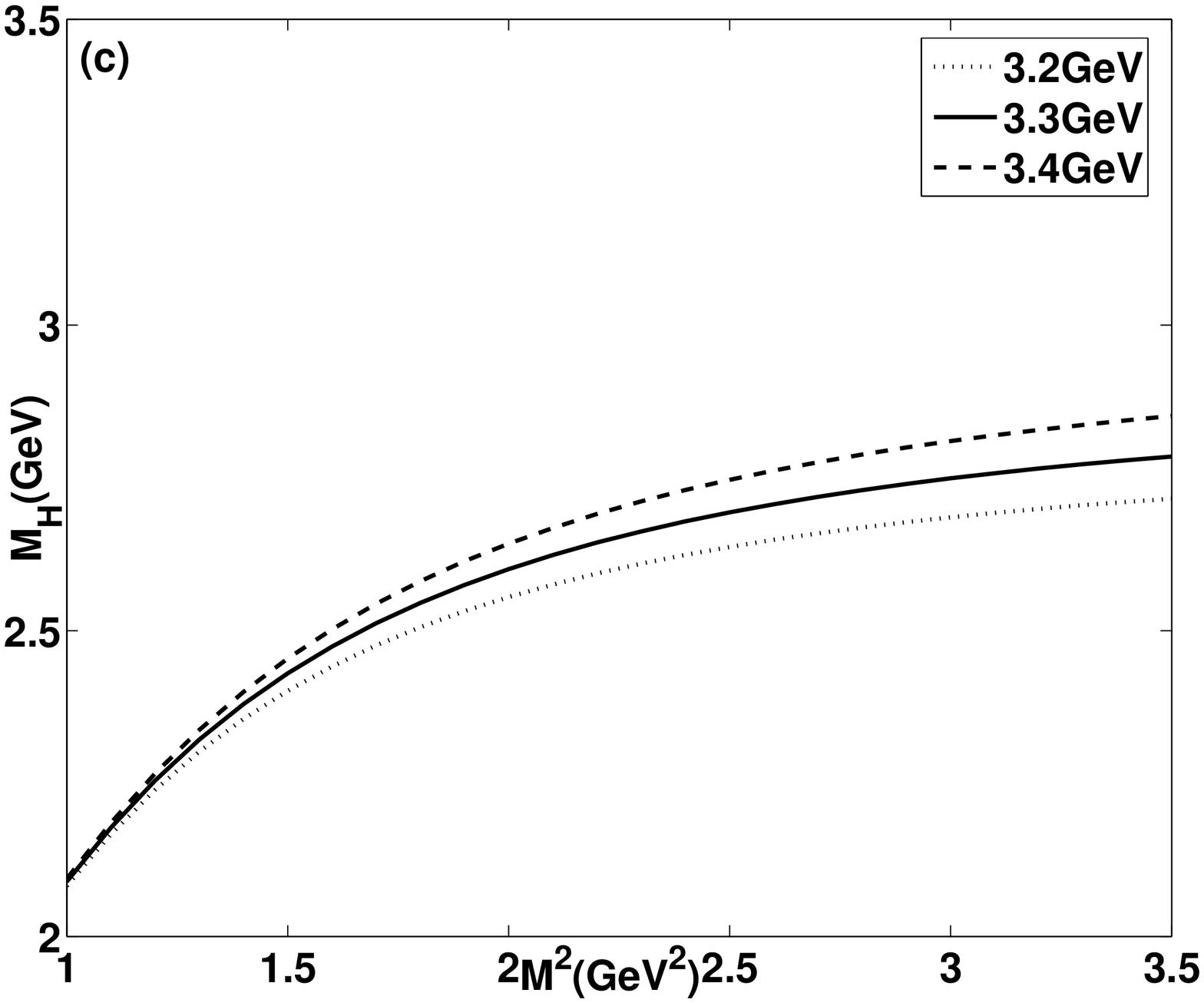}\epsfysize=3.5truecm
\epsfbox{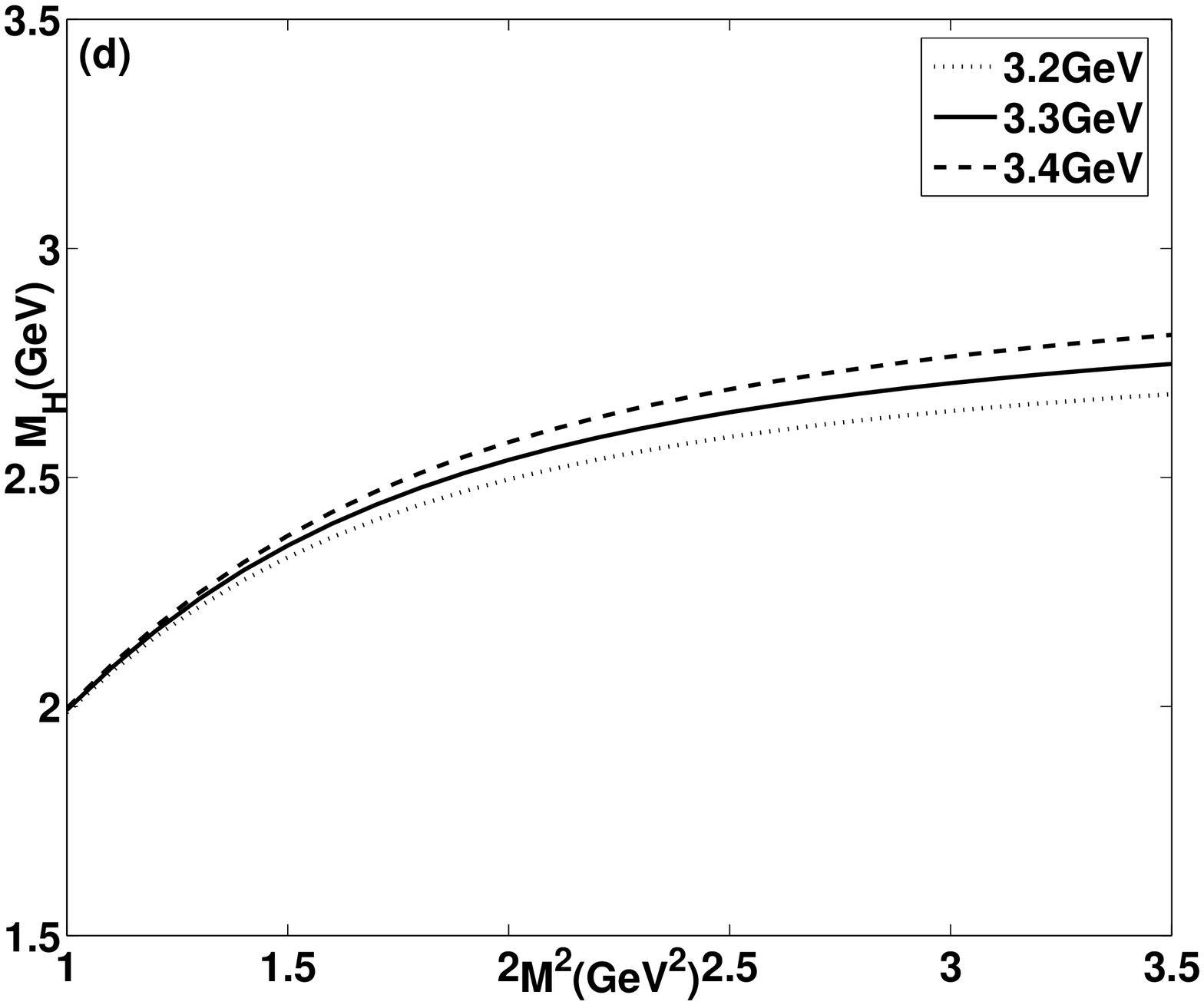}}\centerline{\epsfysize=3.5truecm
\epsfbox{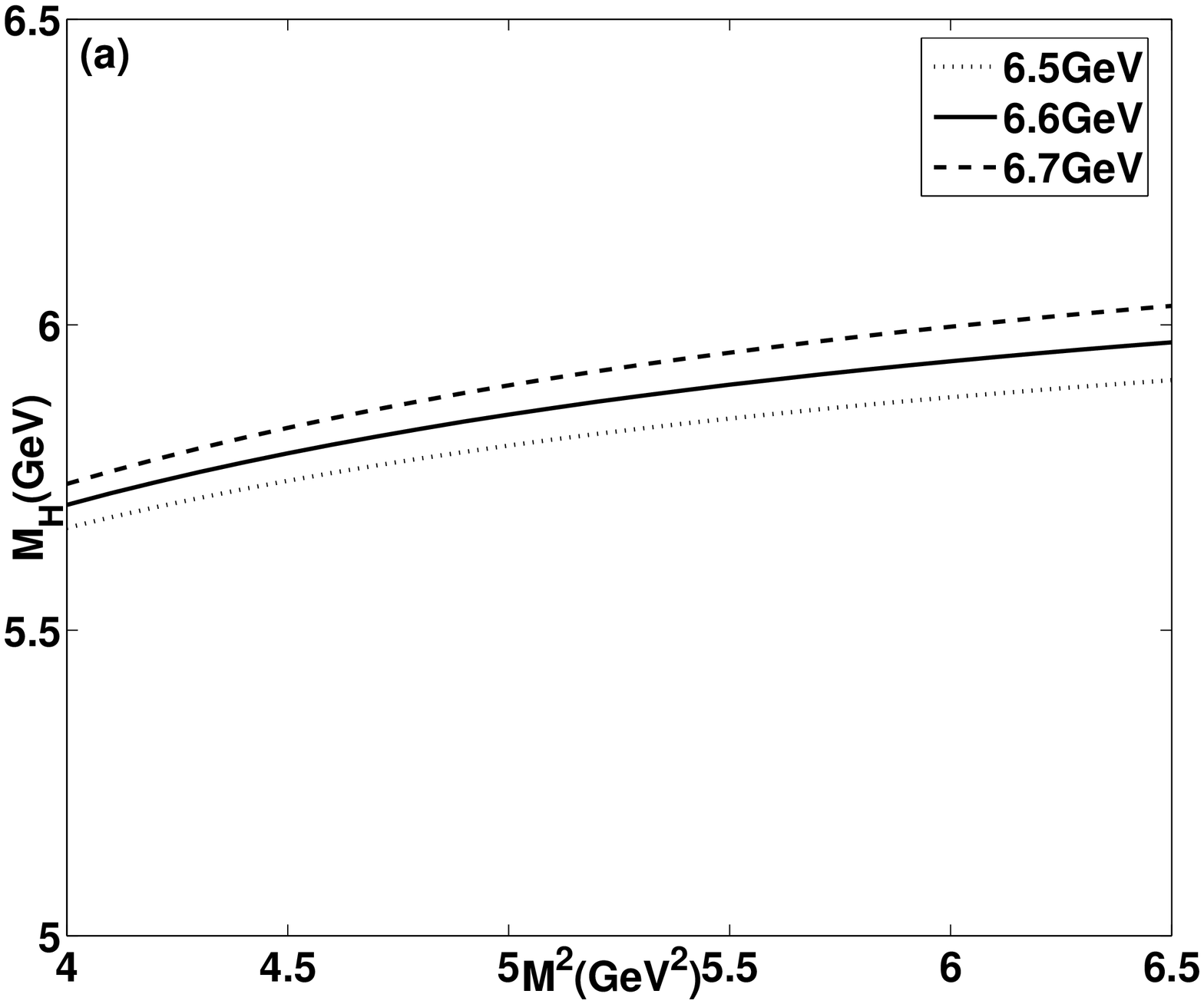}\epsfysize=3.5truecm\epsfbox{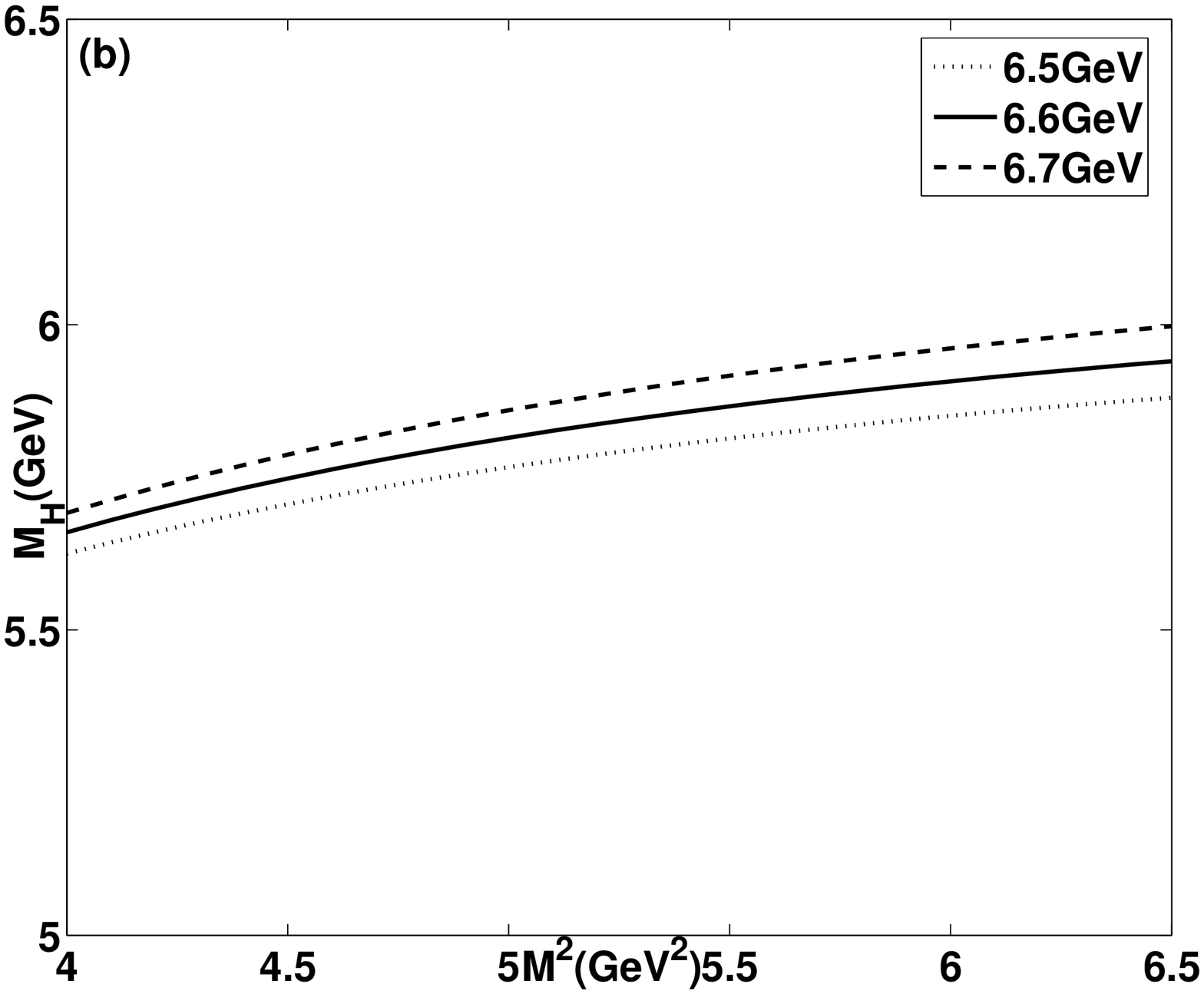}\epsfysize=3.5truecm
\epsfbox{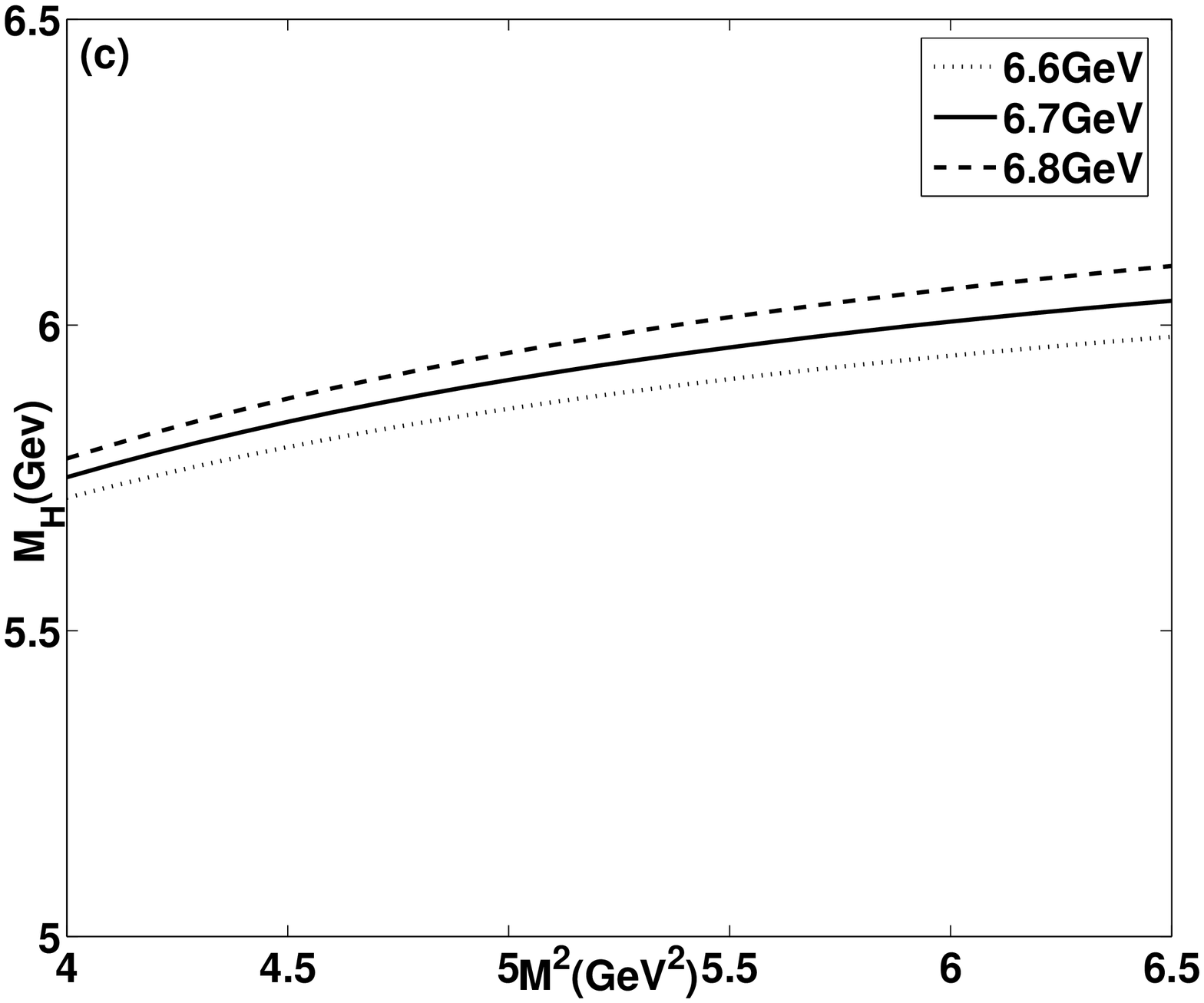}\epsfysize=3.5truecm
\epsfbox{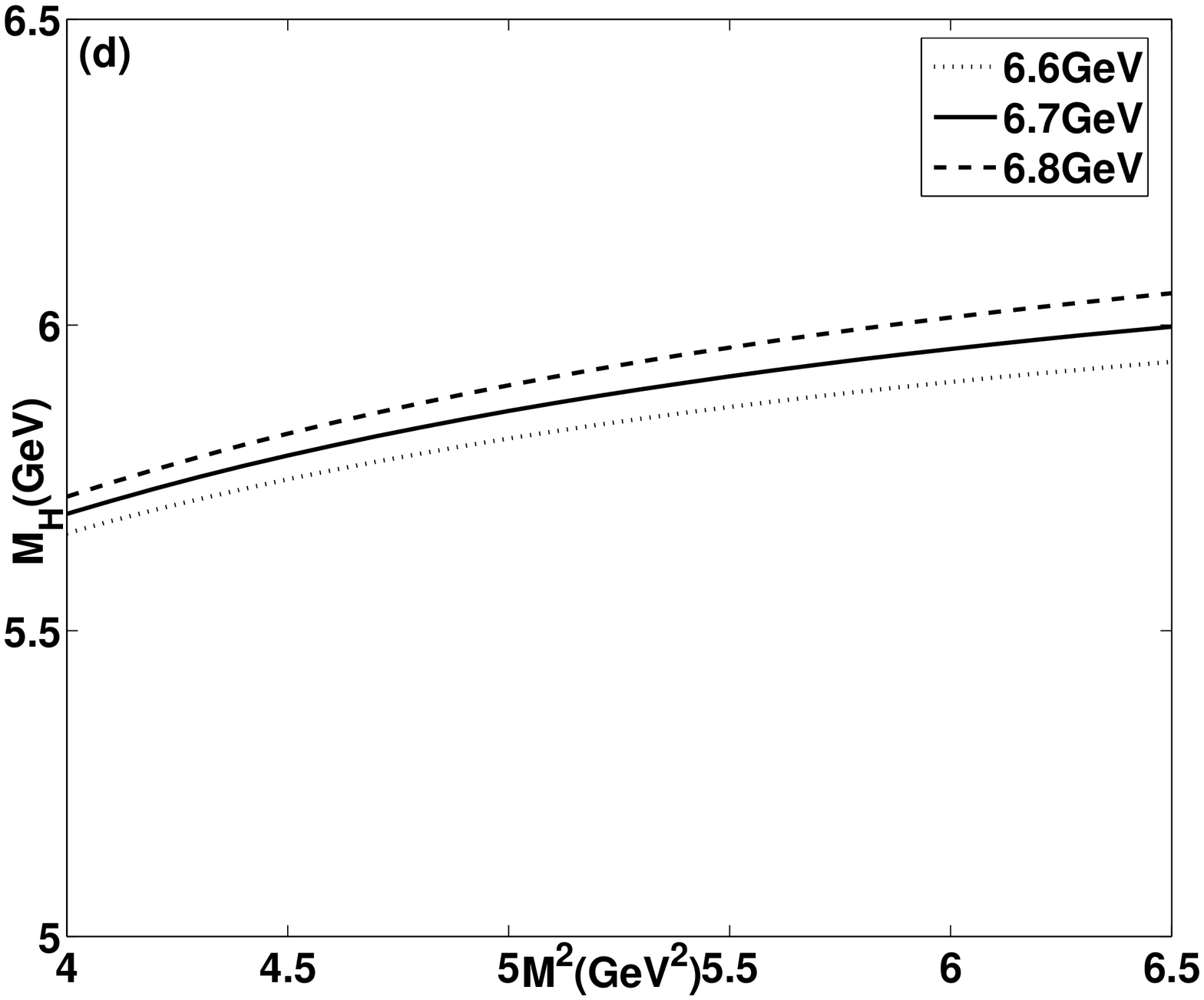}}\caption{The dependence on $M^2$ for
the masses of $\Lambda_{1c}$, $\Lambda_{1c}^{*}$, $\Lambda_{1b}$,
and $\Lambda_{1b}^{*}$. The continuum thresholds are orderly taken
as $\sqrt{s_0}=3.1\sim3.3~\mbox{GeV}$,
$\sqrt{s_0}=3.2\sim3.4~\mbox{GeV}$,
$\sqrt{s_0}=6.5\sim6.7~\mbox{GeV}$, and
$\sqrt{s_0}=6.6\sim6.8~\mbox{GeV}$. (a) and (c) are from the sum
rule (\ref{sum rule m}), (b) and (d) from the sum rule (\ref{sum
rule q}).} \label{fig:2}
\end{figure}

\begin{figure}
\centerline{\epsfysize=3.5truecm\epsfbox{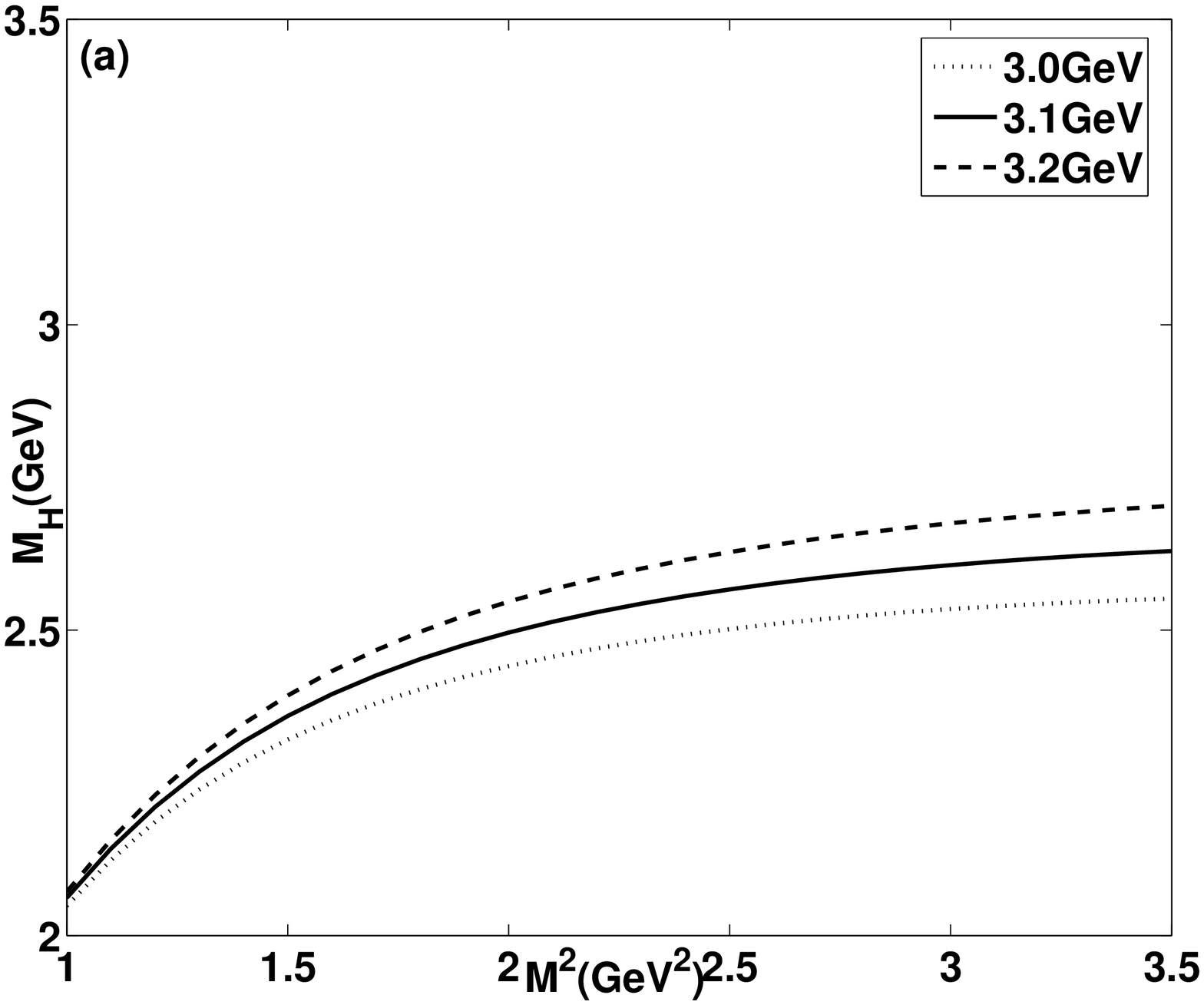}\epsfysize=3.5truecm\epsfbox{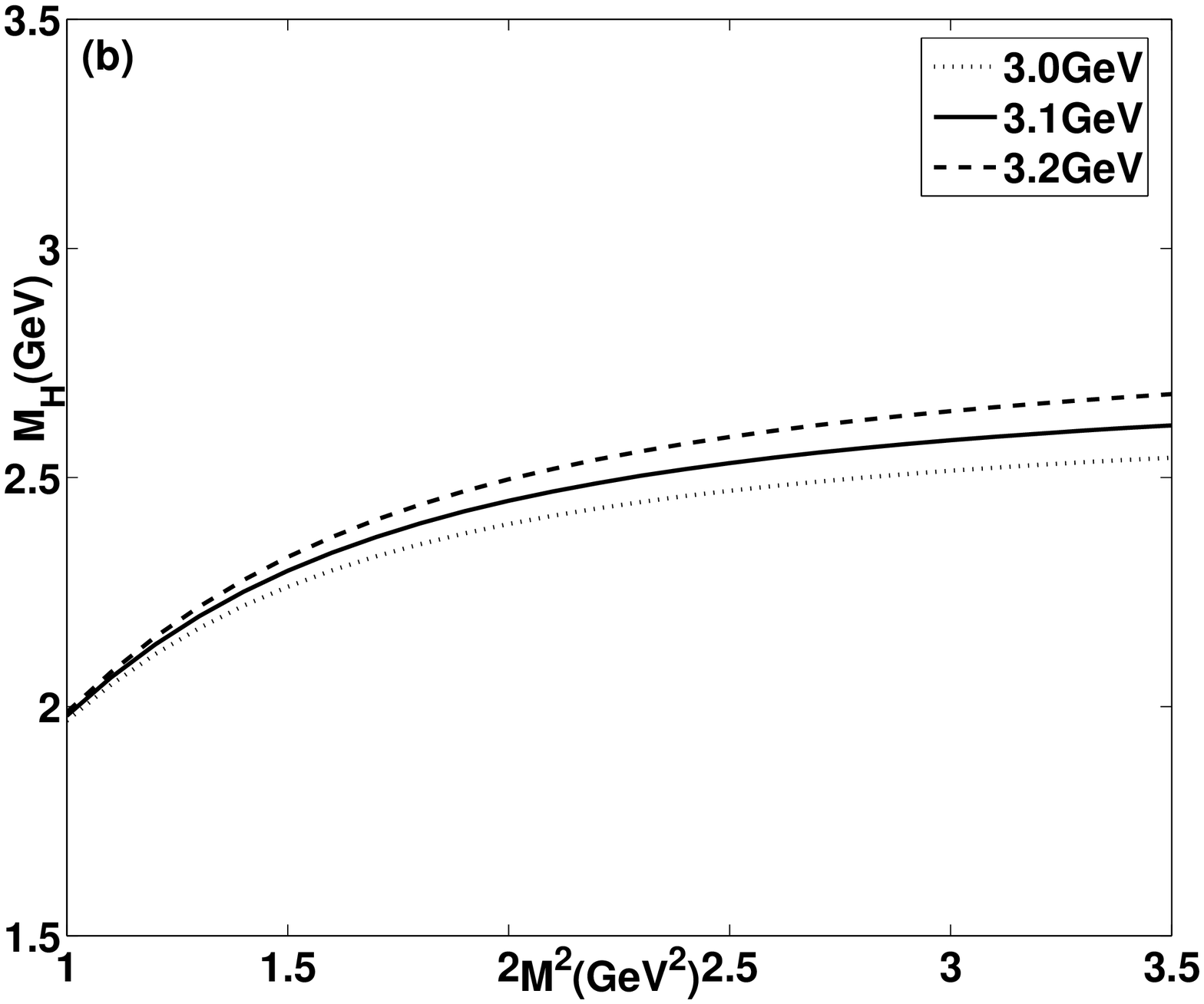}
\epsfysize=3.5truecm\epsfbox{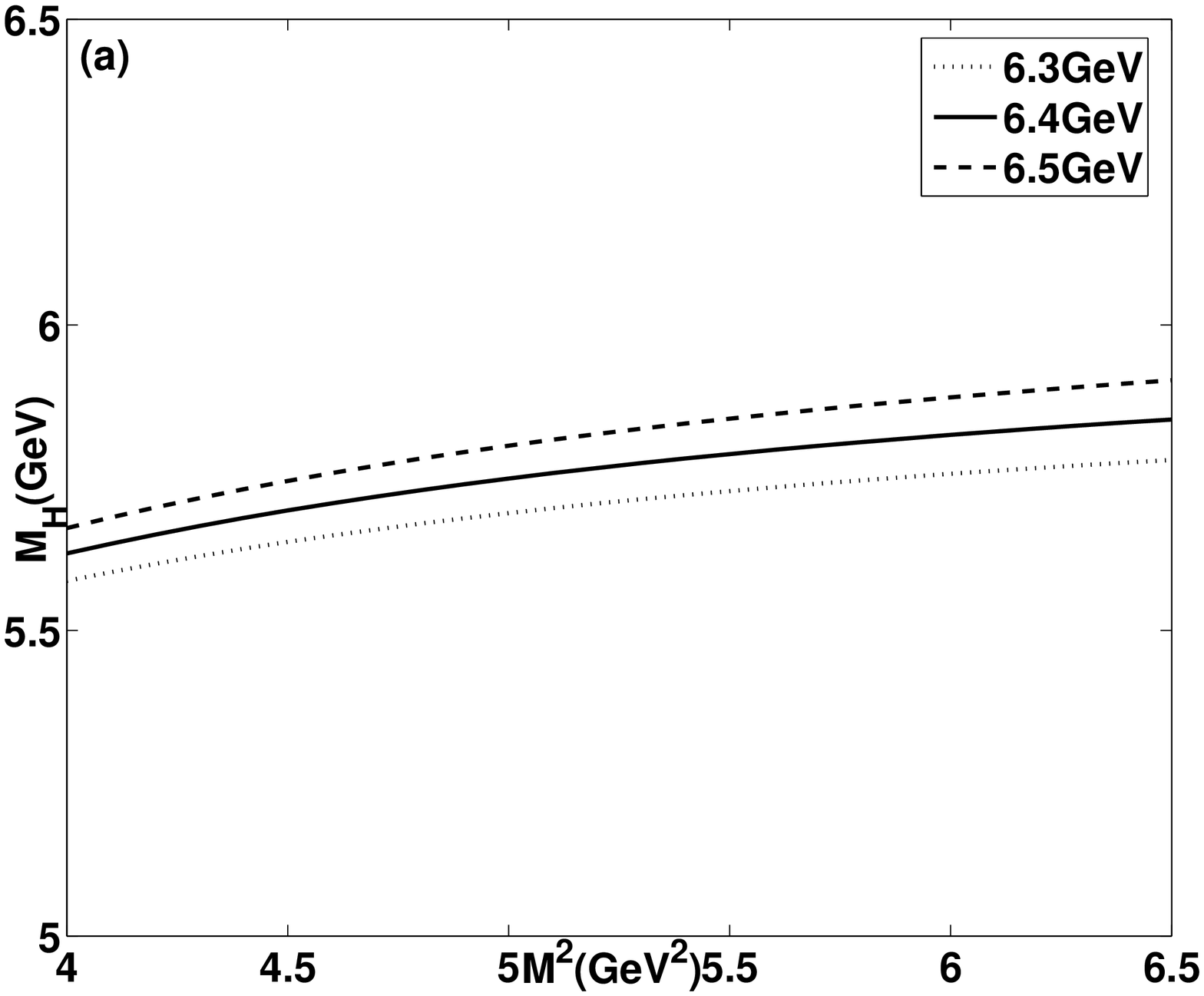}\epsfysize=3.5truecm\epsfbox{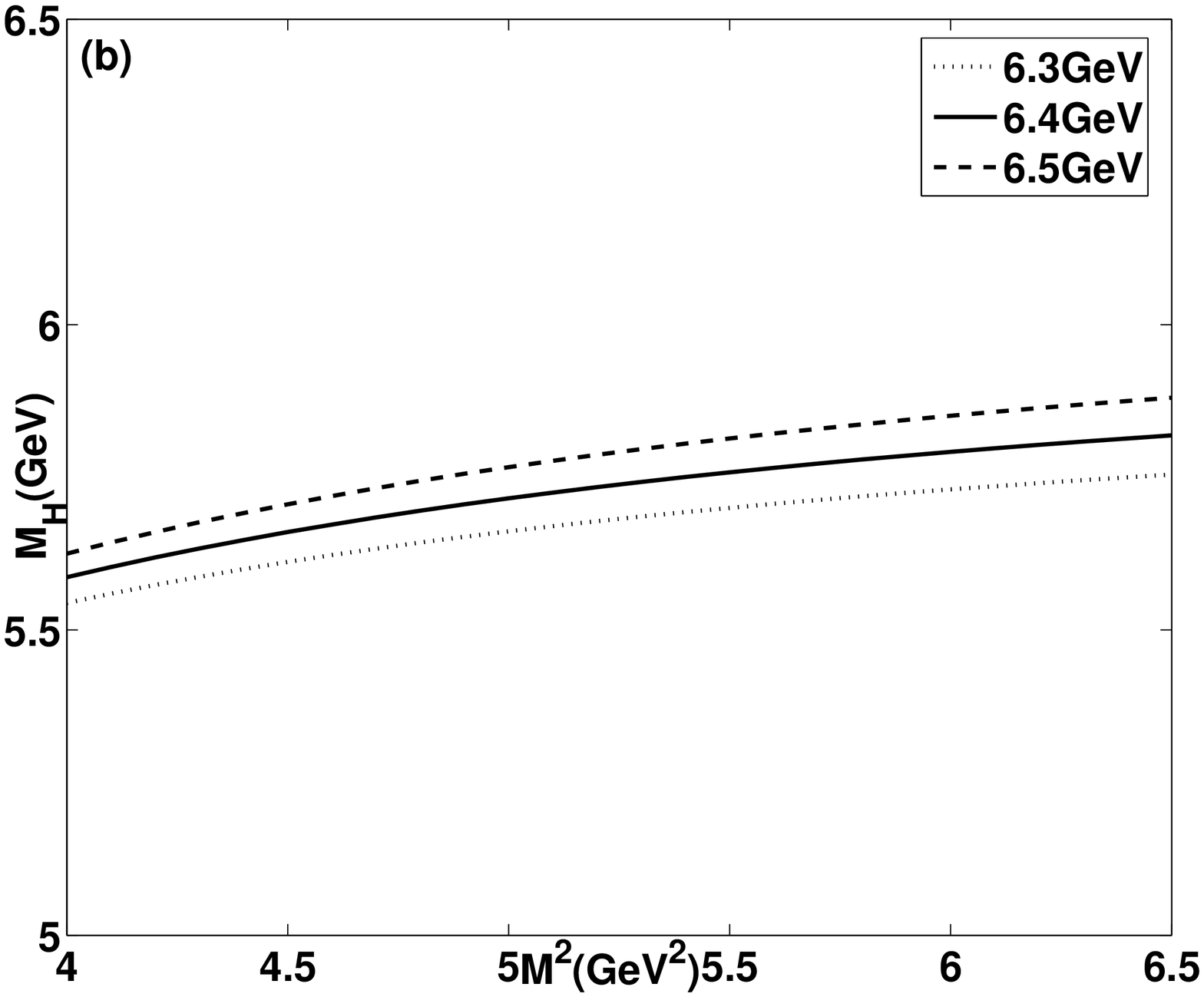}}
\caption{The dependence on $M^2$ for the masses of $\Xi_{c}$ and
$\Xi_{b}$. The continuum thresholds are taken as
$\sqrt{s_0}=3.0\sim3.2~\mbox{GeV}$, and
$\sqrt{s_0}=6.3\sim6.5~\mbox{GeV}$, respectively. (a) are from the
sum rule (\ref{sum rule m}), (b) from the sum rule (\ref{sum rule
q}).} \label{fig:3}
\end{figure}

\begin{figure}
\centerline{\epsfysize=3.5truecm
\epsfbox{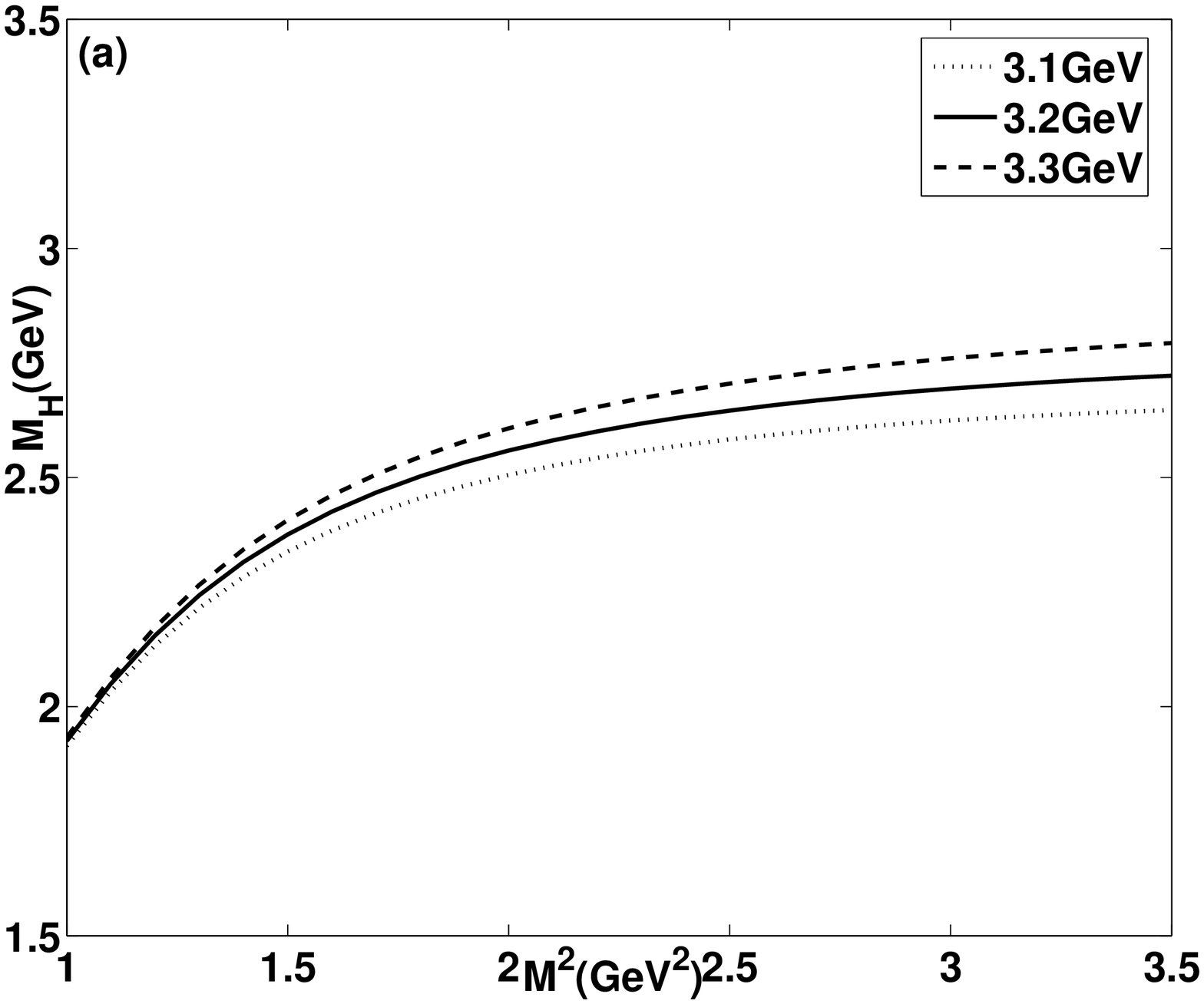}\epsfysize=3.5truecm
\epsfbox{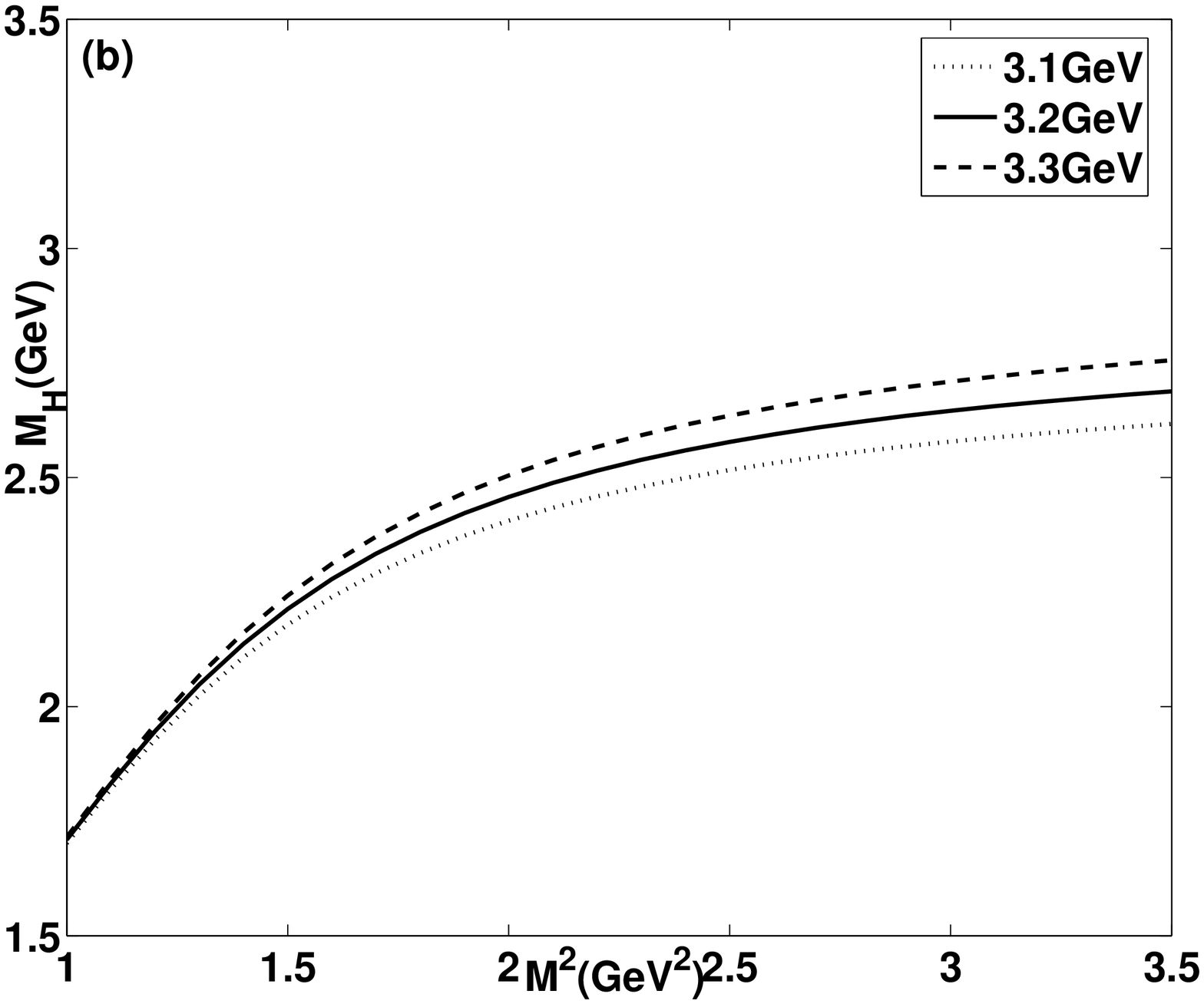}\epsfysize=3.5truecm
\epsfbox{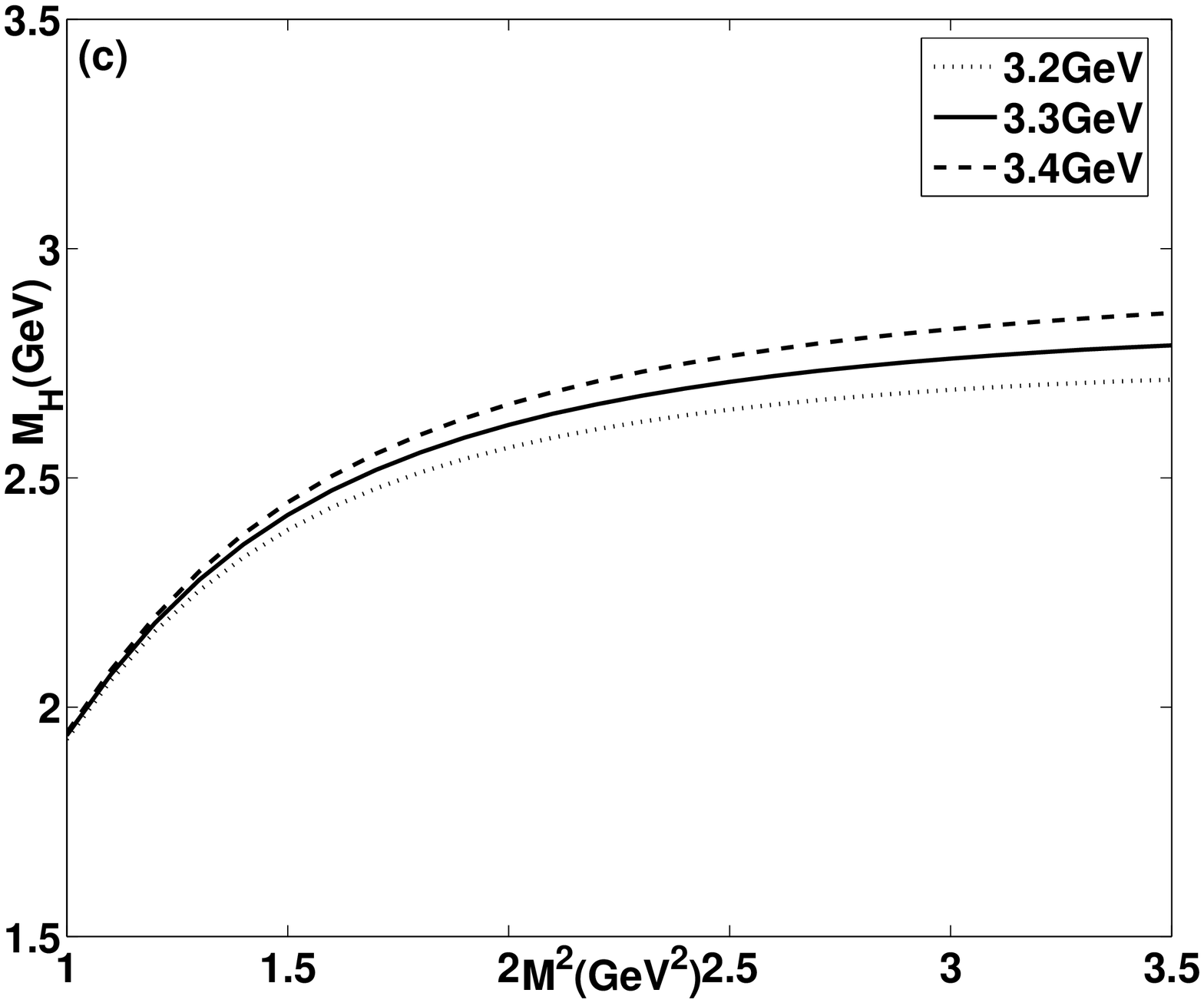}\epsfysize=3.5truecm
\epsfbox{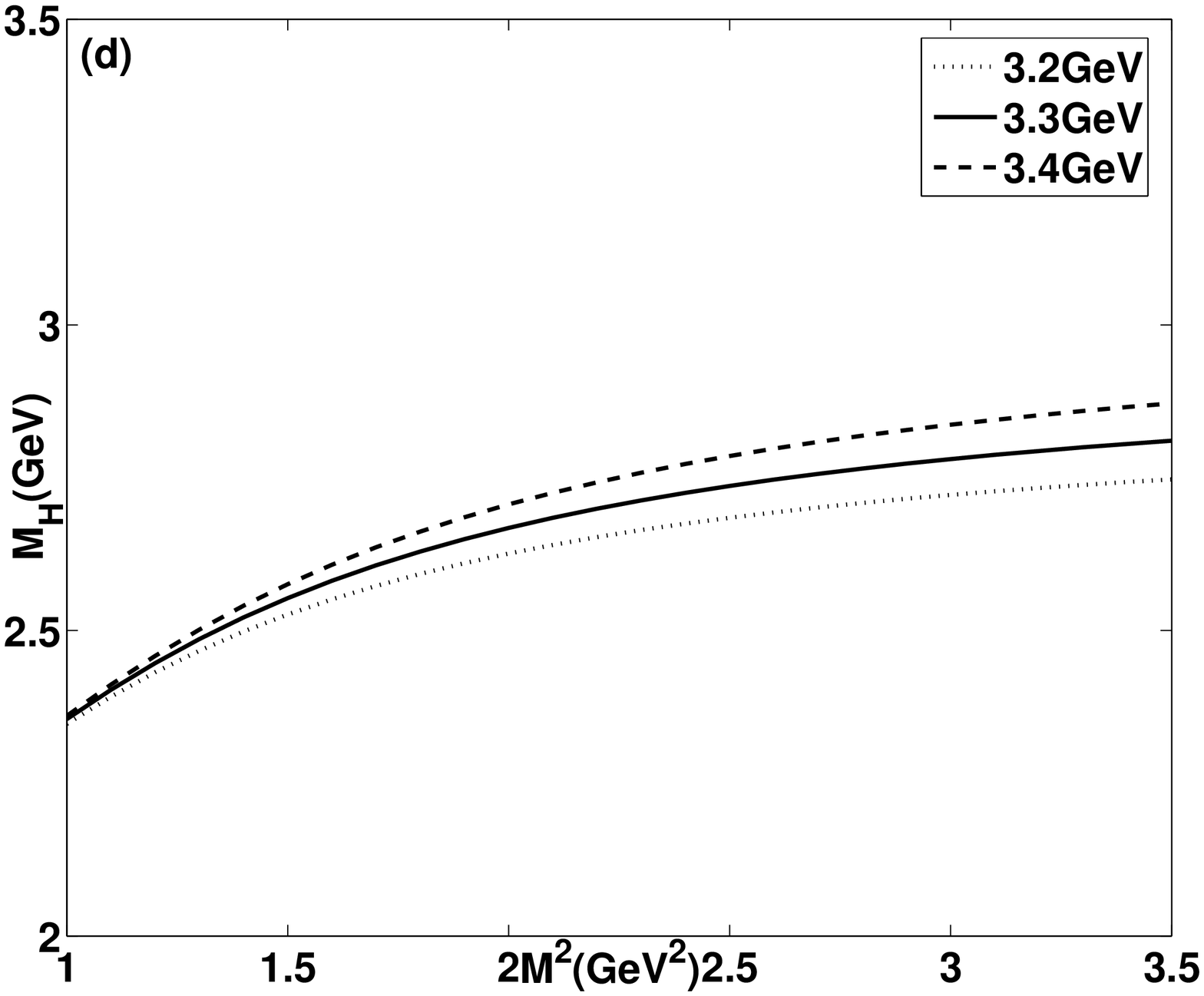}}\centerline{\epsfysize=3.5truecm
\epsfbox{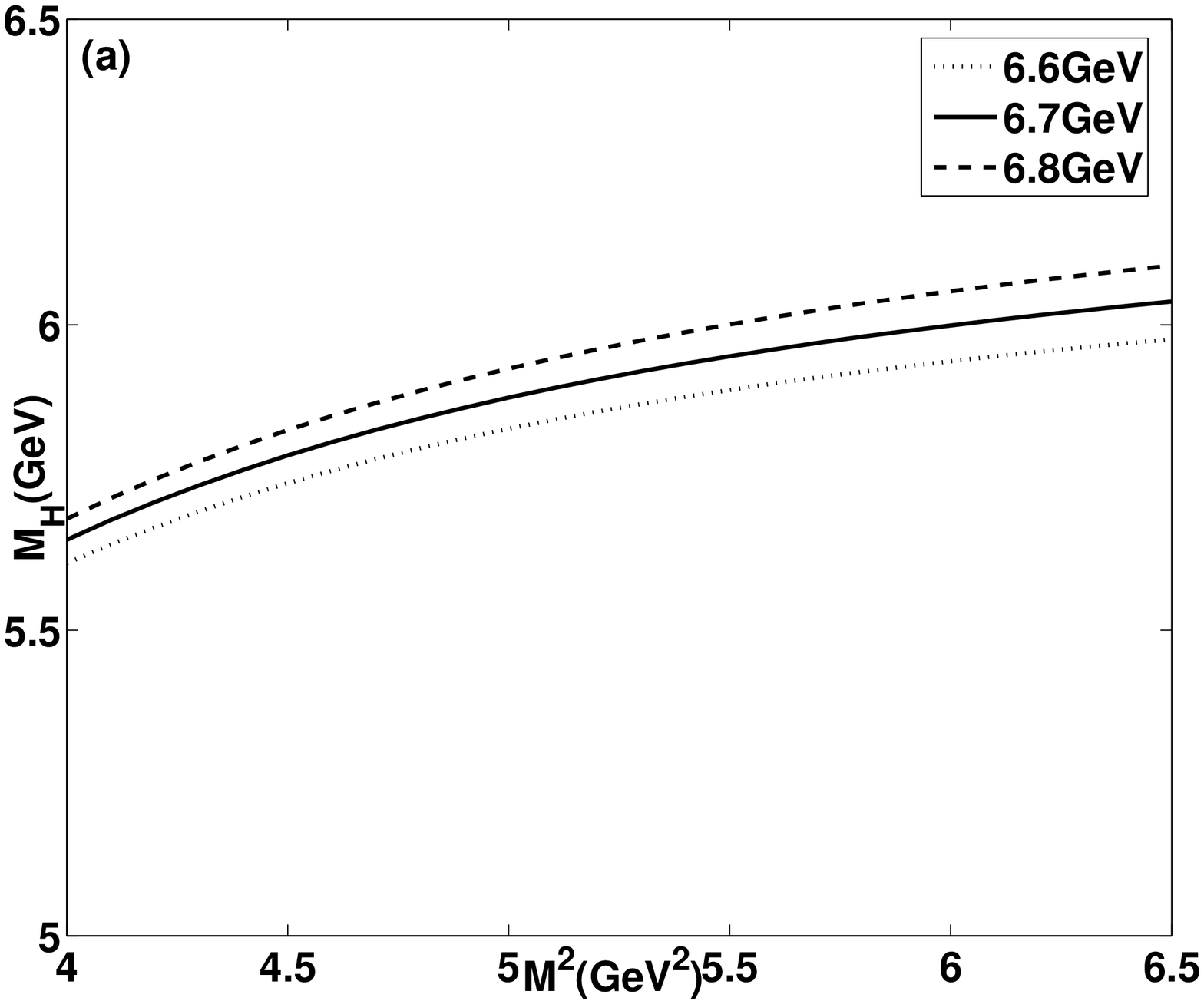}\epsfysize=3.5truecm
\epsfbox{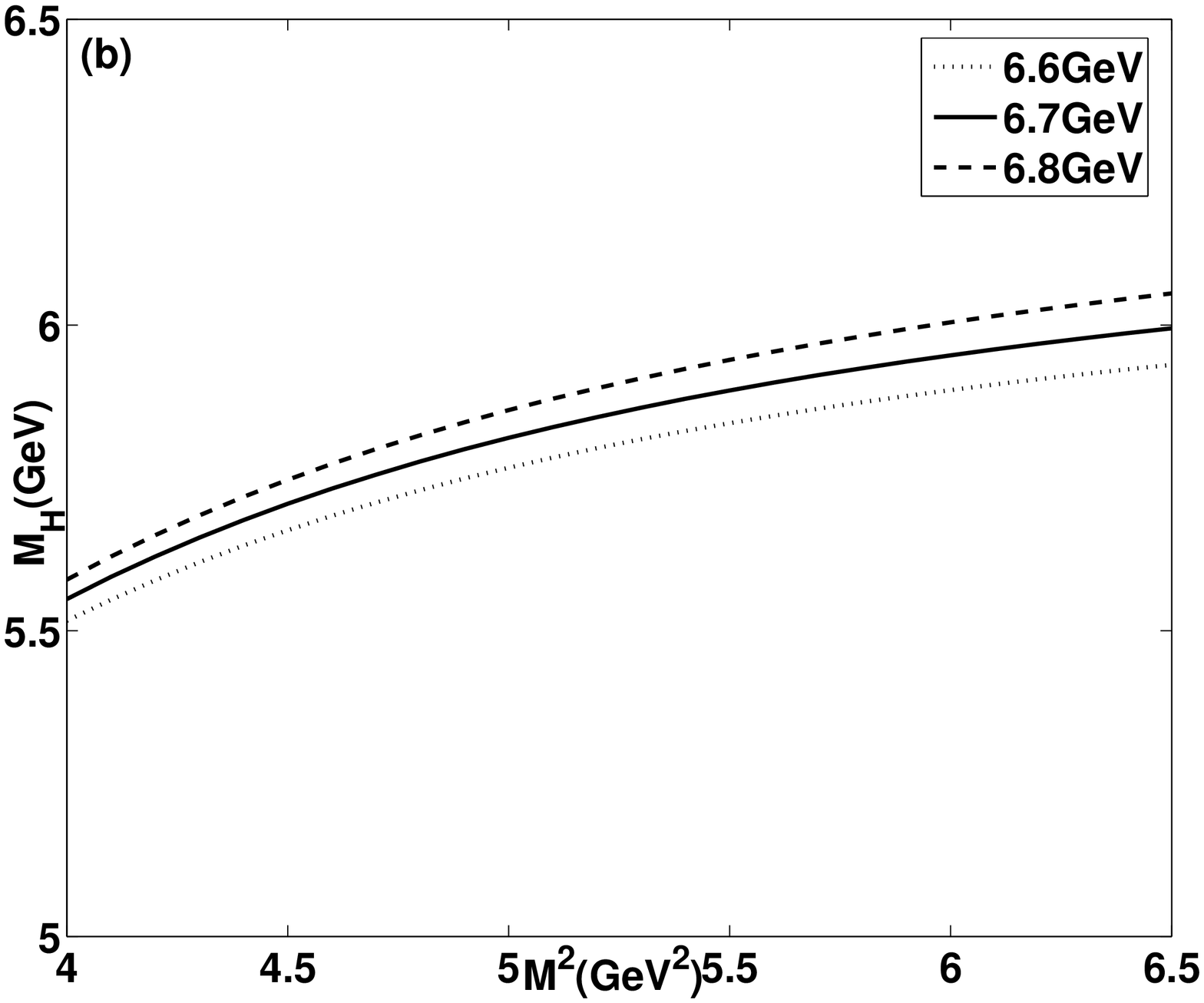}\epsfysize=3.5truecm
\epsfbox{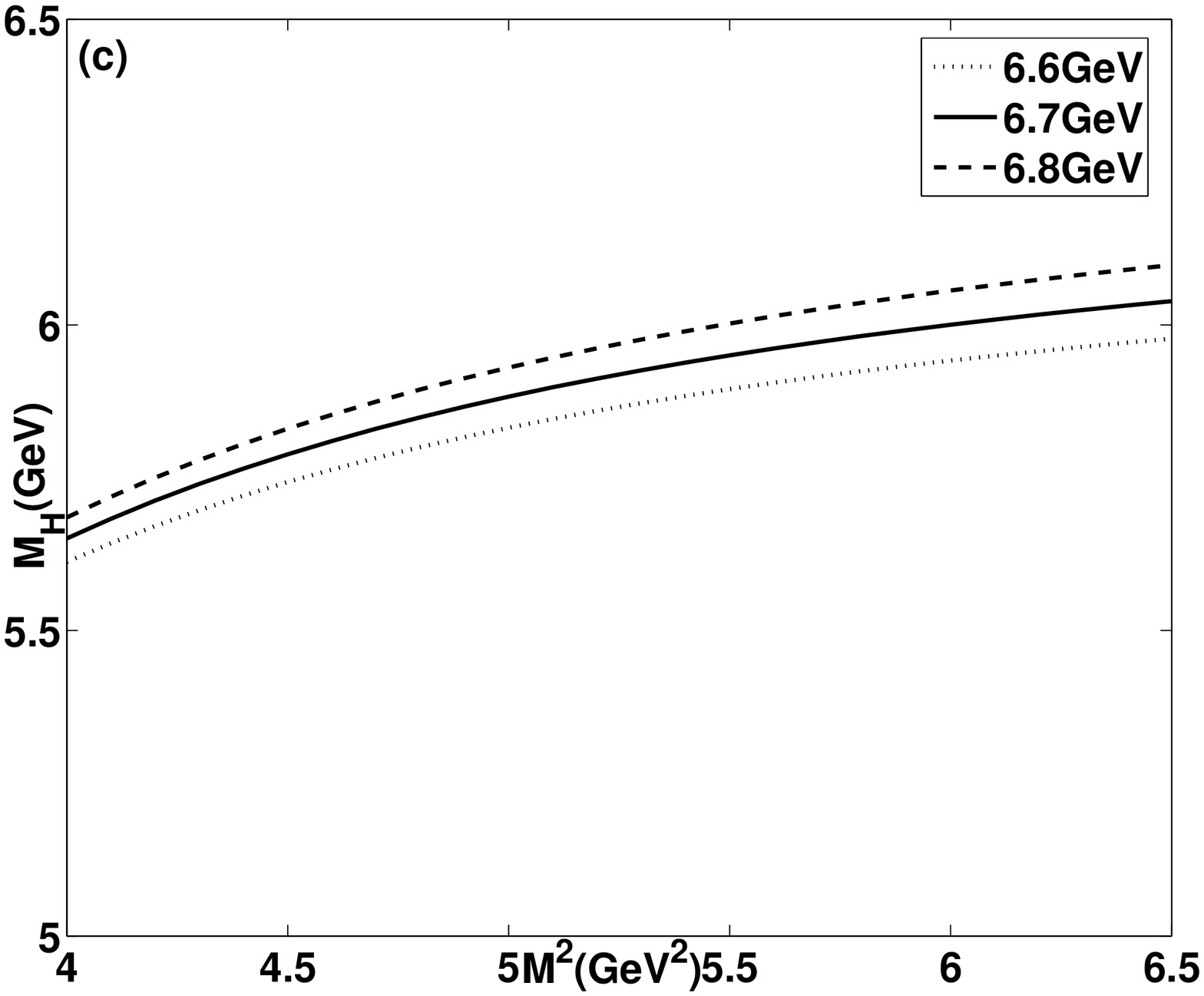}\epsfysize=3.5truecm
\epsfbox{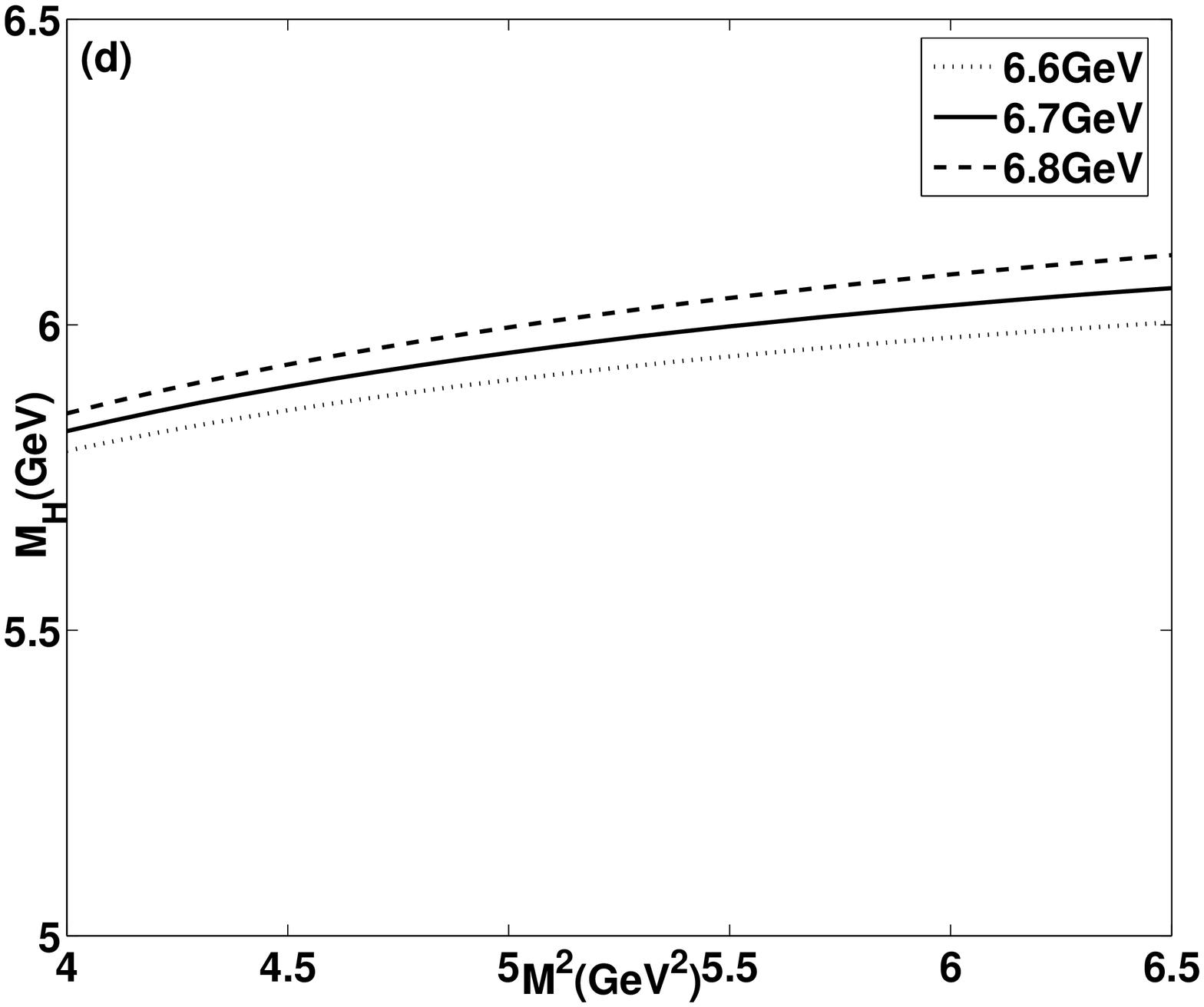}} \caption{The dependence on $M^2$ for the
masses of $\Xi_{c}^{'}$, $\Xi_{c}^{'*}$, $\Xi_{b}^{'}$, and
$\Xi_{b}^{'*}$. The continuum thresholds are orderly taken as
$\sqrt{s_0}=3.1\sim3.3~\mbox{GeV}$,
$\sqrt{s_0}=3.2\sim3.4~\mbox{GeV}$,
$\sqrt{s_0}=6.6\sim6.8~\mbox{GeV}$, and
$\sqrt{s_0}=6.6\sim6.8~\mbox{GeV}$. (a) and (c) are from the sum
rule (\ref{sum rule m}), (b) and (d) from the sum rule (\ref{sum
rule q}).} \label{fig:4}
\end{figure}

\begin{figure}
\centerline{\epsfysize=3.5truecm
\epsfbox{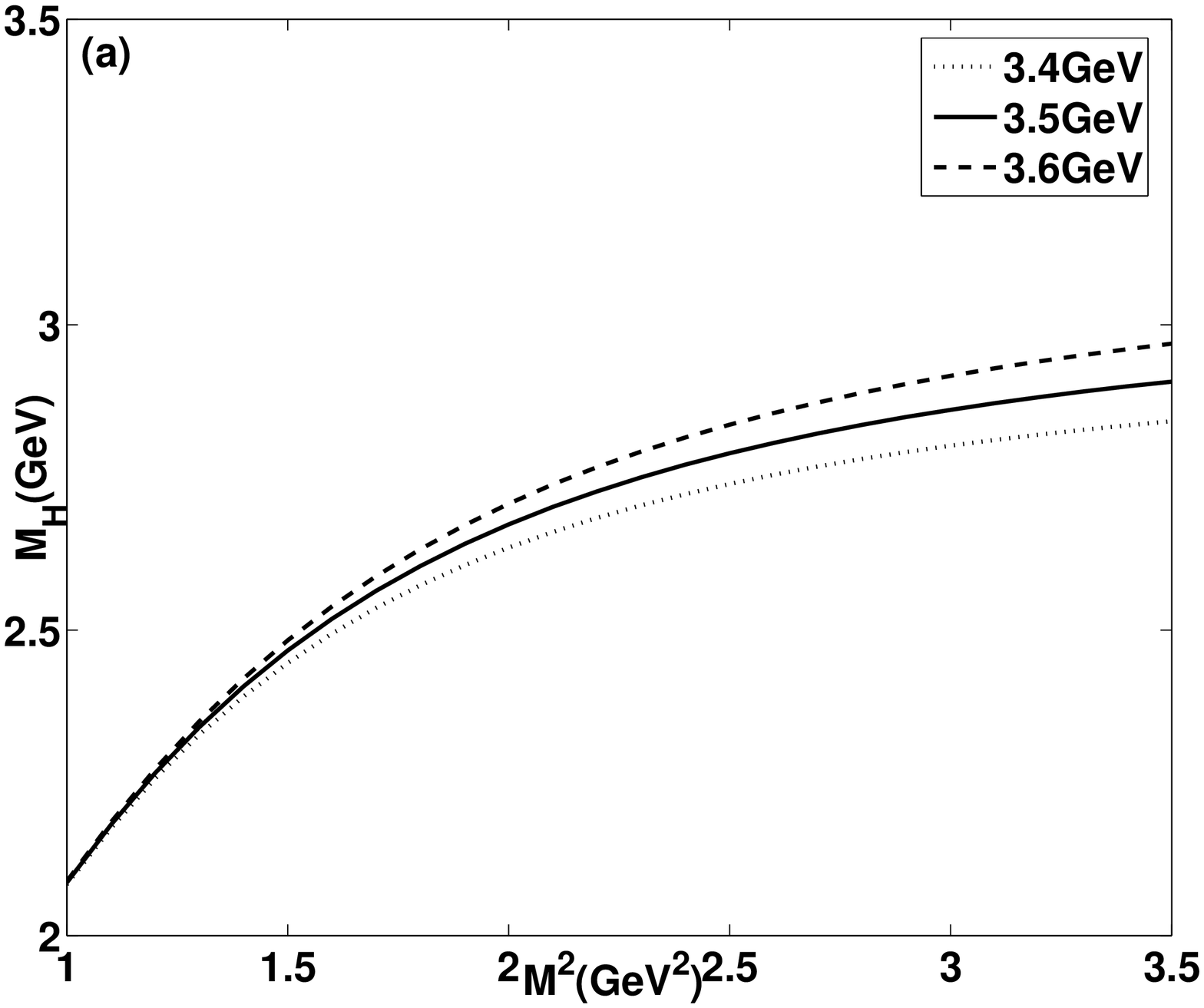}\epsfysize=3.5truecm
\epsfbox{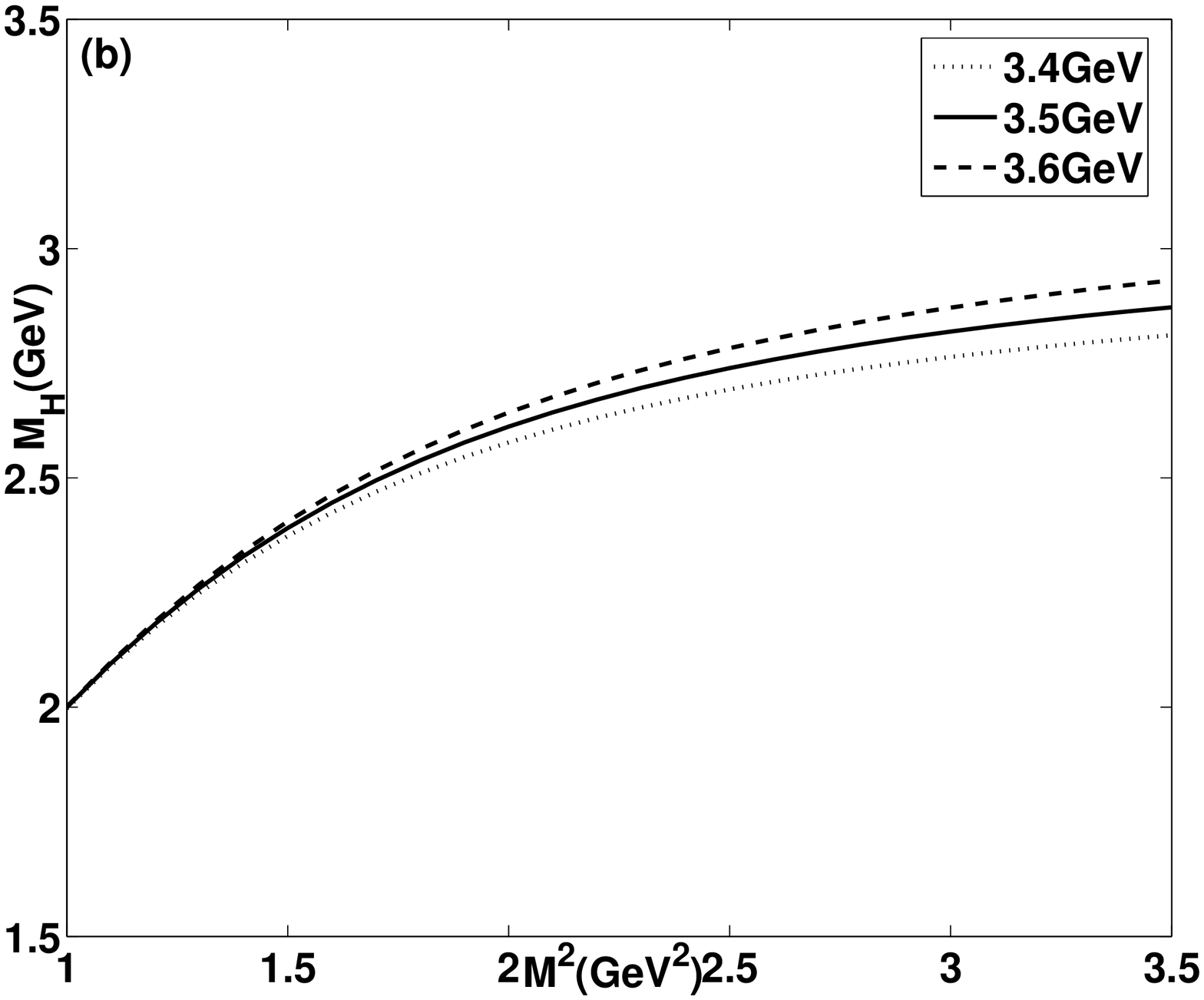}\epsfysize=3.5truecm
\epsfbox{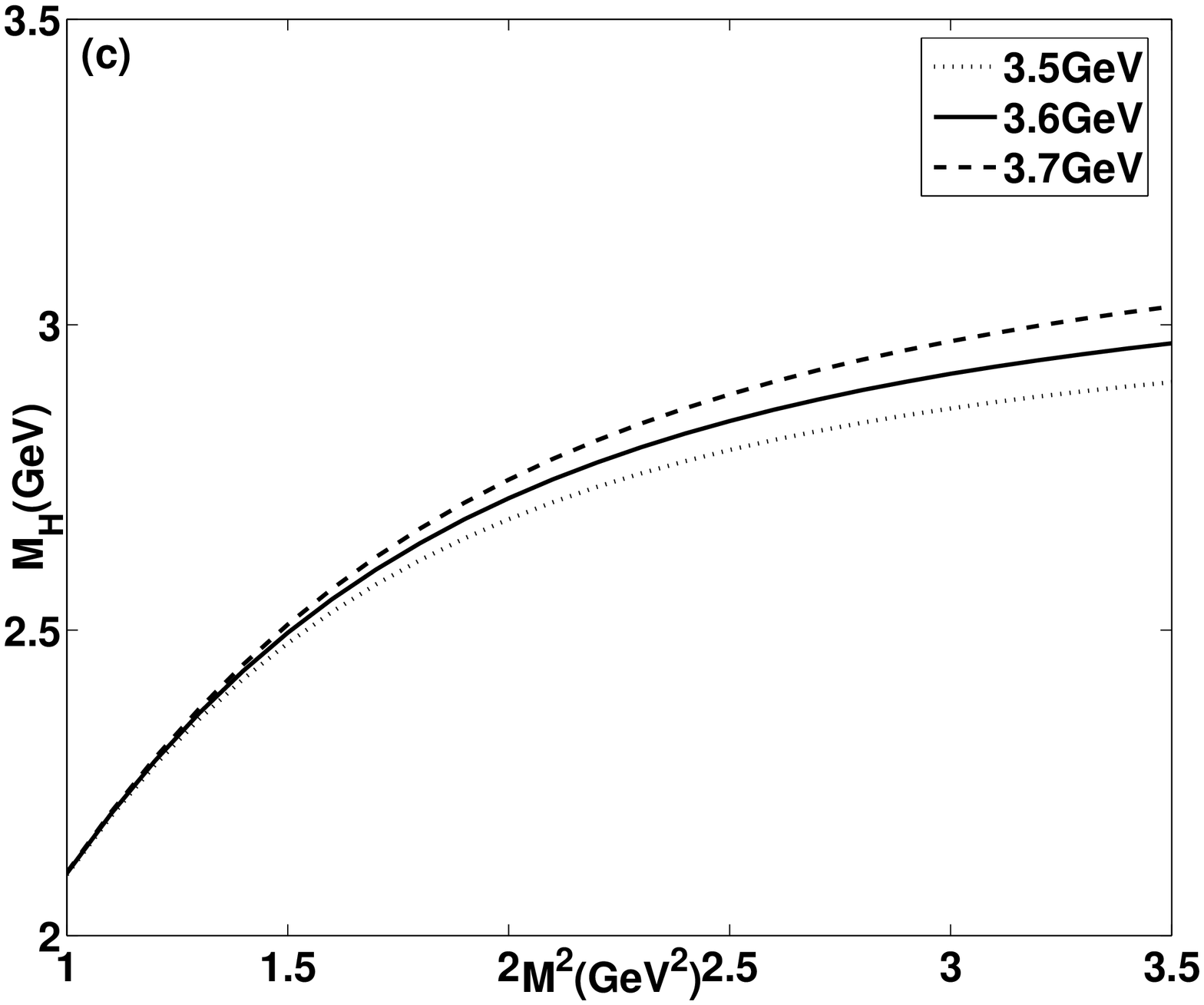}\epsfysize=3.5truecm
\epsfbox{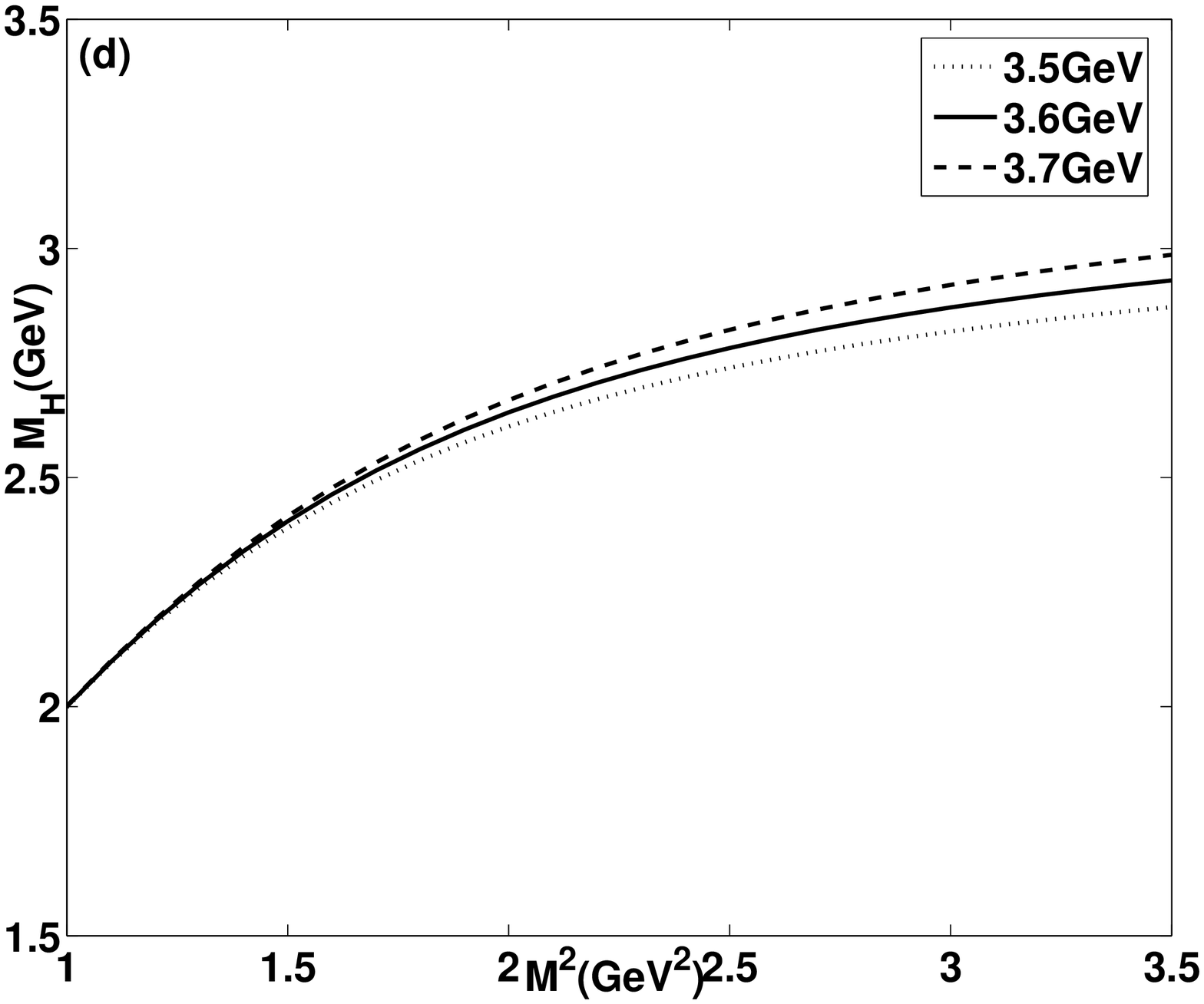}}\centerline{\epsfysize=3.5truecm
\epsfbox{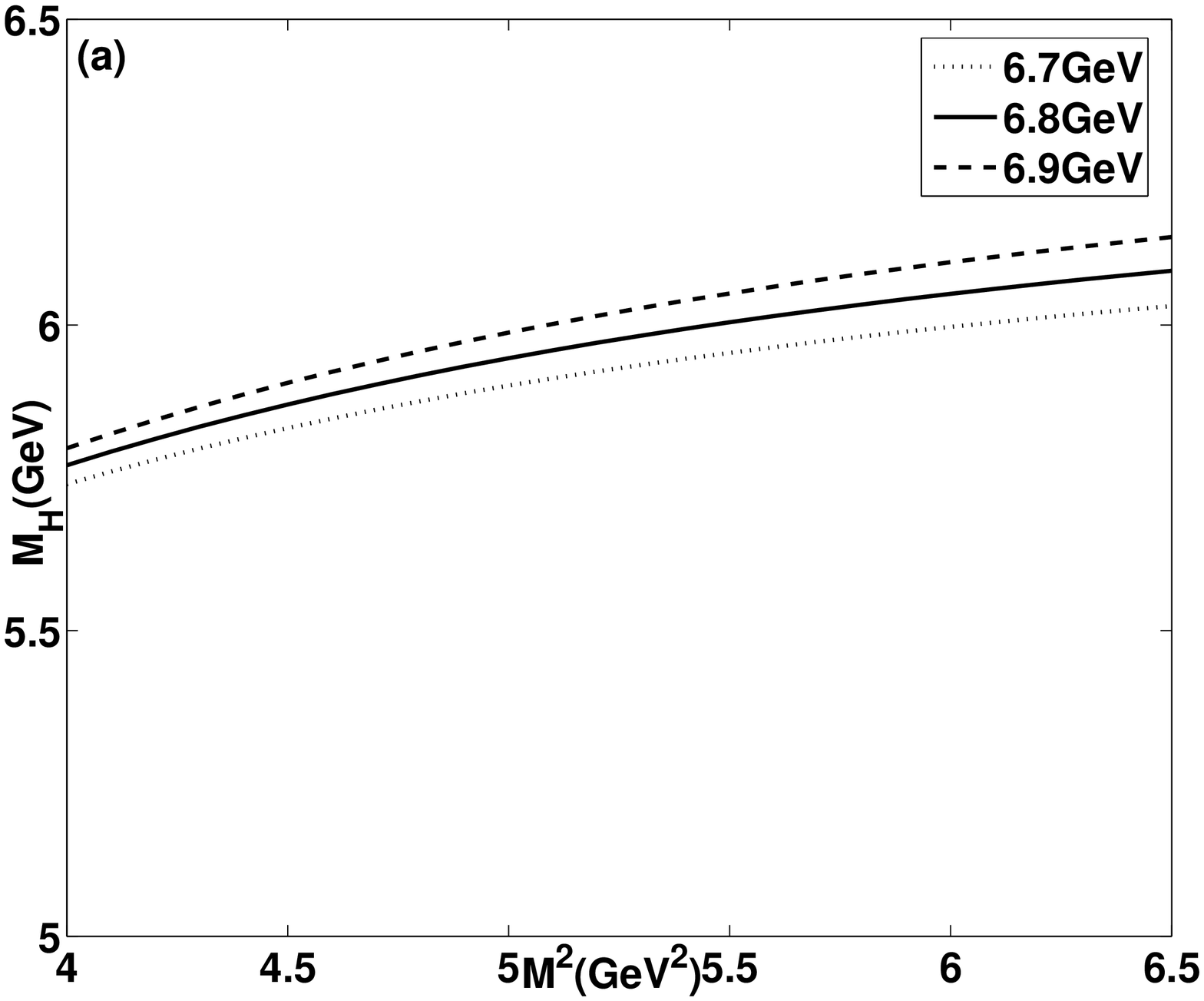}\epsfysize=3.5truecm
\epsfbox{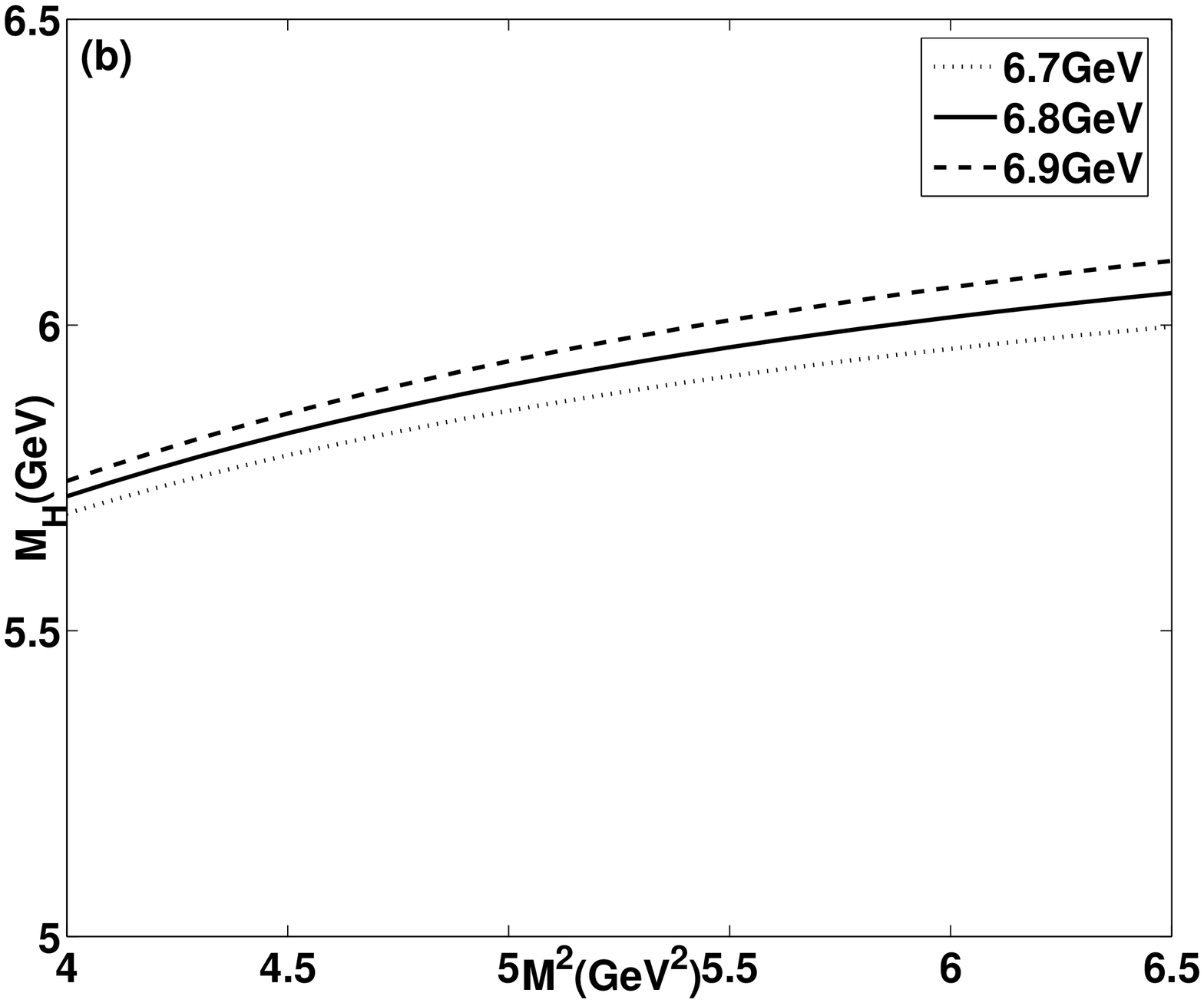}\epsfysize=3.5truecm
\epsfbox{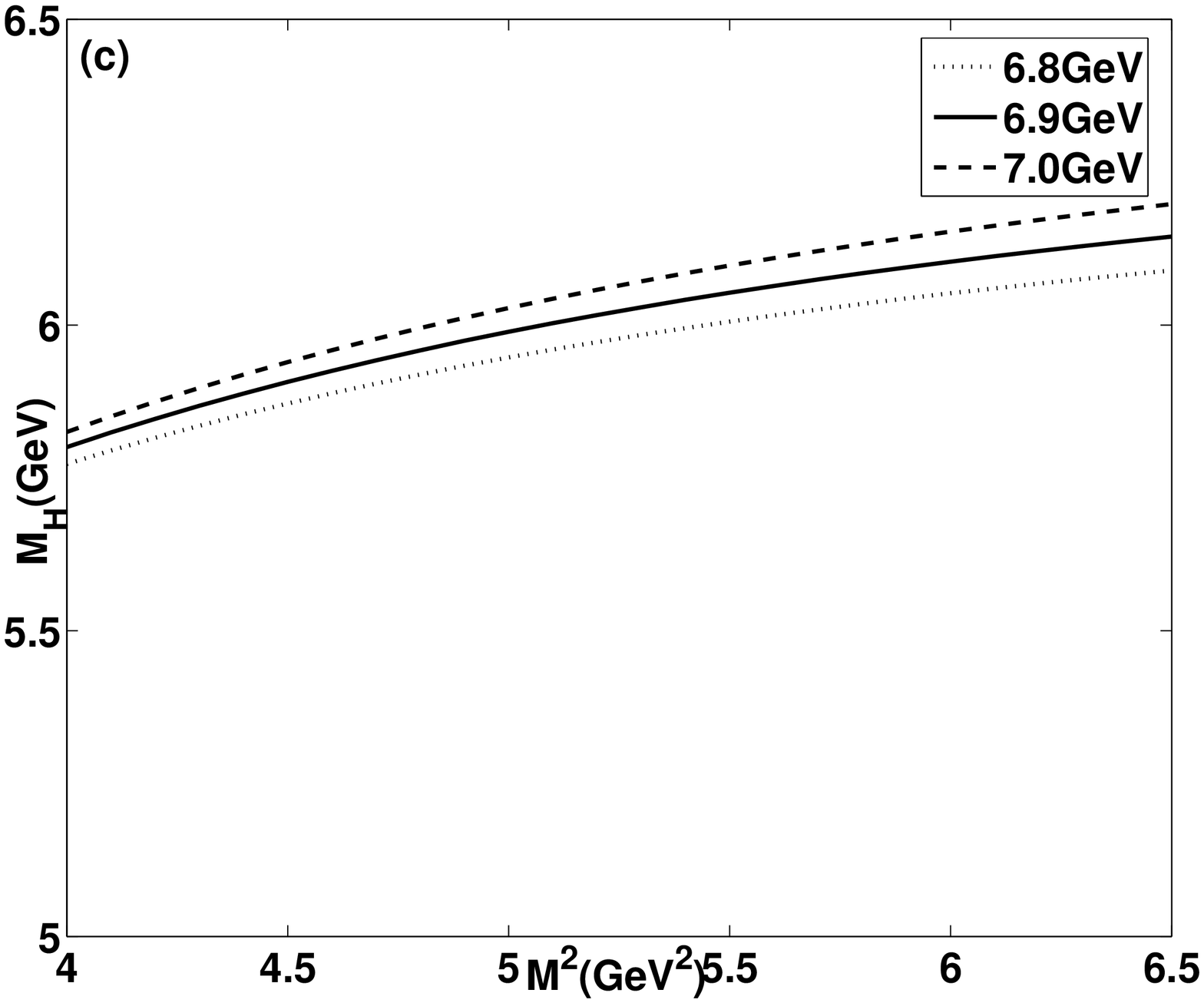}\epsfysize=3.5truecm
\epsfbox{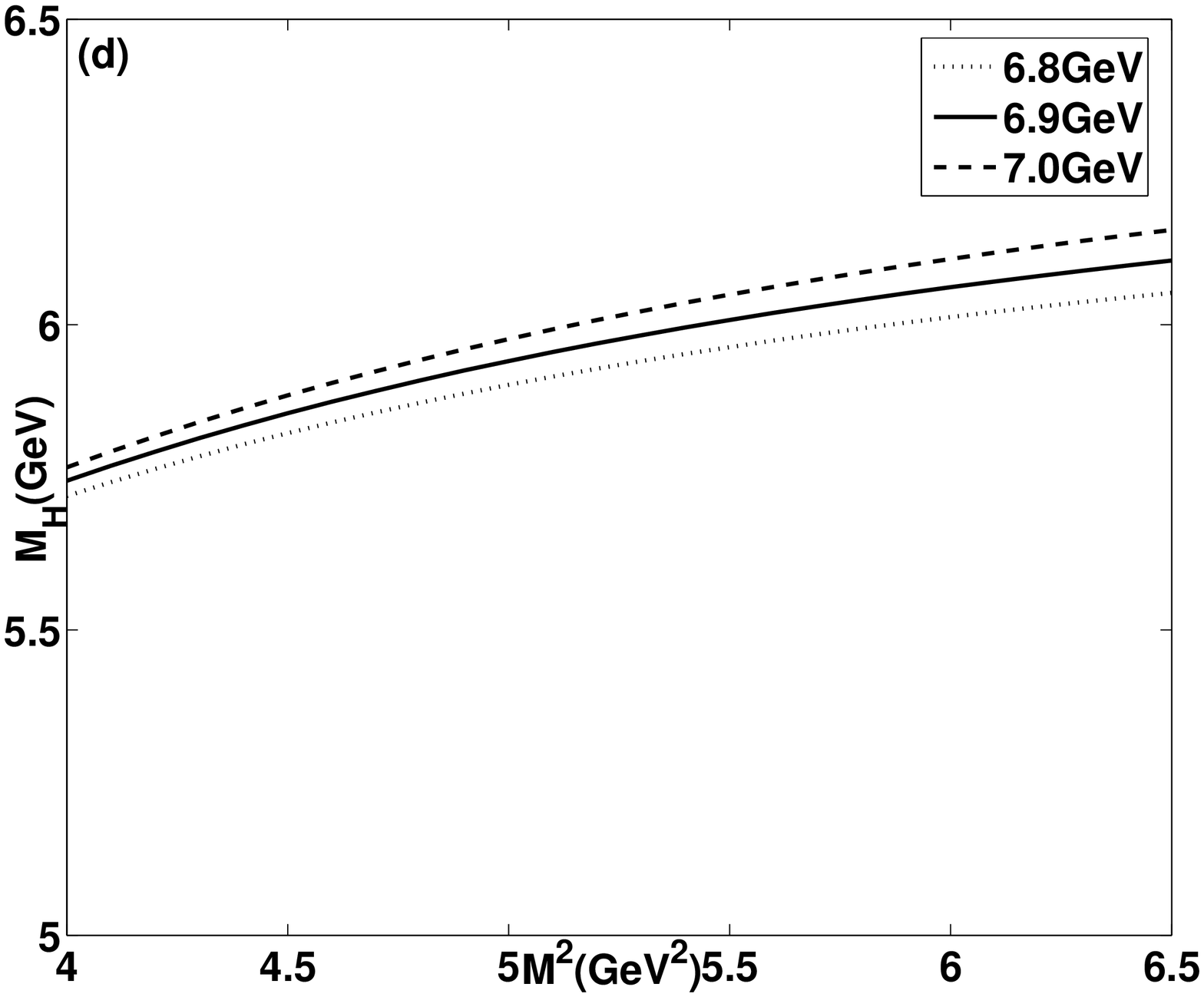}} \caption{The dependence on $M^2$ for the
masses of $\Xi_{1c}$, $\Xi_{1c}^{*}$, $\Xi_{1b}$, and
$\Xi_{1b}^{*}$. The continuum thresholds are orderly taken as
$\sqrt{s_0}=3.4\sim3.6~\mbox{GeV}$,
$\sqrt{s_0}=3.5\sim3.7~\mbox{GeV}$,
$\sqrt{s_0}=6.7\sim6.9~\mbox{GeV}$, and
$\sqrt{s_0}=6.8\sim7.0~\mbox{GeV}$. (a) and (c) are from the sum
rule (\ref{sum rule m}), (b) and (d) from the sum rule (\ref{sum
rule q}).} \label{fig:5}
\end{figure}

\begin{figure}
\centerline{\epsfysize=3.5truecm
\epsfbox{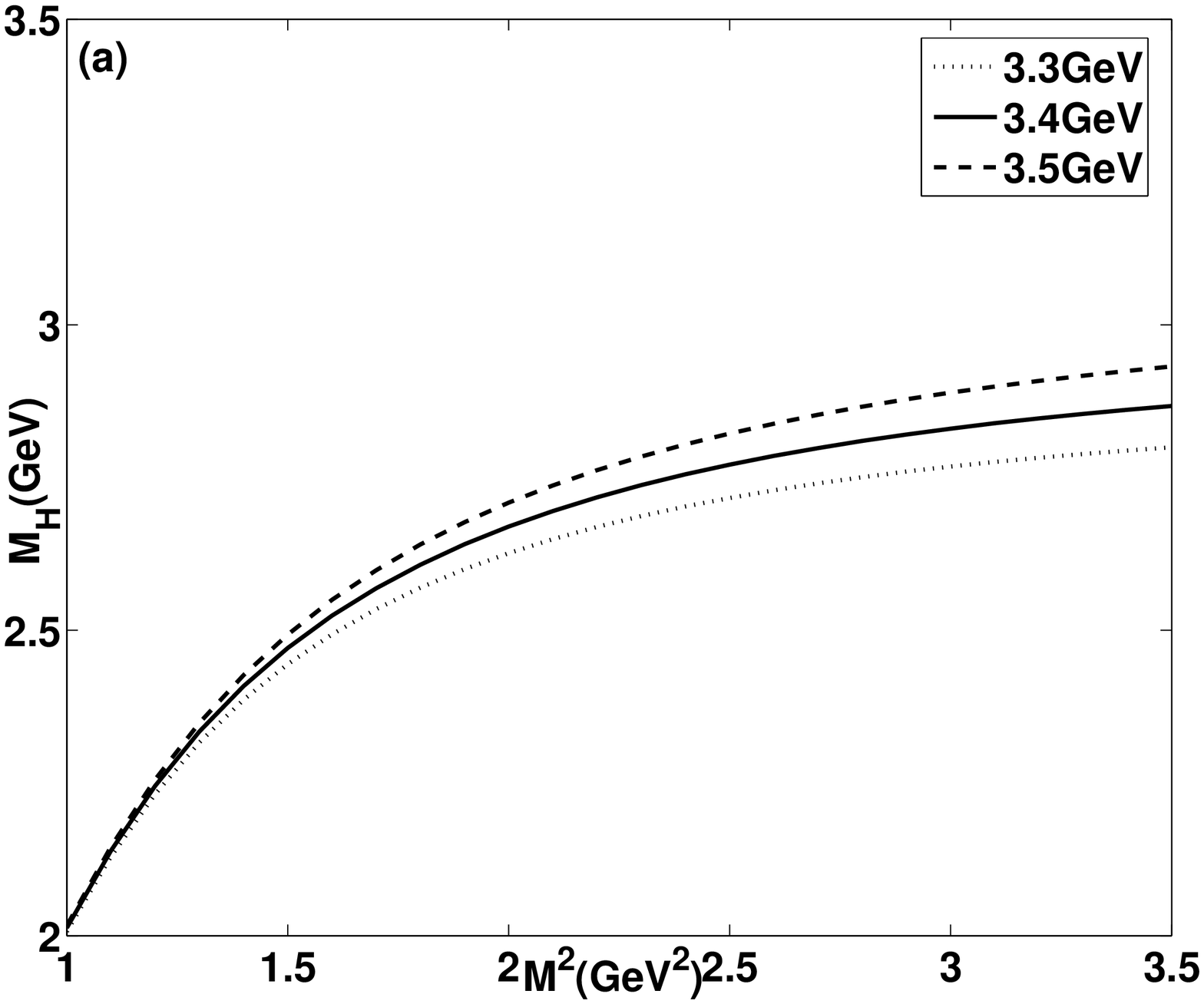}\epsfysize=3.5truecm
\epsfbox{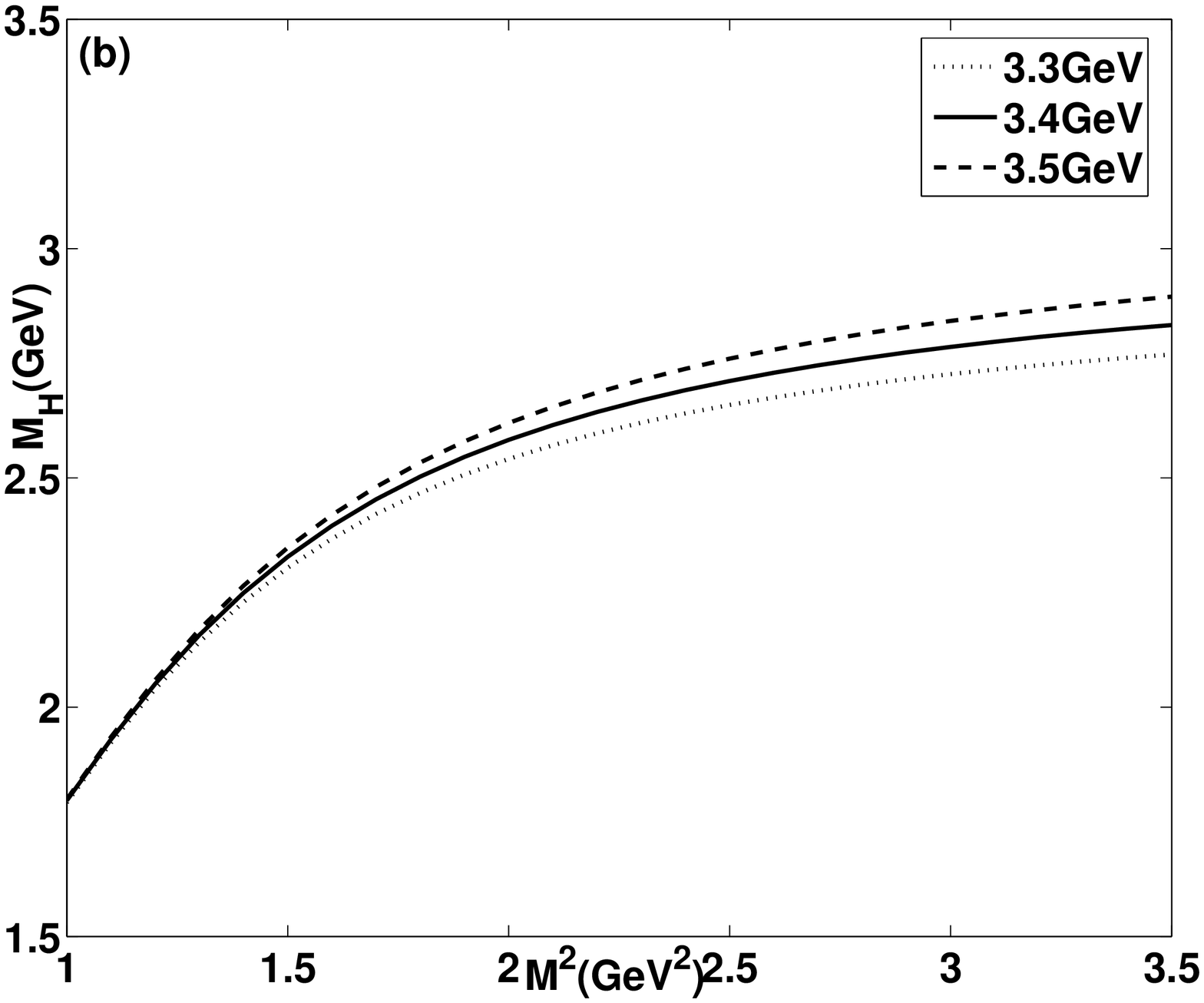}\epsfysize=3.5truecm
\epsfbox{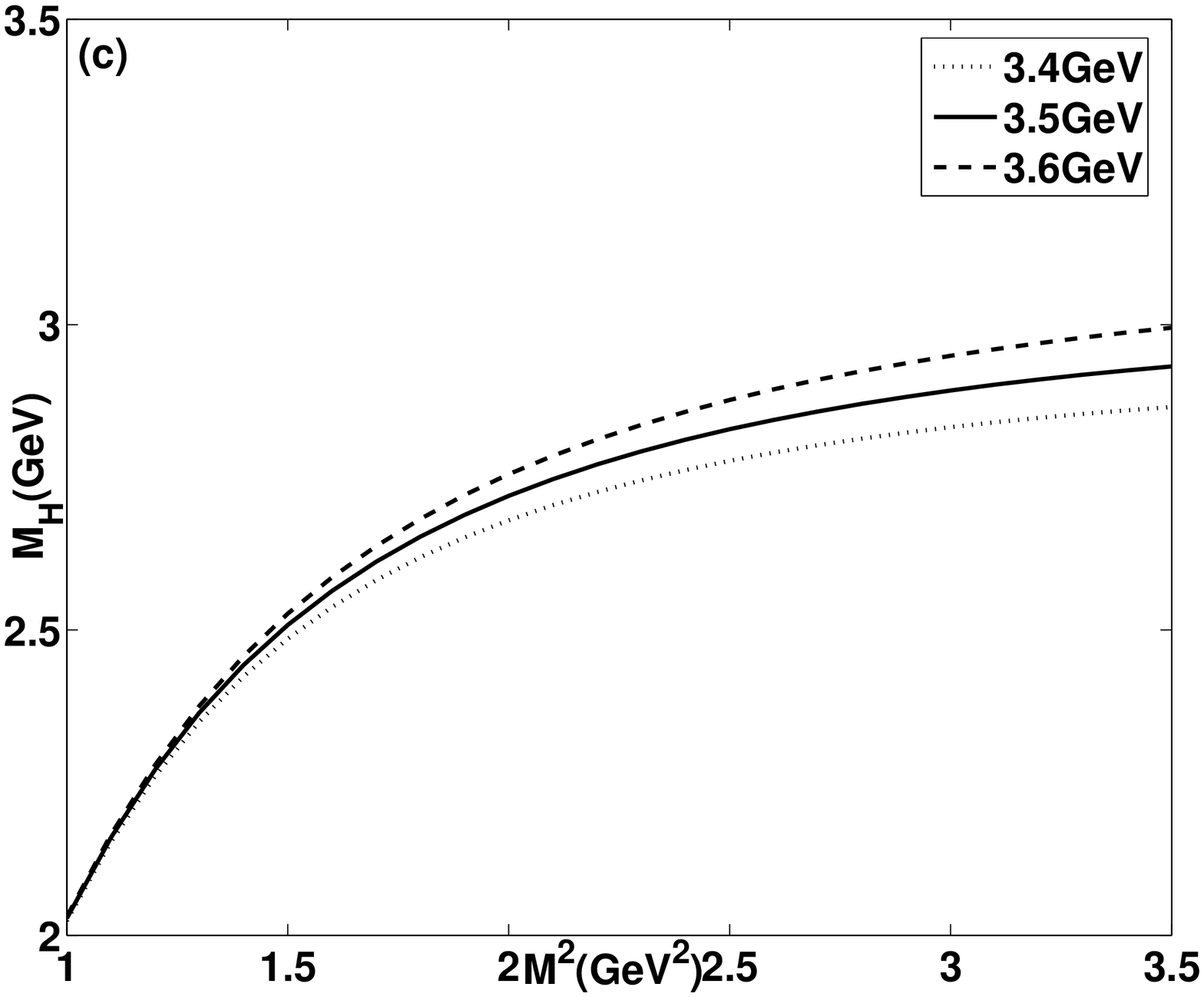}\epsfysize=3.5truecm
\epsfbox{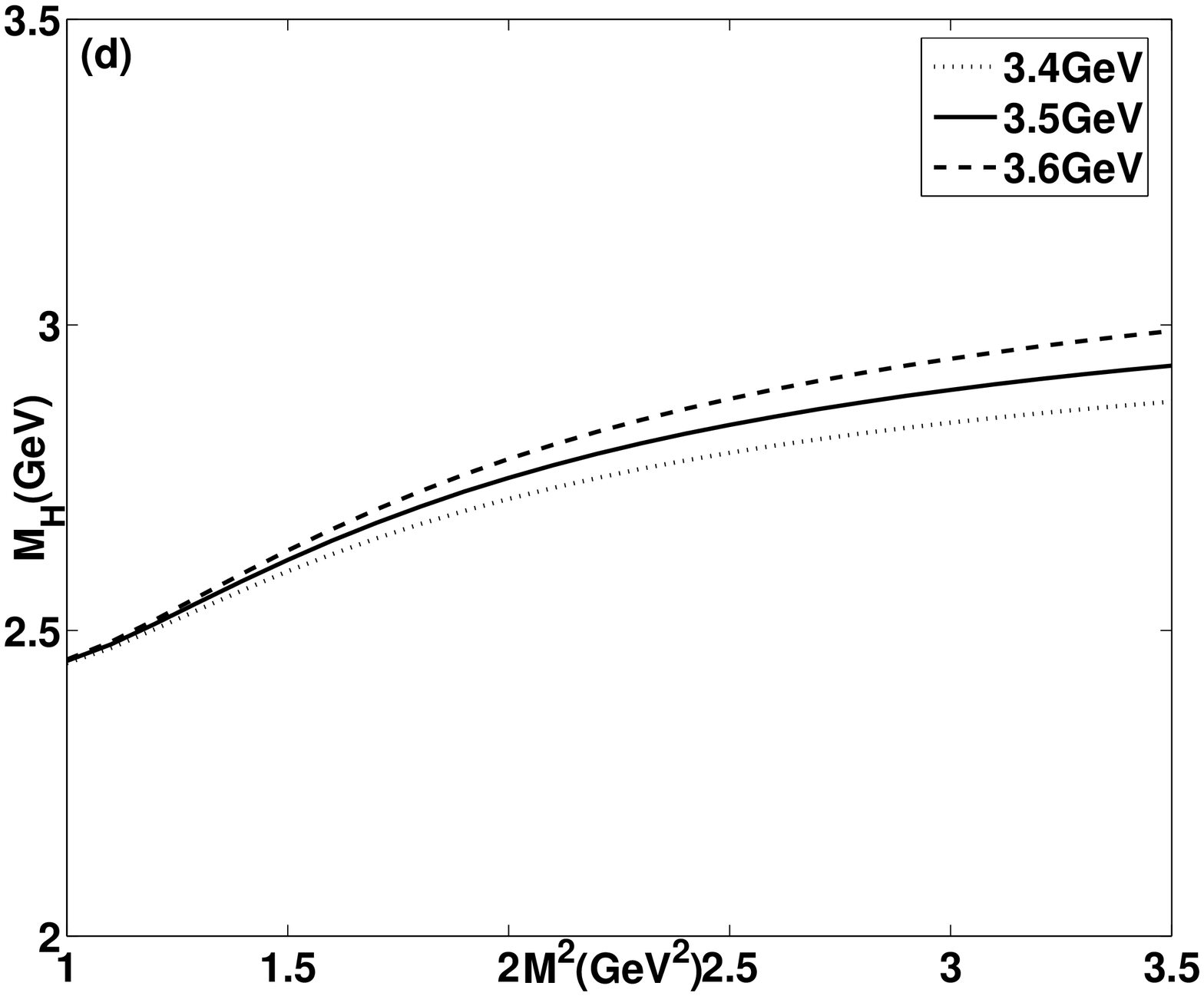}}\centerline{\epsfysize=3.5truecm
\epsfbox{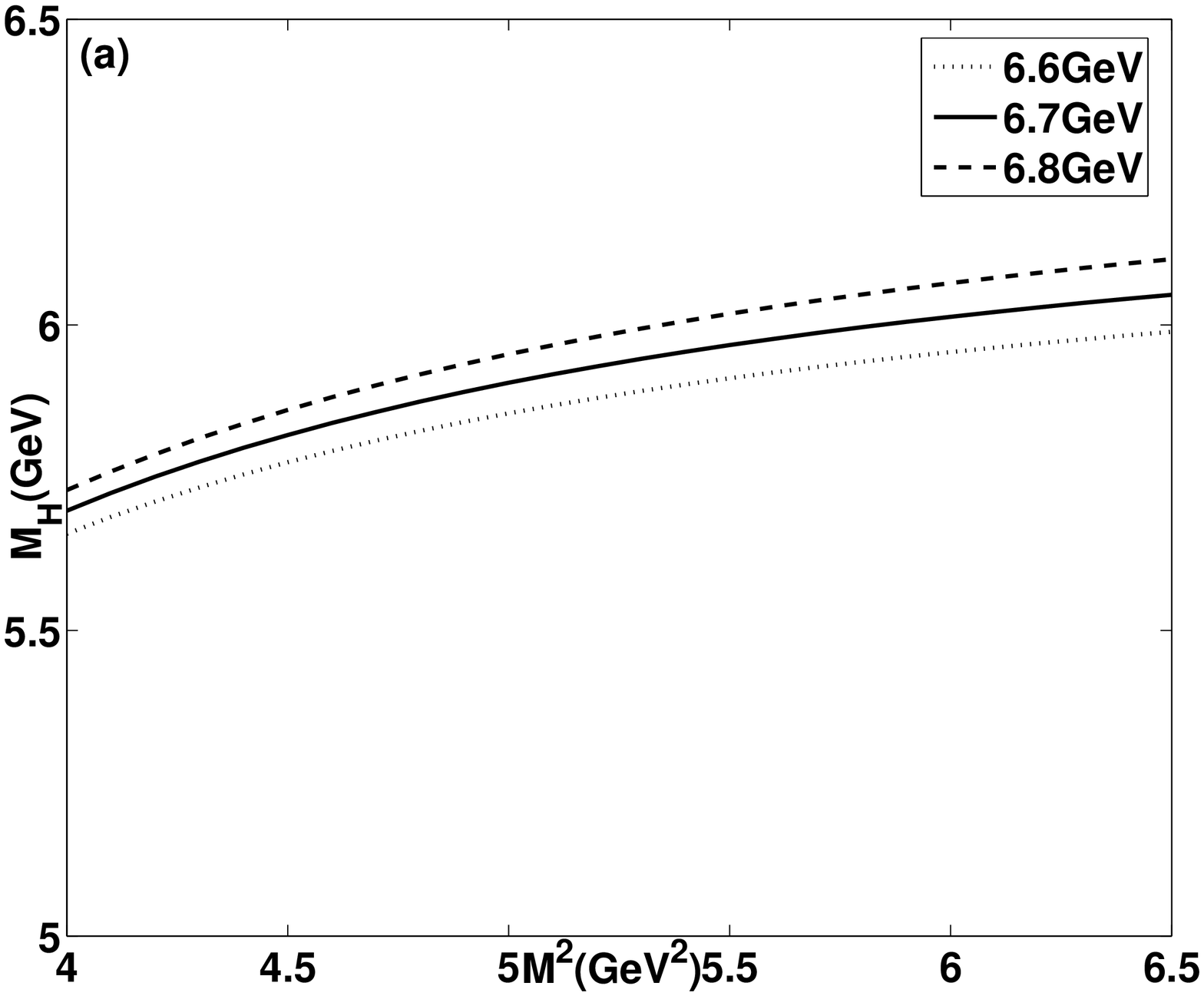}\epsfysize=3.5truecm
\epsfbox{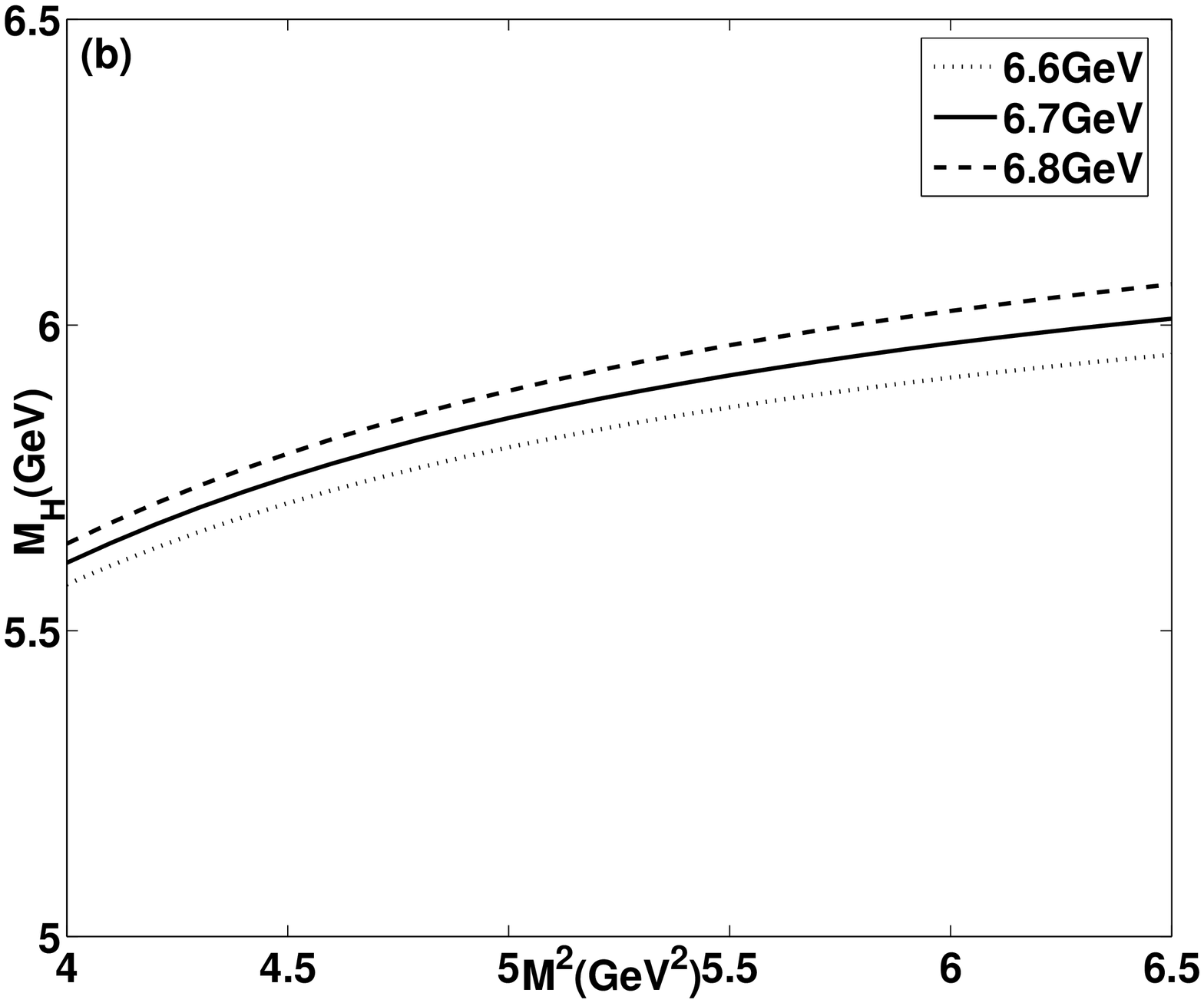}\epsfysize=3.5truecm
\epsfbox{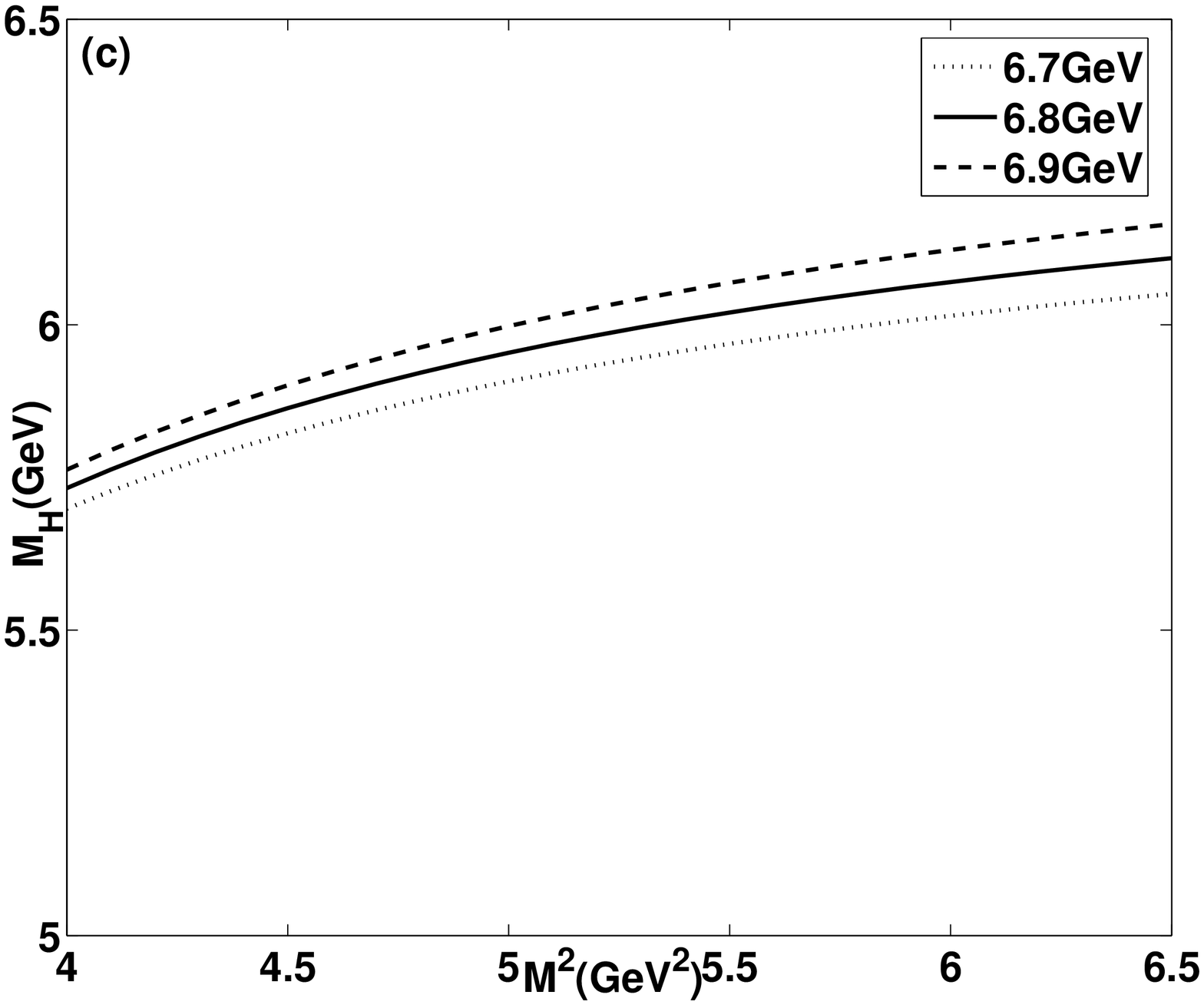}\epsfysize=3.5truecm
\epsfbox{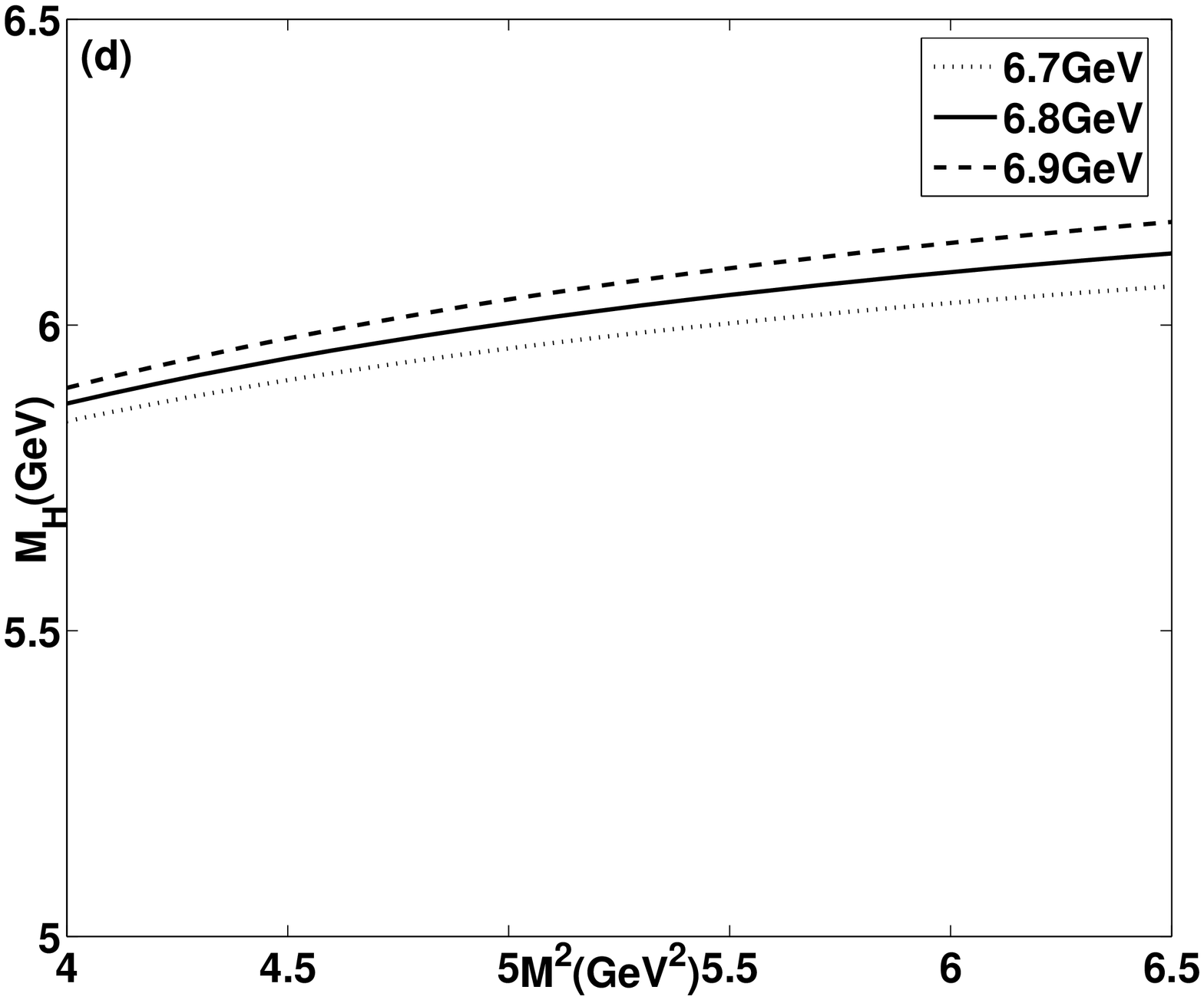}} \caption{The dependence on $M^2$ for
the masses of $\Omega_{c}$, $\Omega_{c}^{*}$, $\Omega_{b}$, and
$\Omega_{b}^{*}$. The continuum thresholds are orderly taken as
$\sqrt{s_0}=3.3\sim3.5~\mbox{GeV}$,
$\sqrt{s_0}=3.4\sim3.6~\mbox{GeV}$,
$\sqrt{s_0}=6.6\sim6.8~\mbox{GeV}$, and
$\sqrt{s_0}=6.7\sim6.9~\mbox{GeV}$. (a) and (c) are from the sum
rule (\ref{sum rule m}), (b) and (d) from the sum rule (\ref{sum
rule q}).} \label{fig:6}
\end{figure}

\begin{table}\caption{ The mass spectra of charmed and bottom baryons (mass in
unit of$~\mbox{MeV}$ except for ``Our works")}
 \centerline{\begin{tabular}{ c  c  c  c  c  c  c  c c c c}  \hline\hline
Baryon                          &     $J^{P}$            &  $S_{\ell}$  &  $L_{\ell}$  &   $J_{\ell}^{P_{\ell}}$  &  Experiment                                                                  & Our works~(\mbox{GeV})                & Refs. \cite{quark model}                       & Ref.\cite{mass formular}                & Ref.\cite{lattice}      & Ref.\cite{zhu}           \\
\hline
 $\Lambda_{b}$                  &  $\frac{1}{2}^{+} $    &     0        &      0       &        $0^{+}$           & $5619.7\pm1.2$ \cite{lamdar-b}                                               & $5.69\pm0.13$\cite{jrzhang}          & 5622                                            &   5620                                  &    5672                 & $5637_{-56}^{+68}$       \\
                                &                        &              &              &                          & $5624\pm9$ \cite{PDG}                                                        &                                      &                                                 &                                         &                         &                          \\
\hline
 $\Lambda_{c}^{+}$              &  $\frac{1}{2}^{+} $    &     0        &      0       &        $0^{+}$           & $2286.46\pm0.14$ \cite{PDG}                                                  & $2.31\pm0.19$\cite{jrzhang}          & 2297                                            &   2285                                  &    2290                 & $2271_{-49}^{+67}$       \\
\hline
$\Sigma_{b}^{+}$                &  $\frac{1}{2}^{+} $    &     1        &      0       &        $1^{+}$           & $5807.8_{-2.2}^{+2.0}\pm1.7$ \cite{sigma-b}                                  & $5.73\pm0.21$\cite{jrzhang}          & 5805                                            &   5820                                  &    5847                 & $5809_{-76}^{+82}$       \\
$\Sigma_{b}^{-}$                &                        &              &              &                          & $5815.2\pm1.0\pm1.7$ \cite{sigma-b}                                          &                                      &                                                 &                                         &                         &                          \\
\hline
$\Sigma_{b}^{*+}$               &  $\frac{3}{2}^{+} $    &     1        &      0       &        $1^{+}$           & $5829.0_{-1.8}^{+1.6}$$_{-1.8}^{+1.7}$ \cite{sigma-b}                        & $5.81\pm0.19$\cite{jrzhang}          & 5834                                            &   5850                                  &    5871                 & $5835_{-77}^{+82}$       \\
$\Sigma_{b}^{*-}$               &                        &              &              &                          & $5836.4\pm2.0_{-1.7}^{+1.8}$ \cite{sigma-b}                                  &                                      &                                                 &                                         &                         &                          \\
\hline
 $\Sigma_{c}(2455)^{0}$         &  $\frac{1}{2}^{+} $    &     1        &      0       &        $1^{+}$           & $2453.76\pm0.18$ \cite{PDG}                                                  & $2.40\pm0.31$\cite{jrzhang}          & 2439                                            &   2453                                  &    2452                 & $2411_{-81}^{+93}$       \\
\hline
 $\Sigma_{c}(2520)^{0}$         &  $\frac{3}{2}^{+} $    &     1        &      0       &        $1^{+}$           & $2518.0\pm0.5$ \cite{PDG}                                                    & $2.56\pm0.24$\cite{jrzhang}          & 2518                                            &   2520                                  &    2538                 & $2534_{-81}^{+96}$       \\
 \hline
 $\Lambda_{c}(2593)^{+}$        &  $\frac{1}{2}^{-} $    &     0        &      1       &        $1^{-}$           & $2595.4\pm0.6$ \cite{PDG}                                                    & $2.53\pm0.22$                        & 2598                                            &                                         &                         &                          \\
\hline
 $\Lambda_{c}(2625)^{+}$        &  $\frac{3}{2}^{-} $    &     0        &      1       &        $1^{-}$           & $2628.1\pm0.6$ \cite{PDG}                                                    & $2.58\pm0.24$                        & 2628                                            &                                         &                         &                          \\
 \hline
 $\Xi_{c}^{0}$                  &  $\frac{1}{2}^{+} $    &     0        &      0       &        $0^{+}$           & $2471.0\pm0.4$ \cite{PDG}                                                    & $2.48\pm0.21$                        & 2481                                            &   2468                                  &    2473                 & $2432_{-68}^{+79}$       \\
\hline
 $\Xi_{c}'^{0}$                 &  $\frac{1}{2}^{+} $    &     1        &      0       &        $1^{+}$           & $2578.0\pm2.9$ \cite{PDG}                                                    & $2.50\pm0.29$                        & 2578                                            &   2580                                  &    2599                 & $2508_{-91}^{+97}$       \\
\hline
 $\Xi_{c}(2645)^{0}$            &  $\frac{3}{2}^{+} $    &     1        &      0       &        $1^{+}$           & $2646.1\pm1.2$ \cite{PDG}                                                    & $2.64\pm0.22$                        & 2654                                            &   2650                                  &    2680                 & $2634_{-94}^{+102}$      \\
\hline
 $\Xi_{c}(2790)^{0}$            &  $\frac{1}{2}^{-} $    &     0        &      1       &        $1^{-}$           & $2791.9\pm3.3$ \cite{PDG}                                                    & $2.65\pm0.27$                        & 2801                                            &                                         &                         &                          \\
\hline
 $\Xi_{c}(2815)^{0}$            &  $\frac{3}{2}^{-}$     &     0        &      1       &        $1^{-}$           & $2818.2\pm2.1$ \cite{PDG}                                                    & $2.69\pm0.29$                        & 2820                                            &                                         &                         &                          \\
\hline
 $\Omega_{c}^{0}$               &   $\frac{1}{2}^{+}$    &     1        &      0       &        $1^{+}$           & $2697.5\pm2.6$ \cite{PDG}                                                    & $2.62\pm0.29$                        & 2698                                            &   2710                                  &    2678                 & $2657_{-99}^{+102}$      \\
\hline
 $\Omega_{c}(2768)^{0}$         &  $\frac{3}{2}^{+} $    &     1        &      0       &        $1^{+}$           & $2768.3\pm3.0$ \cite{PDG}                                                    & $2.74\pm0.23$                        & 2768                                            &   2770                                  &    2752                 & $2790_{-105}^{+109}$     \\
\hline
 $\Lambda_{1b}$                 &  $\frac{1}{2}^{-} $    &     0        &      1       &        $1^{-}$           &                                                                              & $5.85\pm0.15$                        & 5930                                            &                                         &                         &                          \\
\hline
 $\Lambda_{1b}^{*}$             &  $\frac{3}{2}^{-} $    &     0        &      1       &        $1^{-}$           &                                                                              & $5.90\pm0.16$                        & 5947                                            &                                         &                         &                          \\
\hline
 $\Xi_{b}^{0}$                  &  $\frac{1}{2}^{+} $    &     0        &      0       &        $0^{+}$           &  $5792.9\pm2.5\pm1.7$ \cite{ksi-b1}                                          & $5.75\pm0.13$                        & 5812                                            &  5810                                   &    5788                 & $5780_{-68}^{+73}$       \\
\hline
 $\Xi_{b}^{'}$                  &  $\frac{1}{2}^{+} $    &     1        &      0       &        $1^{+}$           &                                                                              & $5.87\pm0.20$                        & 5937                                            &  5950                                   &    5936                 & $5903_{-79}^{+81}$       \\
\hline
 $\Xi_{b}^{'*}$                 &  $\frac{3}{2}^{+} $    &     1        &      0       &        $1^{+}$           &                                                                              & $5.94\pm0.17$                        & 5963                                            &  5980                                   &    5959                 & $5929_{-79}^{+83}$       \\
\hline
 $\Xi_{1b}$                     &  $\frac{1}{2}^{-} $    &     0        &      1       &        $1^{-}$           &                                                                              & $5.95\pm0.16$                        & 6119                                            &                                         &                         &                          \\
\hline
 $\Xi_{1b}^{*}$                 &  $\frac{3}{2}^{-}$     &     0        &      1       &        $1^{-}$           &                                                                              & $5.99\pm0.17$                        & 6130                                            &                                         &                         &                          \\
\hline
 $\Omega_{b}$                   &  $\frac{1}{2}^{+}$     &     1        &      0       &        $1^{+}$           &                                                                              & $5.89\pm0.18$                        & 6065                                            &  6060                                   &    6040                 & $6036\pm81$              \\
\hline
 $\Omega_{b}^{*}$               &  $\frac{3}{2}^{+} $    &     1        &      0       &        $1^{+}$           &                                                                              & $6.00\pm0.16$                        & 6090                                            &  6090                                   &    6060                 & $6063_{-82}^{+83}$       \\
\hline\hline
\end{tabular}} \label{table:2}
\end{table}

In conclusion, we have employed the QCD sum rule approach to
calculate the masses of charmed and bottom baryons including the
contributions of the operators up to dimension six in OPE. The final
results extracted from the sum rules are well compatible with the
existing experimental data. Predictions to the spectroscopy of the
unobserved bottom baryons are also presented. Although there has
been enormous progress in experimental aspects and many theoretical
works have been done for the heavy baryons, plenty of problems are
desiderated to resolve. It is worth elucidating that most of the
$J^{P}$ quantum numbers for the heavy baryons have not been
determined experimentally, but are assigned by PDG on the basis of
quark model predictions, which are waiting for experimental
identification, especially for several higher excited states. More
data on bottom baryons are earnestly expected by putting into
operation the Large Hadron Collider, which may supply a gap of
experimental data in the near future. From the theoretical point of
view, it might be meaningful to reanalyze the QCD sum rules in full
theory for the heavy baryons, taking into account the QCD
$O(\alpha_s)$ corrections to improve the results, which are not
involved in this work.

\begin{acknowledgments}
J.R.Z. is very grateful to Marina Nielsen for communications and helpful discussions.
M.Q.H. would like to thank the Abdus Salam ICTP for warm hospitality.
This work was supported in part by the National Natural Science
Foundation of China under Contract No.10675167.
\end{acknowledgments}


\end{document}